\documentclass{article}

\usepackage{graphicx,vmargin,xspace}
\setpapersize{USletter}
\def\deg{$^{\circ}\,$}

\def\s{\small \bf Planned contents: \\ }

\def\solm{M$_{\odot}\,$}
\def\msun{M$_{\odot}\,$}
\newcommand{\etal}{{\it et~al.}\xspace}
\newcommand     {\atx}  [1]     {{\bf #1}}

\newcommand     {\rtx}  [1]     {{\small{\it #1}}}

\newcommand     {\keywords}  [1]     {{\bf #1}}

\begin{document}

\title{Modeling W44 as a Supernova Remnant in a Density Gradient, with a 
Partially Formed Dense Shell and Thermal Conduction in the Hot Interior}

\author{R. L. Shelton$^{1,2}$, Donald P. Cox$^{1}$,  Witold 
Maciejewski$^{3}$, 
Randall Smith$^{1,4,5}$, \\ Tomasz Plewa$^{6,7}$, Andrew Pawl$^{1}$, and 
Micha\l\ 
R\'o\.zyczka$^{7,8}$}

\maketitle

\begin{abstract}

We show that many observations of W44, a supernova 
remnant in the galactic plane at a distance of about 2500 pc, are 
remarkably consistent with the simplest realistic model.  The model 
remnant is evolving in a smooth ambient medium of fairly high 
density, about 6 cm$^{-3}$\ on average, with a substantial density 
gradient.  At the observed time it has an age of about 20,000 years, 
consistent with the age of the associated pulsar, and a radius of 
11 to 13 pc.  Over most of the outer surface, radiative cooling has 
become important in the post shock gas;  on the denser end there has 
been sufficient compression of the cooled gas to develop a very thin 
dense half shell of about 450 \msun, supported against further 
compression by nonthermal pressure.  The half shell has an expansion 
velocity of about 150 km s$^{-1}$, and is bounded on the outer 
surface by a radiative shock with that speed.  

The deep interior of the remnant has a substantial and fairly uniform 
pressure, as expected from even highly idealized adiabatic models;  
our model, however, is never adiabatic.  Thermal conduction, while 
the remnant is young and hot, reduces the need for expansion 
cooling, and prevents formation of the intensely vacuous 
cavity characteristic of adiabatic evolution.   It radically alters the interior
structure from what one might expect from familiarity with the Sedov 
solution.  At the time of 
observation, the temperature in the center is about 6$\times 10^6$\ 
K, the density about 1 cm$^{-3}$.  The temperature decreases 
gradually away from the center, while the density rises.  Farther out 
where cooling is becoming important, the pressure drops 
precipitously and the temperature in the denser gas there is quite 
low.  Our model is similar to but more comprehensive than the recent one 
by Harrus \etal (1997).  Because their model lacked thermal conduction, 
ours is more successful in 
providing the thermal x-rays from the hot interior, including
a better match to the spectrum, but neither 
provides the sharpness of the central peaking without further complications.  

By using a 2d hydrocode to follow the evolution in a density gradient, 
we are able to verify that the spatial and velocity structure of the H{\small I} 
shell are a good match to the observations, without the complications 
suggested by Koo and Heiles (1995), and to demonstrate that the remnant's 
asymmetry does not substantially affect the distribution of x-ray 
emitting material.  A 1d hydrocode model is then used to explore the
effects of nonequilibrium ionization on the x-ray spectrum and intensity.
We calculate the radio continuum 
emission expected from the compression of the ambient magnetic 
field and cosmic rays into the dense shell (the van der Laan 
mechanism, 1962a) and find it to be roughly consistent with observation,
though the required density of ambient cosmic ray electrons is about
4 times greater than that estimated for the solar neighborhood.
 We estimate 
the optical emission that should be present from fluorescence of 
UV, emitted by the forming shell and the radiative shock and absorbed 
in the cold shell and the ambient medium, and the associated 
63 $\mu$m [OI] emission.  Both are in agreement with the
intensity and spatial structures found in recent observations.  Neither
requires interaction with a dense molecular cloud for its generation.  We
calculate the gamma rays that should be emitted by cosmic ray electrons
and ions in the shell, interacting with the cold material, and find each
capable of generating about 25\% of the flux 
reported by EGRET for the vicinity.

We provide several analytic tools for the assembly of models of 
this type.  We review the early evolution and shell formation analyses 
and their generalizations to evolution in a density gradient.  We also calculate the 
density and temperature that should be present in the hot interior of 
a remnant with thermal conduction.  We supply the van der Laan 
mechanism in a particularly useful form for the calculation of radio 
continuum from radiative remnants. Finally, we demonstrate a simple 
technique for estimating the optical emission expected.  These tools 
are employed to choose parameters of models which we then explore with our 
1d and 2d hydrocodes, providing, respectively, 
the detailed x-ray spectra and dynamical characteristics.

\end{abstract}


affiliations:\\
1: Department of Physics, University of Wisconsin-Madison, 1150 University 
Ave., Madison, WI 53711 USA\\
2:  Laboratory for High Energy Astrophysics, NASA/Goddard Space Flight 
Center,
Greenbelt, MD 20771\\
3: Department of Astronomy, University of Wisconsin-Madison, 750 Charter St., 
Madison, WI 53711 USA\\
4:  Laboratory for Astronomy and Solar Physics, Code 685, 
NASA/Goddard Space FlightCenter, Greenbelt, MD 20771\\
5: Harvard-Smithsonian Center for Astrophysics, 60 Garden St, Cambridge, MA 02138\\
6: Max-Planck-Institut f\"ur Astrophysik, Karl-Schwarzschild-Strasse, 85-740 
Garching, GERMANY\\
7: N. Copernicus Astronomical Center,  Bartycka 18, 00-716 Warszawa,
POLAND\\
8: also, Member, Interdisciplinary Center for Mathematical Modelling,  
Pawi\'nskiego 5a, 02-106 Warszawa, POLAND\\

e-mail addresses:\\
RLS = shelton@spots.gsfc.nasa.gov\\
DPC = cox@wisp.physics.wisc.edu\\
WM = witold@uwast.astro.wisc.edu\\
RKS = rsmith@kracko.harvard.edu\\
TP = tomek@camk.edu.pl\\
MR = mnr@camk.edu.pl\\

\section{Introduction}
\label{section:Intro}

W44 (3C392) is a supernova remnant (SNR), 
located near the Galactic plane ($l=34.7$\deg, $b=-0.4$\deg), which
has a centrally peaked distribution of thermal x-ray surface brightness, 
coupled with
a shell-like radio continuum morphology.
This combination makes it a  member of 
the SNR class which also includes
W28, 3C400.2, Kes 27, MSH 11-61A, 3C391, and CTB1 (Rho 1995).

In this paper we demonstrate that many observed features of this 
remnant are consistent with the simplest realistic model, a remnant 
evolving in a fairly smooth ambient medium of moderate density with a 
substantial density gradient, and with thermal conduction active in its 
interior.  It has a radiative shock and dense shell on one side which is 
emitting copious amounts of optical and IR radiation, and is sweeping 
up and compressing ambient cosmic rays and magnetic field to produce 
radio synchrotron and gamma rays.  This interpretation differs substantially 
from earlier attempts to model the remnant's x-ray emission or 21 
cm characteristics as arising from a population of evaporating clouds 
in the interior or the disruption and acceleration of a wind blown shell, 
but is similar in spirit to the recent investigation by 
Harrus \etal (1997).  Our model differs from the latter in that it 
includes thermal conduction within the hot interior of the remnant, 
assesses the effects of the density gradient, and is considerably more 
detailed dynamically and more ambitious in comparison with 
observations.  Our interpretation also differs from past efforts to understand
the radio continuum, the infrared emission, and associated OH masers:  we
find no evidence for interaction of the remnant with dense molecular clouds 
in the vicinity.

Section \ref{section:Obs} opens with a discussion of the complexity 
of the region in which W44 is located, followed by a revision in 
the estimated distance (using a recent rotation curve and the
standard galactocentric distance of the Sun).  It then
surveys the observations of W44, including 
radio continuum, 21 cm, \mbox{X-rays}, characteristics of the pulsar 
and its weak nonthermal nebula, optical, IRAS and ISO observations, 
magnetic field measurements via polarization of OH masers, and the
gamma rays seen with EGRET. 

Section \ref{section:AnalMod} supplies the analytic 
tools for estimating remnant parameters consistent with the 
observations.  It reviews the Sedov and 
cooling phases for uniform density, provides a means for estimating 
the effects of thermal conduction on the characteristics of the 
remnant interior, surveys the anticipated shell characteristics, 
introduces the radio continuum expected from the van der Laan 
mechanism, and provides simple tools for estimating the 
H$\alpha$, 63 $\mu$m, and gamma ray intensities.  It then shows how,
using the approximations of Maciejewski and Cox (1998), the
remnant parameters can 
be estimated in the presence of a density gradient, closing with a 
summary of what can be learned from analytical modeling alone.

Section \ref{section:PhysAss} presents the physical assumptions of 
the hydrocode modeling and describes their implementation and 
justification.  This includes the use of a nonthermal pressure term, 
the thermal radiation emissivity and spectrum, and the assumptions 
regarding dust, thermal conduction, and synchrotron emission.

Section \ref{section:2d} presents the results of 2d modeling,  
exploring the density, temperature and pressure structures of the 
hot interior and the mass, velocity, and nonthermal emissivity structures 
of the dense shell.  Extensive comparison is made with both the 21 cm 
and radio continuum observations.  The non-equilibrium ionization 
could not be followed, but we verified that the structure in the 
hot interior is nearly spherically symmetric and should be reasonably 
approximated by the 1d model, which is able to
track the ionization of the plasma.  

Section \ref{section:1d} presents the results of 1d modeling, aimed 
principally at obtaining the x-ray surface brightness distribution and 
its spectrum, and comparison with those observations.

Finally, Section \ref{section:Discussion} provides the discussion and 
conclusions.

\section{Survey of Observations}
\label{section:Obs}

The essential observed properties of W44 itself are summarized in 
Table \ref{obstable} at the end of this section. Detailed description 
beyond the survey below can be found in recent papers by 
Giacani \etal (1997), de Jager \& 
Mastichiadis (1997) and Harrus \etal (1997).

\subsection{Environment}
\label{section:Environ}

W44 lies in a complex region of the sky, in close proximity to 
several molecular clouds (Sato, 1986), in a giant molecular cloud 
(GMC) complex (Dame 1983, Dame \etal 1986), and at the base of 
the Aquila Supershell (Maciejewski \etal 1996). 
The ambient density in which the remnant evolves is fairly high 
(averaging about 6 cm$^{-3}$, from our results below), presumably 
because of its GMC environment.  Filamentation of the surface emission
shows that the medium is moderately irregular on scales somewhat 
smaller than the remnant.  It has been suggested that the 
remnant may be impacting directly on a very dense molecular cloud; the 
interpretation of the clearly associated OH masers by
Claussen, \etal (1997), for example, is based on that 
assumption.  Our model does not include this possible interaction and 
appears not to require it to understand any of the other observations.  
Considering the superbubble and its funnel-like connection to the 
GMC complex in the galactic plane, it is tempting
to suppose that W44 might be the latest of a series of SNe in the 
general region, the earlier ones having contributed energy to the superbubble.  
Thus far, there is no evidence for any transport of energy from 
W44 except by cooling radiation and, possibly, escaping cosmic rays;
 its high density environment has 
quite effectively localized the mechanical and thermal components of its 
energy. 
Any contribution it may make to the superbubble lies in the future.  

\subsection{Distance Determination}
\label{section:Dist}

In order to estimate the distance to the supernova remnant,
Caswell \etal 1975 measured the 21 cm absorption profile 
toward W44.  They found components at velocities of $+12, +22, 
+30$, and $+42$ km sec$^{-1}$.  (In confirmation, Goss, Caswell, 
and Robinson (1971) observed 1667 MHz OH absorption 
components at $+12.5, +22, +30$,
and $+42$ km sec$^{-1}$.)  Because there are no higher velocity
components, the $+42$ km sec$^{-1}$ component corresponds to
the near kinematic distance.  Using the Schmidt (1965) galactic
rotation curve and 10 kpc as the distance between the Earth and the
galactic center, Caswell \etal estimated the distance to W44 as 3 kpc.
We have recalculated the distance using the Clemens (1985) galactic
rotation curve and 8.5 kpc as the distance between the Earth and
the galactic center.  The new estimate is 2.5 to 2.6 kpc.

\subsection{Radio Continuum Observations}
\label{section:RadContObs}

W44 appears as an elongated shell-type remnant
in radio continuum maps.  The size, measured from the Giacani \etal
 (1997) 1442.5 MHz image is $25^{'} \times 35^{'}$. In the less 
detailed map from the 11 cm survey of Reich \etal (1990), the 
remnant appears noticeably brighter over the eastern half. 

The radio continuum image presented by Jones, Smith, \& Angellini 
(1993) shows a very highly filamentary structure indicative of a 
severely edge brightened distribution of emission. The filaments are
concentrated on the eastern side and are contained within a sharp 
boundary, while the western side contains only one very bright
feature and has a muted boundary. 

Kundu and Velusamy (1972) show that W44 has a very regular
linear polarization pattern.  In the brighter half
of the source, the polarization reaches $20\%$, and 
the magnetic field vector is almost parallel 
to the long axis of the remnant.  The bright filaments tend to
lie along this axis as well.

The flux densities have been measured over a wide range of radio 
frequencies (e.g. by Kundu and Velusamy, 1972; 
Dickel and DeNoyer, 1975; Clark and Caswell, 1976; 
Clark, Green, and Caswell, 1975; Kassim, 1992).  The spectral index is
about -0.33 (Kovalenko \etal 1994) to -0.4 (Giacani, \etal 1997), 
the latter reporting a flux density of about
400 Janskys at 330 MHz.  Table \ref{obstable} shows 500 Janskys at 
100 MHz as representative.

\subsection{H{\small I} 21 cm Emission from W44}
\label{section:HIObs}

Koo \& Heiles (1995) have made a 21 cm velocity resolved map of W44
and found emission up to $v_{\rm LSR}$ = 210 km s$^{-1}$. By
subtracting off the systemic velocity of the clouds in this region
($\sim$43 km/s, Sato 1986), they were able to estimate the shell's
expansion velocity as $150 \pm 15$ km s$^{-1}$.  The observations at
low velocity near the limb were badly confused with ambient material,
leaving an accessible velocity range only above 130 km s$^{-1}$\ LSR.
The atomic hydrogen mass in the shell integrated over this range
was about 73 \msun. Thus, if the neutral material is confined to a
thin shell, the average H{\small I} column density of the receding
side of the remnant is about $3 \times 10^{19}$ atoms cm$^{-2}$.  Koo
and Heiles searched for the near side of the SNR, but found no
emission with $v_{\rm LSR} < -70$ km s$^{-1}$.

By fitting the profile of an expanding shell to their
position-velocity diagrams, Koo and Heiles (1995) estimated the full
angular size of the H{\small I} distribution from their observations
over a restricted velocity range.  They concluded that the H{\small I}
distribution is somewhat smaller than and interior to the radio
continuum shell.  This conclusion will be disputed in section
\ref{section:2dHI} below where the spatial distribution and
position-velocity diagrams of the H{\small I} data are compared
directly with our model.

\subsection{X-ray Observations}
\label{section:XObs}

The ROSAT PSPC (Rho \etal, 1994) and {\it{Einstein}} IPC
(Smith \etal, 1985) x-ray maps show centrally concentrated
thermal emission whose dim outer edge extends as far as the radio
continuum
shell in the north, south, and west (Rho \etal 1994).  (This
appearance, however, may be due in part to scattering of the x-rays by
the substantial column density of dust along the line of sight.)  The
X-ray emitting region may similarly extend to the eastern edge where
it is absorbed by a foreground molecular cloud which covers
irregularly from $\alpha(1950) \stackrel{\sim}{=} 18^h53^m40^s$ to
$18^h54^m55^s$ and $\delta(1950) \stackrel{\sim}{=}
01^{\rm{o}}10^{\rm{'}}$ to $01^{\rm{o}}20^{\rm{'}}$ (Wootten, 1977).
Rho \etal (1994) separately analyzed 11 spatial zones in the
ROSAT PSPC image.  From fitting the pulse height distribution with
equilibrium ionization thermal models, they found values for the
temperature of the X-ray emitting gas from 3.9 to 7.6$\times 10^6$ K.
The fitted values for the column density of intervening absorbers
ranged from 1.6 to 2.1 $\times 10^{22}$\ cm$^{-2}$.  From the spatial
distribution of these results they conclude that there is no
significant radial trend in temperature.  Working from the fitted
results is somewhat misleading, however, as Figure 7 of
Rho \etal (1994)  clearly
shows that the spectral hardness is significantly greater in the center
where the x-rays are brightest.

Smith \etal (1985), Jones \etal (1993),
and Rho \etal (1994) calculated the x-ray luminosity
for the remnant interior from the {\it{Einstein}}, EXOSAT, and 
ROSAT data, respectively.  Their estimates should be 
reduced by about 25 to 30$\%$ at our revised distance.
The implied emission measure within the 
remnant requires an rms electron density of about 0.4 to 0.75 cm$^{-3}$.

The details of the 
observed x-ray surface brightness distribution and spectrum 
are presented in 
section \ref{section:1d} where they are compared with the 
results of our 1d model.

\subsection{The Pulsar and its Synchrotron Nebula}
\label{section:PulsarObs}

An apparently associated 267 millisecond pulsar (PSR B1853+01) 
lies projected on the southern part of the remnant (Wolszczan, 
Cordes, and Dewey 1991).  Its age ($\sim$20,000 years 
$\pm$20\%) provides an estimate for that of W44, and is typical of 
radiative remnants.  The pulsar powers a small synchrotron nebula, 
observed in 4.0 to 9.5 keV ASCA data (Harrus, Hughes, and 
Helfand, 1996), and in 4860 and 8440 MHz VLA data (Frail, 
Giacani, Goss, and Dubner, 1996).  The  x-ray 
flux from synchrotron is much smaller than the thermal component.

\subsection{Optical}
\label{section:OptObs}

Giacani \etal (1997) have taken H$\alpha$ and [S{\small II}] images of 
W44.  All observed
optical emission is within the boundaries of the radio shell, but
at the sensitivity of the observation is not found everywhere within it.  
There appears to be noticeably higher foreground extinction in the south.  
The brightest H$\alpha$ features have a surface brightness 
$\sim 1 \times 10^{-16}$ ergs cm$^{-2}$ s$^{-1}$ arcsec$^{-2}$ 
while the faintest features visible in the image have about a tenth that.
The brightness of the [S{\small II}] lines is not reported but the exposure time
was the same and the image showed comparable structure leading us to 
believe that the intensity is similar.  Much of the optical emission is 
filamentary and in excellent correlation with radio continuum filaments.  
In some places there is fairly bright H$\alpha$ that is smoother and not well
correlated with the radio.  On the eastern half of the remnant, where the
radio is generally brighter, the observable optical emission is much lower.
The H$\alpha$ features correlate with [S{\small II}] emission, which often traces 
radiative shocks.  Conversely, the spatial correlation between the 
optical and x-ray emission is poor, even on 
very large scales, differing entirely from what would be expected if  
the H$\alpha$ and [S{\small II}] emission arose from evaporating clouds 
which were loading the interior with mass and enhancing the density 
of hot gas.

\subsection{Infrared, OH Masers, and Gamma Rays}
\label{section:IRObs}

Giacani \etal (1997) show IRAS observations in 60 and 100 $\mu$m in a 
large field around W44, which they say indicate emission compatible with 
an origin in shock heated dust, while ISO observations presented by Reach 
and Rho (1996) indicate that, in all fields they examined, the continuum is 
characteristic of dust heated by the diffuse interstellar radiation field and is 
dominated by pre-shock and unrelated clouds.  The IRAS map, however, 
shows weaker emission interior to the northern and mid-southern sections 
of the remnant not explored by ISO, that may well arise from hot dust 
within the remnant. (The ISO sampling crossed the southern edge of the 
northern enhancement, and Reach and Rho report that the continuum is 
brighter there and peaked at the edge where there is a radio ridge, but make 
no further comment about the heating mechanism.)

Reach \& Rho (1996) also observed bright [O{\small I}] 63$\mu$m emission, 
particularly enhanced on the edge of the remnant where their cut 
crossed a radio filament, but everywhere higher within the remnant than 
outside.  They quote their peak surface brightness as 
$10^{-3}$ erg cm$^{-2}$ sr$^{-1}$ (presumably also s$^{-1}$), and 
estimate the total luminosity of the remnant in this line at $\sim 1000 $ 
L$_{\rm \odot}$.  This line is an important coolant in shocked gas, 
but after an unsuccessful attempt to model it as shock emission, they 
interpreted it as evidence that W44 is interacting with a dense
molecular cloud, even though the region explored does not appear from the 
IRAS maps to be abutting a dense cloud, being well northwest of that 
explored by Wootten (1977), for example.  In Section 
\ref{section:Alphaetc}, we show that the 
[O{\small I}] 63$\mu$m emission is in fact quantitatively consistent with our 
model, 
involving no interaction with dense molecular material.

As mentioned previously, the OH masers observed by 
Claussen \etal (1997), were also interpreted 
as being collisionally excited by  the SNR shock passing through a 
molecular cloud.  They point out that DeNoyer (1983) contradicted
the early evidence for such interaction, but then cite the [O{\small I}] observations
just discounted as the strongest independent evidence for it.

Claussen \etal (1997) were further 
influenced by the presence of an unidentified EGRET gamma ray
source error circle overlapping the eastern limb of W44, suggesting
that it may be due to cosmic rays from W44 interacting with an abutting 
dense cloud.  The Wooten cloud is in fact projected on that location and 
Esposito \etal (1996) show that the source brightness, 
$50 \times 10^{-8}$\ photons cm$^{-2}$\ s$^{-1}$\ above 100 MeV, 
would require a cosmic ray intensity within the
cloud nearly 100 times ambient, even if the entire cloud is being irradiated.

\begin{table}
\begin{center}
\begin{tabular}{|c|c|c|c|}
\hline \hline 
Characteristic & Value & Observation & Notes            \\ \hline \hline 
SNR Angular Size & 25$^{\rm{'}} \times 35^{\rm{'}}$ & Radio Continuum &   
\\
Observed Shape  & Pearlike      & Radio Continuum &     \\ 
Elongation & Northwest-Southeast &  &   \\ 
Evolutionary Stage & Partial Shell Formation & H{\small I} Emission  & 1 \\ 
Model Axis & Northeast-Southwest &  & 2 \\ 
Distance     & 2.5 to 2.6 kpc & H{\small I}, OH Absorption & 3 \\
Radius & $\sim$\ 11 to 13 pc   &                      &        \\
Age          & $\sim 20,000$ years & Pulsar               &        \\ 
         &            &                      &        \\
Radio Continuum Flux Density & $\sim 500$ Jansky @  100 MHz &   &    \\ 
Spectral Index & $\sim$\ 0.33 to 0.4 & Radio Continuum &   \\ 
Shell Expansion & 150 $\pm 15$ km sec $^{-1}$ & H{\small I} Emission &    \\
Shell N$_{\rm{H}}$ & $\sim 3 \times 10^{19}$ cm$^{-2}$   & H{\small I} 
Emission   &    \\
         &            &                      &        \\
Optical Configuration & filamentary, within radio remnant & H$\alpha$, [S{\small 
II}]  &    \\ 
Peak H$\alpha$ surface brightness &  
     $\sim 1 \times 10^{-16}$ ergs cm$^{-2}$ s$^{-1}$ arcsec$^{-2}$ &     &   \\ 
Peak 63 $\mu$m surface brightness & $10^{-3}$ erg cm$^{-2}$ sr$^{-1}$ s$^{-
1}$ 
          & ISO &    \\
63 $\mu$m luminosity & $\sim 1000 $ L$_{\rm \odot}$   &       &    \\
         &            &                      &        \\
Gamma-Ray Source Box & includes eastern limb &  &    \\ 
Gamma-Ray Flux above 100 MeV &  
     $50 \times 10^{-8}$\ photons cm$^{-2}$\ s$^{-1}$\  &  EGRET   &   \\ 
         &            &                      &        \\
X-Ray Morphology & Centrally Bright, No Shell & 0.2 to 4 keV X-ray & 4 \\ 
                 & Extends Dimly to Radio Boundary  &  & 5 \\ 
X-Ray Emission Mechanism &  Thermal         & Observable Lines   &        \\
Central Temperature & $\sim 4 - 8 \times 10^6$ K & X-ray &        \\  
Interior $n_{\rm e}$     & $\sim 0.4$ to $0.75$ cm$^{-3}$ &  X-ray   & 6       \\
Central Pressure      & $\sim 1.4 \times 10^{-9}$ dyn cm$^{-2}$           &  X-ray   
&        \\
Absorption N$_{\rm{H}}$ & $\sim 1.6 - 2.1 \times 10^{22}$ cm$^{-2}$  
& X-ray  &      \\ 
         &            &                      &        \\
Pulsar Nebula & Unimportant in 0.2 to 2 keV & 4 - 9.5 keV, 20 cm & \\ \hline 
\hline
\end{tabular}
\end{center}
\caption{Quantities derived from the observations: 
\label{obstable}
Note 1:  The dense shell formation is more fully developed 
on the northeast rear section; 
Note 2:  The model axis is assumed perpendicular to the elongation in 
radio continuum emission because the shell is partial; 
Note 3:  This is a new distance estimate.   
Note 4:  Harrus \etal 1996 present a 0.5 to 4.0 keV image which shows
the central brightening, as does Rho \etal's (1994) ROSAT data. 
Note 5:  The dim envelope of the x-ray emission extends to the boundary in most 
directions.
Note 6:  The reported electron densities have been scaled by 
1.07 to reflect a change in the assumed distance from 3000 pc to 2600 
pc.}
\end{table}

\section{Analytic Modeling}
\label{section:AnalMod}

\subsection{Overview of the The Basic Model}
\label{section:OverviewBasic}

In combination, the radio continuum and 21 cm observations provide 
a persuasive qualitative picture of this object.  The strong 
filamentation of the radio continuum is characteristic of that from the 
very thin dense shell on the surface of a radiative remnant.  The radio 
continuum maps, with their concentration toward (and sharp edge on) 
the east, and elongation southeast to northwest, give the impression of 
an emitting surface that is incomplete, somewhat like half an egg shell 
with the open end roughly toward the west or southwest.  The 21 
cm emission is also suggestive of an incomplete shell, with the 
observable part in recession.  

For our model, we adopt this qualitative picture, assuming that 
evolution is occurring in a medium that is denser on the northeast 
far side, that shell formation has occurred, but only on the 
denser end, and that the filamentation is  indicative of 
irregularity (corrugation) in the shell's surface, as 
explored for optical filamentation by Hester (1987).  A schematic of the model
is shown in Figure~\ref{schematic}.

Our choice of northeast to southwest, orthogonal to the observed major 
axis, for the projected axis of the density gradient requires a bit of 
explanation.  We had initially expected that the density gradient would 
provide the elongation of the remnant, and that we should be taking the 
gradient along the elongation.  When we became more familiar with the 
Kompaneets Approximation (Maciejewski and Cox, 1998), however, 
we discovered that in an exponential density distribution a remnant remains 
nearly spherical much longer than we had expected. 
It expands more rapidly in directions of lower density, but the effect is 
to displace the center of the sphere down the gradient from the explosion 
site, not to elongate the envelope.  Eventually we realized that a partial 
shell capping one end of a sphere could appear elongated, but that the 
elongation would be transverse to the density gradient.  For this reason 
we have adopted the observed semimajor axis as the remnant radius, 
and taken the model gradient orthogonal to it.

The remnant characteristics, summarized in Table \ref{obstable} can be 
combined in a variety of ways to demonstrate that W44 must indeed be a 
supernova remnant, with an explosion energy of about 10$^{51}$\ 
ergs.  For brevity we will adopt this result and from it derive the 
remnant characteristics assuming the age to be 20,000 years, the 
nominal age of the pulsar.  There are two free parameters, the density at 
the explosion site and the scale height of the density distribution.

Determination of these two parameters follows from three requirements: 
(1) that shell formation has occurred on the dense end and extended 
about half way  around the remnant to the equator, that (2) there be 
sufficient mass already in the half shell to explain the amount seen in 21 
cm, but (3) with a shell expansion velocity as large as that seen.

In the remainder of this section, we determine these parameters and make 
analytic predictions for the basic structure (the remnant radius, shell 
expansion velocity,  edge temperature, central pressure, central density 
and temperature, shell density and magnetic field), and for anticipated 
characteristics in observed emission modes (thermal \mbox{X-rays}, 
radio synchrotron, H$\alpha$, 63 $\mu$m, and gamma rays).

\subsection{Review of Shell Formation in a Homogeneous Medium}
\label{section:ShellForm}

In our notation, $n$\ and $n_{\rm e}$\ refer to the densities  of nuclei and 
electrons, respectively.  Then $n =  n_{\rm \scriptscriptstyle H} + 
n_{\rm \scriptscriptstyle He}  = 1.1 n_{\rm \scriptscriptstyle H}$, where  
$n_{\rm \scriptscriptstyle H}$\ and $n_{\rm \scriptscriptstyle  He}$\ are the 
densities of  hydrogen and helium  nuclei, and $n_{\rm \scriptscriptstyle  
He}/n_{\rm \scriptscriptstyle H} = 0.1$\ is assumed.  In  this notation, the mass 
density is $\rho = m n$,  where the average mass per nucleus is $m =  
(1.4/1.1) m_{\rm \scriptscriptstyle H}$,  $m_{\rm \scriptscriptstyle H}$\ being the 
mass of  hydrogen.  The perfect gas law, assuming equal electron  and ion 
temperatures is then $p = \chi n  {k_{\rm \scriptscriptstyle B}} T$, where $\chi 
=  (n+n_{\rm e})/n$.  For neutral gas, $\chi$\ is 1, for  fully  ionized it is 
$(2.3/1.1)$.

The Sedov self similar structure and evolution yield  several useful results 
for middle aged remnants if the effects of cooling have not yet been large.  
The  shock radius is 
$R_{\rm \scriptscriptstyle S} = [2.025  E_{\rm o} t^2 / (mn_{\rm o})]^{1/5}$ 
where $E_{\rm o}$\ is the explosion energy and $n_{\rm o}$\ 
is the ambient 
value of $n$, assumed uniform.  The shock velocity is  then 
$v_{\rm \scriptscriptstyle S} =  dR_{\rm \scriptscriptstyle S}/dt = (2/5)  
R_{\rm \scriptscriptstyle S}/t$.  From the jump  conditions, the post shock 
density is $4 n_{\rm o}$, the  pressure is $p_{\rm \scriptscriptstyle S} = [3/4] m 
n_{\rm o}  
v_{\rm \scriptscriptstyle S}^2$, and from the perfect gas  law in our notation, 
the post shock temperature is $T_{\rm \scriptscriptstyle S} =  (3/16) m 
v_{\rm \scriptscriptstyle S}^2 / (\chi  {k_{\rm \scriptscriptstyle B}})$.  In 
addition, the  pressure in the center of the remnant is  
$p_{\rm \scriptscriptstyle C} \approx 0.31  p_{\rm \scriptscriptstyle S}$.

Kahn (1975, 1976) introduced a very useful approximation technique for 
estimating the cooling timescale for supernova remnants.  In that 
approximation, the cooling coefficient takes the form $L = \alpha_{\rm Kahn} 
T^{-1/2}$. It has the special property that in the absence of thermal 
conduction, the time for a parcel of hot gas to cool from an initial state is 
independent of its history and equals
\begin{equation}
\label{e:Deltat}
\Delta t_{\rm cool} = \frac{p_{\rm \scriptscriptstyle S}}{Ln^2}.
\end{equation}
For a newly shocked parcel of gas, $p, T$, and $n$\ are known and thus, so is 
$\Delta t_{\rm cool}$.  To find the actual time $t_{\rm finish}$\ at which 
the parcel will finish cooling, the post shock cooling time must be added to 
the time $t_{\rm shock}$\ at which the parcel is first shocked ($t_{\rm 
finish} = t_{\rm shock} + \Delta t_{\rm cool})$.  To find the time at which 
the first parcel cools, the minimum is found for the function $t_{\rm 
finish}(t_{\rm shock})$.  This minimum is the cooling time for the remnant, $ 
t_{\rm cool}$.  Details appear in Cox and Anderson (1982) and Cox (1986).

The cooling time, however, should not be confused with the shell formation 
time $t_{\rm shell}$. The first gas which cools in a remnant does so rapidly. 
It drops to a very low pressure relative to its surroundings, and is only 
later compressed and repressurized by flows of cooling gas from both sides, 
these driven by the pressure differential. It is therefore not correct to 
assume that ``cooling" and ``compression" (i.e. dense shell formation) 
proceed simultaneously.  From experiments with hydrocodes, we estimate that  
$t_{\rm shell} \approx (7/6) t_{\rm cool}$.  The result, and numerical forms 
with $\alpha_{\rm Kahn} =  1.3 \times 10 ^{-19}$\ cgs, are
\begin{eqnarray}
          t_{\rm cool} & \approx & 0.208 \left[(m^{9/14}  E_{\rm o}^{3/2}) /   
[\alpha_{\rm Kahn}^{5/2} (\chi  {k_{\rm \scriptscriptstyle B}})^{5/4} n_{\rm 
o}^4] 
\right]^{1/7}\\
          t_{\rm cool} & \approx & 45.5 \times 10^3  \mbox{ years } 
E_{51}^{3/14} / n_{\rm o}^{4/7} \\
          t_{\rm shell} & \approx & 53 \times 10^3 \mbox{ years } 
E_{51}^{3/14} / n_{\rm o}^{4/7}.
\end{eqnarray}

We are now in a position to estimate the remnant's dynamical 
characteristics at the moment when shell formation reaches the equator, 
and to make an initial estimate of the ambient density, by solving for the 
epoch at which $t_{\rm shell}\!(E_{\rm o}, n_{\rm avg})$\ equals the remnant 
age, 
about 20,000 years.   Including the possibility that the age or 
explosion energy are probably not exactly our nominal values, the 
values for shock radius, shell velocity, and central pressure when the 
shell formation time equals the age, and the required average density are 
(with $t_{20} = t / (20,000 \mbox{ yr})$)

\begin{eqnarray}
          R_{\rm \scriptscriptstyle S} & \approx & 12 \mbox{ pc } 
                    {E_{51}}^{1/8}{t_{20}}^{3/4}\\
          v_{\rm shell} & \approx & 0.75 v_{\rm 
\scriptscriptstyle S} \approx 176 
                     \mbox{ km s}^{-1} 
                    {E_{51}}^{1/8}/{t_{20}}^{1/4}\\
          p_{\rm \scriptscriptstyle C} & \approx & 0.31 p_{\rm \scriptscriptstyle S} 
\approx 
                     1.5 \times 10^{-9} 
                     \mbox{ dyn cm}^{-2}
                    {E_{51}}^{5/8}/{t_{20}}^{9/4}\\
          n_{\rm o} & \approx & 5.5 \mbox{ cm}^{-3} 
                    {E_{51}}^{3/8}/{t_{20}}^{7/4}
\label{equations:shell}
\end{eqnarray}
These are the (approximate) anticipated conditions at the moment when
shell formation first occurs, a condition which we believe applies at
the equator of W44.  The remarkable fact is that even at the nominal
values of explosion energy and age, the radius and central pressure
are easily within the observational uncertainties.  The shell velocity
is somewhat high, but we expect a lower velocity toward the denser end
where the shell is more completely developed, and which in our
projection is closer to the line of sight.  Further insight derives
from evaluating the temperature that would have been at the edge of
the remnant had there been no cooling, considering it to be an
estimate of the temperature of hot gas at the outer edge of the hot
interior (just interior to the dense shell) at this epoch.  The result
is $T_{\rm outer} \approx 7.6 \times 10^5 \mbox{ K }
{E_{51}}^{1/4}/{t_{20}}^{1/2}$.  As found by Harrus \etal (1997), the
remnant edge at this epoch is somewhat too cool to provide a strong
shell source of X-rays.

There is another consistency check in the H{\small I} column density,
about $3 \times 10^{19}$\ cm$^{-2}$ from Table \ref{obstable}.  If all
the mass in a sphere is swept into a thin shell at the edge, the
perpendicular column on one side 
should be $n_{\rm o} R_{\rm \scriptscriptstyle
S}/3$.  This is about $7 \times 10^{19}$\ cm$^{-2}$ for our
parameters, a consistent value considering the uncertainty in the
observations, the recency and incompleteness of shell formation, and
the unknown fraction of the mass that is molecular.  If the observed
H{\small I} arises from any other cause, it would be exceedingly
unlikely for it to have the column density and expansion velocity
expected for the cooled SNR shell.  Making the details of the shell
mass work out correctly depends only on making the correct choice for
the density scale height, a topic we defer until after considering the
effects of thermal conduction on the central density and temperature
of the remnant.

\subsection{The Central Density and Temperature}

\label{section:TCtheory}

As we have suggested in the introduction and shown elsewhere (Slavin
and Cox, 1992, Cui and Cox, 1992, and Shelton, Smith, and Cox, 1995),
thermal conduction in the hot interior of a remnant prevents formation
of the vacant interior characteristic of nonconductive evolution.
Transporting heat away from the remnant interior lowers the specific
entropy and therefore the expansion rate needed to accommodate the
falling pressure, thereby moderating the extraordinary reduction in
central density found in models ignoring conduction.  The central
pressure remains uniform at about 0.3 of that at the edge, but the
temperature and density are also made nearly uniform as well.  In this
section we show that this effect can provide the conditions found in
the x-ray emitting interior of W44.

There is considerable concern, of course, that thermal conduction
might be very heavily suppressed, and/or anisotropic in supernova
remnants due to the magnetic field.  We have addressed this point
previously (Slavin and Cox, 1992, and Cox, 1996) but, in light of its
importance to our present application, discuss it somewhat further in
section \ref{section:CondImpl}.

Experience with hydrocodes has suggested that the central density
$n_{\rm \scriptscriptstyle C}$\ at the epoch of shell formation, in a
remnant with unsuppressed conduction, is roughly $0.1 n_{\rm o}$.  We
have previously shown (Shelton, Smith, and Cox, 1995) that this factor
of 10 density contrast between the interior and the ambient medium can
be understood analytically, and elaborate on that result now,
exploring its dependence on the magnitude of the thermal conductivity.

Cui and Cox (1992) developed a similarity solution for conduction that
made it possible to estimate a remnant's central adiabat, temperature,
and density analytically for times in the vicinity of the cooling
epoch.  There was unfortunately a typographical error; the density
exponent in their equations (A17) and (39) should be 208/225.  In
addition, the units are incorrectly described just before (39) but are
correct in their appendix.  The generalization of the coefficient B in
their equation (A14), to find the dependence on the magnitude of the
conductivity, is
\begin{equation}
  B = 29.2 \beta m^{12/25} n_{\rm o}^{208/225} / [(\chi {k_{\rm 
\scriptscriptstyle 
B}})^{7/2} E_{\rm o}^{26/75}]
\end{equation}
where the classical conductivity is taken as $\beta  T^{5/2}$, with $\beta 
\sim 6 \times 10^{-7}$\ cgs.

     With this result, and assuming the Kahn cooling function, the
ratio of the central to ambient density turns out to be independent of
explosion energy or ambient density.  The value and form are,
\begin{equation}
  n_{\rm \scriptscriptstyle C}/n_{\rm o} = 0.334 \, q^{18/71}(t_{\rm 
cool}/t)^{18/25} \, 
\left[ (t/t_{\rm cool})^{1/25} -  (0.44 q^{5/14})^{1/25}\right]^{18/71},
\label{equation:cox1} 
\end{equation}
with $q \equiv 145 m \alpha_{\rm Kahn} \beta/(\chi  {k_{\rm \scriptscriptstyle 
B}})^3$\ (defined so that $q = 1$\ for  full conduction and 
normal cooling rates).  For 
$t 
= (7/6) \, t_{\rm cool}$, as at shell formation, and $q = 1$, we  find 
$n_{\rm \scriptscriptstyle C}/n_{\rm o} \approx 0.13$, nicely in  agreement with 
hydrocode experience.  Equation  \ref{equation:cox1} probably does not 
successfully  describe time dependence, but should provide a  guide to the 
dependence on conductivity.  It predicts  the central 
density to be proportional to  $\beta^{0.253}$.  A factor of 10 reduction in  
conductivity should lead to only a factor of 1.8  reduction in the central 
density and increase in central temperature. 

With $n_{\rm o} \approx 5.5$\ cm$^{-3}$, and $q = 1$, we expect a
central density within the remnant cavity of $n_{\rm
\scriptscriptstyle C} \approx 0.7$\ cm$^{-3}$, just the level found
from the x-ray surface brightness of W44 (compare with the x-ray
results for the central density in Table \ref{obstable}).  As the
central pressure was shown in the previous section to be at the
correct level, the temperature will be appropriate also, roughly $7
\times 10^6$\ K.  We can therefore be confident that conduction will
lead to x-ray emission similar in characteristic temperature and
emission measure to that seen.  We defer to the hydrocodes, however,
to learn whether the spectral details in the presence of
nonequilibrium ionization, and the radial profile of the surface
brightness are also well approximated.

\subsection{The Radio Continuum}
\label{section:RC}

\subsubsection{Evaluation of The Contribution of the van der Laan Mechanism}
\label{section:vdL}

There is considerable interest in the possibility that the bulk of the cosmic 
rays arise from diffusive shock acceleration in supernova remnants, an 
interest which is encouraged by their powerful nonthermal emission 
(Ellison \etal 1994).  
Before assuming that diffusive shock acceleration plays the dominant role in 
an evolved
supernova remnant which is bounded by a radiative shock, however, it is 
important to consider the synchrotron emission that would arise from swept 
up ambient cosmic ray electrons, adiabatically 
compressioned into the dense shell and radiating in the 
compressed ambient magnetic field (van der Laan, 1962).  

 Neglecting small inefficiencies due to misalignment and
isotropization, a co-moving relativistic particle in a region
compressed by a factor $x$\ would have its energy betatron-enhanced
(2d adiabatic compression) by about a factor of $x^{1/2}$.  The net
effect for a supernova remnant with a cooled shell of mass $M$\ and
original density $\rho_{\rm o}$\ is that cosmic ray electrons
previously occupying a volume $M/\rho_{\rm o}$\ of interstellar space
find themselves compressed into a shell whose volume is $x$\ times
smaller, enhanced in their energies by a factor of $x^{1/2}$, and
radiating in a magnetic field about $x$ times higher.  The
characteristic frequency of synchrotron emission of an electron of
energy $E$\ in a field of strength $B$\ is proportional to $B E^2$,
and is therefore a factor of $x^2$\ higher.  Similarly, the total
power per electron, proportional to $(BE)^2$, is raised by a factor
$x^3$.  For a distribution of electrons radiating synchrotron with an
emission spectrum $A \nu^{-\alpha}$\ per electron before compression,
the spectrum per electron after compression is $A^{\prime}
\nu^{-\alpha}$, with $A^{\prime} = x^{1+2\alpha}A$, for a total gain
at fixed frequency of $G(x) = x^{1+2\alpha}$.

If one is willing to make assumptions about the isotropy of the particle 
distribution before, during, and after the acceleration, then this estimate 
can be made somewhat more precise.  For example, assuming an initially 
random distribution of pitch angles and the post acceleration 
re-randomization, the overall gain per electron from van der Laan (1962) is
\begin{equation}
G(x) = x^{1+\alpha} \left[\frac{1}{2} + \frac{1}{2}
\frac{x}{(x-1)^{1/2}} {\rm
sin}^{-1}(\frac{x-1}{x})^{1/2}\right]^{2\alpha}.
\label{equation:gainVdL} 
\end{equation}
Compared with the previous estimate, the numerical difference is slight 
(here $G(x) \approx (\pi/4)^{2\alpha}x^{1+2\alpha}$\ at large $x$). 

Therefore, except for the required units conversions, the flux density of the 
remnant's radio continuum from this mechanism is 
\begin{equation}
S(\nu) =   G(x) V_{\rm o} \epsilon_{\rm o}(\nu)/d^2, 
\label{equation:fluxVdL} 
\end{equation}
where $V_{\rm o} = M_{\rm shell}/\rho_{\rm o}$\ is the shell volume
before compression, $\epsilon_{\rm o}(\nu)$\ is the synchrotron
emissivity of the ambient ISM at frequency $\nu$, $x$ is the shell
compression factor, and $G(x)$\ is the gain as given above.  The next
few paragraphs assemble these parameters.

>From the observations summarized in Table \ref{obstable} and the
subsequent modeling, we have estimated the shell radius at 11 to 13
pc, the shell formation process complete
over half that surface, the one sided surface
density at $3 \times 10^{19}$\ cm$^{-2}$, and the initial density at 
6 cm$^{-3}$.  Thus, the mass of material now in the half shell
is about 300 M$_{\rm \odot}$\ and its initial volume about $V_{\rm o}
= 1600$ pc$^3$.

W44 has been studied between about 10$^2$ and 10$^4$ MHz.
Unfortunately, prior to their being swept up by the remnant, the
responsible electrons radiate only below 10 Mhz where their study is
difficult.  Thus, the ambient emissivity must be evaluated at the
lowest observable frequencies and then extrapolated into the frequency
range observed in the remnant.  Webber \etal (1980) and Rockstroh and
Webber (1978) conclude that the data available appear to be consistent
at the lowest frequencies with an ambient interstellar synchrotron
emissivity of roughly
\begin{equation}
\epsilon_{\rm o} \sim 1.5 \times 10^{-40}\,[\nu/30({\rm MHz})]^{-.57} \, 
\mbox{W(m}^3 
\mbox{sr Hz)}^{-1},
\label{equation:synchRate}
\end{equation}
with apparently real place to place 
variability.  Extrapolation to 
100 MHz then yields 
\begin{equation}
\epsilon_{\rm o}(\mbox{100 Mhz}) \sim 0.8  \times 10^{-40}\, 
\mbox{W(m}^3 
\mbox{sr Hz)}^{-1}
\label{equation:synchRate100} 
\end{equation}

The fact that the emission spectrum quoted above, with $\alpha =
0.57$, is steeper than the observed spectrum for W44 requires further
elaboration.  It is attached to the conclusion of Webber \etal (1980)
that the ambient electron particle index (1+2$\alpha$)
 is 2.14 between 70 and 2500
MeV.  But they further conclude that below 70 MeV the particle
spectrum flattens toward an index of 1.6.  In the shell of W44, the
emission at 100 MHz derives from electrons with energies about 250
MeV, but prior to compression into the shell the energies of the same
electrons would have been 7 to 10 times lower, or roughly 30 MeV.  As
these electrons are in fact in the flattened part of the spectrum, a
spectral index lower than 0.57 but greater than 0.3 (corresponding to
particle index 1.6) is expected.  Therefore the observed spectral
index of W44, 0.33 to 0.4, is nicely consistent.  There should be a
gradual steepening of the radio spectrum at high energies, but 
the observability is
complicated by the distribution of compression factors within the
remnant shell.

This spectral curvature complicates the simple analysis we have
undertaken.  We have elected to use an intermediate effective index,
$\alpha_{\rm eff} = 0.5$, in evaluating the gain, both because it
represents the overall increase in the number of suitable radiating
electrons better than the observed indices of W44 or the ambient ISM,
and because it leads to a very nice summary formula for the predicted
flux density.  We expect this simplification to provide an estimate
which is reliable to within a factor of two except for remnants with
very flat radio spectra.  Our adopted approximation for the gain is
therefore

\begin{equation}
G(x) \approx \frac{\pi}{4} x^2. 
\label{equation:gainEff} 
\end{equation}

In attempting to estimate the compression factor in the shell, we make
use of the fact that with either analytic approximations or hydrocode
models, it is the total pressure in the dense shell that is its best
known characteristic.  Then, making assumptions about the thermal,
magnetic, and cosmic ray contributions, the corresponding compression
factor can be determined.

In the Sedov solution, the pressure at the densest part of the edge is
about 3.3 times higher than the central pressure.  When cooling sets
in, the thermal pressure near the edge drops precipitously at first
and then recovers somewhat as neighboring material compresses the
cooled gas.  On the dense side of W44, the pressure in the dense shell
should have recovered in this way, but to a somewhat lower value than
for a Sedov remnant which had had no cooling.  Cox (1972) suggested
that the shell velocity would probably be about the same as the post
shock velocity just before cooling, leading to a shell pressure  that is 3/4ths
of the precooling edge pressure.  With this guidance, we estimate that
the shell pressure in W44 is now about twice the central pressure
$p_{\rm \scriptscriptstyle C}$, where from Table \ref{obstable}, we
have $p_{\rm \scriptscriptstyle C} \sim 1.4 \times 10^{-9}$\ dyn
cm$^{-2}$.

For an overall compression factor $x$, the thermal pressure in the
dense shell will be $(\chi {k_{\rm \scriptscriptstyle B}} n_{\rm o}
T_f) x $, where $T_f$\ is the temperature in the shell.  The final
temperature is not well known, but assuming $T_f \sim 1000$\ K, and
$\chi = 1$\ provides a reasonable estimate in the neutral gas.  With
$n_{\rm o} \approx 6$\ cm$^{-3}$, the thermal pressure in the shell is
negligible compared to the nonthermal terms treated next.

In the shock and subsequent compression into the shell, it is expected
that the tangential component of the magnetic field will increase in
proportion to the density.  As a rough approximation (also used in our
hydrocodes), we take the field as effectively transverse everywhere,
in which case the magnetic pressure is $ x ^2 B_{\rm o\perp}^2/(8 \pi)
$. We consider values of $B_{\rm o\perp}$\ in the range 2 to 5 $\mu$G
to be plausible.

The pressure due to the compressed ambient cosmic ray electrons is
negligible, but the ions could well be important.  Their ambient
pressure is comparable to that of the magnetic field and their
non-relativistic component will have its pressure increasing
approximately as $x^2$, like the magnetic pressure, so long as they do
not isotropize during the compression.  Thus, their contribution to
the shell pressure, after strong compression, could be comparable to
the magnetic pressure.  It can be substantially less if the cosmic ray
ions are not efficiently localized (i.e. they leak out) or greater if
they are accelerated by more than just the compression.  In what
follows, the cosmic ray pressure is taken as $p_{\rm CR} = \beta
p_{\rm B}$, where $\beta$\ is expected to be in the range 0 to 1.

The total pressure is thus approximately 
$(1+\beta)x^2 B_{\rm o\perp}^2/(8\pi)$\ which must 
equal $p_{\rm shell}$, yielding an estimated compression factor of 
\begin{equation}
x \approx  \frac{90}{(1+\beta)^{0.5}}  \left( \frac{3 
\mu{\rm{G}}}{B_{\rm o\perp}}\right)
\label{equation:X} 
\end{equation}
which ranges from 40 to 130 over the ranges of $\beta$\ and 
$B_{\rm o}$.  The implied range of shell density is 220 to 800 cm$^{-2}$.  
The magnetic field within the shell is less uncertain, ranging from 185 to 
260 $\mu$G, depending only on $\beta$.

This result can be generalized to any remnant with known shell expansion 
speed and preshock density, realizing that what we have done is to equate 
the nonthermal pressure in the shell with the ram pressure of the radiative 
shock, $p_{\rm ram} = \rho_{\rm o} v_{\rm shell}^2$, yielding
\begin{eqnarray}
x^2 & \approx & \left(\frac{8\pi}{1+\beta}\right) 
\left(\frac {\rho_0 v_{\rm shell}^2}{B_{\rm o\perp}^2}\right),\\
x & \approx & \left(\frac{90}{(1+\beta)^{0.5}}\right) 
\left(\frac {3 \mu{\rm{G}}}{B_{\rm o\perp}}\right)
\left(\frac{n_{\rm o}}{6 {\mbox{ cm}}^{-3}}\right)^{1/2} 
\left(\frac{v_{\rm shell}}{150 {\mbox{ km s}}^{-1}}\right).\\
\label{equation:Xgen} 
\end{eqnarray}

Thus, with $\alpha_{\rm eff} = 0.5$, the gain can be written 
\begin{equation}
G(x) \approx \frac{\pi}{4}  x^2 \approx 
  \left(\frac{2\pi^2}{1+\beta}\right) 
\left(\frac {\rho_{\rm o} v_{\rm shell}^2}{B_{\rm o\perp}^2}\right),
\label{equation:gainFinal} 
\end{equation}
the initial volume of material in the shell as
\begin{equation}
V_{\rm o} = \frac{M_{\rm shell}}{\rho_{\rm o}} =
{\mbox{1600 pc}}^3  
\left(\frac{M_{\rm shell}}{300 {\mbox{M}}_{\rm \odot}}\right)
\left(\frac{\mbox{6 cm}^{-3}}{n_{\rm o}}\right), 
\label{equation:volFinal} 
\end{equation}
and the resulting synchrotron flux density as 
\begin{equation}
F  \approx  
\left(\frac{2\pi^2M_{\rm shell}v_{\rm shell}^2\epsilon_{\rm o}}{(1+\beta)\ 
B_{\rm o\perp}^2 d^2}\right) 
\label{equation:fluxForm} 
\end{equation}
or
\begin{equation}
F  \approx  
\left(\frac{500 \ {\rm Janskys}}{1+\beta}\right) 
\left(\frac {\epsilon_{\rm o}}{10^{-40} \ {\rm W/(m^2\ Hz \ sr)}}\right)
\left(\frac{M_{\rm shell}}{300 M_{\rm \odot}}\right)
\left(\frac{3 \ \mu{\rm G}}{B_{\rm o\perp}}\right)^2
\left(\frac{v_{\rm shell}}{150 {\mbox{ km s}}^{-1}}\right)^2
\left(\frac{2.5 \ {\rm kpc}}{d}\right)^2
\label{equation:fluxFinal} 
\end{equation}

For comparison, Table \ref{obstable} shows the measured flux density for 
W44 at 100 MHz to be about 500 Janskys, within the uncertainties, in complete 
agreement with this estimate.

\subsubsection{Discussion}
\label{section:RCDisc}

Our estimate of the magnetic field in the shell of $B_{\rm shell}
\approx 200\, \mu$G was made before learning of a measurement of the
field in W44 by Claussen, \etal (1997), who also found of order 200
$\mu$G.  Their measurement was of polarization of OH masers, was more
uncertain than our estimation procedure, and was attributed to quite a
different picture of the source region.  It is therefore somewhat less
satisfying than the numerical coincidence would indicate, but could be
appropriate.  Their interpretation is that a very slow shock is being
driven into an abutting molecular cloud, with a density close to
10$^4$\ cm$^{-3}$\ and an ambient field not much smaller than what
they measure.  But the close association of the OH masers with
apparently tangentially moving regions of radio continuum emission and
the fact that the measured fields can be achieved in the remnant shell
cause one to question whether a molecular cloud interaction is
necessary.  Certainly the small compressions involved 
in their scenario would yield
negligible radio synchrotron emission via the mechanism we are
investigating.

For densities in the range found for the shell, 220 to 800 cm$^{-2}$,
the recombination time is only a few hundred years, so it is
reasonable to expect the dense shell to be mostly neutral and
responsible for the observed 21 cm radiation.

With a shell compression factor of roughly 100, the local shell
thickness (before any distortions) is less than half a percent of the
radius.  This implies that the radio synchrotron emission is very
highly edge brightened and subject to the high degree of
ripple/tangency filamentation described by Hester (1987).  We present
some simulations of this effect in section \ref{section:2dsynch}.

The synchrotron emissivity estimate above hides a fundamental problem
with this approach.  We have assumed that the best estimates of the
ambient cosmic ray electron population, the synchrotron emissivity,
and the tangential field strength are mutually consistent.  As
discussed by Rockstroh and Webber (1978), the characteristic field
strength required for consistency is surely greater than the mean
field, with the rms field being more suitable.  Results of attempts at
consistency depend somewhat on the assumed spectral index, but
typically require a characteristic field of about 9 $\mu$G.  The value
of $\epsilon_{\rm o}$\ appropriate for a 3 $\mu$G field should then be
reduced (for $\alpha_{\rm eff} = 0.5$) by about a factor of 5.
Therefore, the flux density of W44 cannot be matched by the van der
Laan mechanism alone unless the ambient CR electron population is
about a factor of 3 to 5 larger, depending on the shell mass, than
estimated for the solar vicinity.  (Alternatively, the anisotropy of
the synchrotron radiation formulae may have fooled us).  We will
return to this subject in discussing the gamma rays anticipated from
the shell.

The radio continuum flux predicted from the van der Laan mechanism is
similar to that observed, but there is clearly room for contributions
or enhancements by additional processes, so long as those processes do
not produce a markedly different spectrum.  The next three paragraphs
discuss some of the possibilities.

Some injection by the shock of new cosmic ray electrons may be
occuring.  Alternatively, electrons accelerated by the remnant when it
was younger, or by the pulsar, may be caught up in shell formation and
forced to radiate in the vastly enhanced magnetic field there.  One
might suppose that the problem with either possibility is that there
is no obvious reason for the surface brightness and spectrum to be so
close to those found for the van der Laan mechanism and by
observation, suggesting that we consider, as below, factors that could
enhance the radio synchrotron emission over the above estimate without
invoking a totally independent mechanism.  But perhaps there is a
reason for achieving approximately this result with other mechanisms,
namely that our emissivity is basically the equipartition result--if
for a given pressure, ambient field, and shell mass we ask for the
emission from the usual percentage of electrons associated with cosmic
ray ions in equipartition with the shell's magnetic field, we will
have roughly what we have calculated, independent of the details of
the acceleration mechanism.

Ambient cosmic ray electrons may receive substantial diffusive
acceleration on passing through the adiabatic gas subshock, followed
by betatron acceleration during compression into the dense shell.
These ambient electrons, apparently having a slightly shallower index
than that of newly injected and accelerated ones, will maintain their
original spectral index while achieving an average energy boost per
particle.  From elementary theory, the magnitude of the boost is
unknown, but potentially large.

Shell formation offers yet another acceleration possibility.  Electrons which 
attempt to diffuse away from the cold dense shell find themselves continually 
swept back in because there is approaching flow on both sides:  recently 
swept up ambient material cools rapidly and is added to the shell from the 
outside, while material just interior to the shell which was heated long ago 
finally cools and provides a condensation flow to the inside (occasionally 
accompanied by another weak radiative shock).  This two sided flow 
convergence not only reduces particle losses from the densest region, but 
offers an as-yet-unexplored site of first order Fermi acceleration.

\subsection{The H$\alpha$, [O{\small I}] 63$\mu$m, and Related Luminosities}
\label{section:Alphaetc}

A rough estimate of the H$\alpha$\ luminosity of the remnant can be
made following the methods of Cox (1970, 1972a).  There are several
factors to consider.  Hot material inside the remnant and newly
shock-heated material on the outside edge are cooling off,
recombining, and joining the dense shell.  They contribute directly to
H$\alpha$, but also emit EUV which fluoresces in the neighboring
neutral material, shell and ambient, to yield even more H$\alpha$.  A
further complication is that fluorescence in the ambient medium is not
instantaneous; ionization is effectively stored for a recombination
time, which at $n_{\rm o} \sim 6$\ cm$^{-3}$\ is about the age of W44.

Because the recombinations in shell and ambient material can be
distinguished via both spatial distribution and in velocity, we
consider them separately.  Following Cox (1972a), we estimate that in
cooling from high temperature, approximately 4 ionizing photons are
radiated per atom.  Assuming that half of these photons escape to
ionize the ambient medium, the ionization rate for the surroundings is
$2 \, dN/dt$\, and the recombination rate for the shell is $3 \,
dN/dt$\ (there being one extra from the recombination of the cooling
atom).  (Here $dN/dt$\ is the rate at which atoms join the shell.)
The total number of atoms in the remnant is approximately $n_{\rm o} V
\sim 10^{60}$\ (about 1000 M$_{\rm \odot}\,$).  Of these, we expect
that roughly a third (but see below) have cooled during the $\sim$\
3000 years between $t_{\rm cool}$ and $t_{\rm shell}$, contributing $4
\times 10^{48}$\ atoms s$^{-1}$ to $dN/dt$.  The rate at which atoms
are overtaken by the radiative shock (presently covering half the
remnant's surface, of total area A) is $n_{\rm o} (A/2) v_{\rm
\scriptscriptstyle S} \sim 1 \times 10^{48}$\ atoms s$^{-1}$.  We thus
expect a total of $dN/dt \sim 5 \times 10^{48}$\ atoms s$^{-1}$.  Results from
our 2d hydrocode (below) show this estimate to be about a factor of 2
too high at present (20,000 yr), with the higher rate more
representative of the recent past.  Because the ionization level of
the ambient medium averages the rate over a substantial time, we take
the higher rate for it, yielding $1 \times 10^{49}$\ photoionizations
s$^{-1}$ for the last 3000 years.  For the shell fluorescence the
instantaneous value of dN/dt is more appropriate, yielding $0.8 \times
10^{49}$\ recombinations s$^{-1}$ in the dense shell.

>From recombination theory there are approximately 0.6 eV of H$\alpha$\
emission per average (Case B) recombination.  The shell luminosity in
H$\alpha$\ from the radiative shock should thus be about $3 \times
10^{36}$\ erg s$^{-1}$.  From the above analysis, however, the total
number of atoms presently joining the dense shell, including those
from the cooling interior, is 2 to 3 times higher, implying a total
H$\alpha$\ luminosity of about $7 \times 10^{36}$\ erg s$^{-1}$.  Its
spatial distribution and velocity structure should resemble those of
the radio continuum and H{\small I} shell respectively, with one
caveat: because shell formation is presently occurring at the equator
of W44, we might expect the emission to be brightest there, rather
than in the east or northeast where shell formation has already passed
its peak.  This equatorial configuration is in fact apparent in Fig 1b
of the H$\alpha$\ picture of Giacani \etal (1997) (who attributed it
to enhanced absorption in the east), while their Figure 5 demonstrates
the expected strong relationship between the H$\alpha$\ and radio
continuum enhancements and filamentation.

Our luminosity estimate corresponds to an average surface brightness
over the $\sim$\ 500 square arcmin of the remnant of $6 \times 10^{-
15} $ ergs (cm$^2$ s$^1$ arcmin$^2)^{-1}$.  The extinction at
H$\alpha$\ implied by an x-ray absorption column density of about (1.6
to 2.1)$\times 10^{22}$, however, is a factor of 600 to 4400, for an
observable average of (0.1 to 1)$\times 10^{-17} $ ergs (cm$^2$
s$^1$ arcmin$^2)^{-1}$.  Giacani \etal (1997) quote their peak
observed surface brightness as $10^{-16} $ ergs (cm$^2$ s$^1$
arcmin$^2)^{-1}$, with the dimmest observable features at $10^{-17} $
ergs (cm$^2$ s$^1$ arcmin$^2)^{-1}$.  Their brightest features, a
factor of 10 to 100 higher than our average, are strongly enhanced by
filamentation; but their dimmer features, a factor of 1 to 10 above
our estimate, appear much smoother in the equatorial belt.  Given the
fact that we expect concentration in that region, and that the large
extinction is not only uncertain but likely to be variable, the
agreement is as good as can be expected.

     Because the recombination time in the ambient medium is long, its
luminosity will be lower.  There have been a total of roughly
10$^{60}$\ ionizing photons captured in the surroundings.  Neglecting
recombinations, they will ionize a similar number of atoms, which will
then recombine with a timescale of $10^5 \mbox{ years } / n_{\rm o}
\sim 1.8 \times 10^4$\ years, yielding $ 0.2 \times 10^{49}$\
recombinations s$^{-1}$ and a luminosity of $2 \times 10^{36}$\ erg
s$^{-1}$.  Although the total radiation by the shell is expected to be
about 3 times brighter, and enormously higher in surface brightness in
restricted regions, the fluorescence in the ambient medium is not
inconsequential.  The ionized mass should be about 1000 M$_{\rm
\odot}\,$, like the mass within the remnant, and extend roughly 3 pc
outside (even farther but dimmer on the low density end), with an
emission measure exceeding 100 cm$^{-6}$\ pc.  In the absence of
inhomogeneity other than the density gradient, it should be brightest
behind the northeast part of the remnant, where the ambient density is
high.

Of course the emission of the cooling gas is not restricted to
H$\alpha$.  The overall spectrum should be rather similar to that of a
radiative shock with a speed of about 150 km s$^{-1}$.  From Model G
of Raymond (1976), for example, we expect emission from optical lines
of [O{\small III}], [O{\small II}], [O{\small I}], [N{\small II}], and
[S{\small II}], and near IR lines of [C{\small I}] that are comparable
in intensity to H$\alpha$.  In the infrared, predicted strong lines
are [Fe{\small II}] at 26 $\mu$m, [O{\small I}] at 63 $\mu$m, and
[Ne{\small II}] at 12.8 $\mu$m, the first of which is about twice as
bright as H$\alpha$, the others about one third.  The basic picture is
not much different in more recent calculations (although
inclusion of proton impact excitation  
alters the strengths of the infrared lines somewhat).
Essentially all shock
models in the velocity range 130 to 160 km sec$^{-1}$\ yield 1 to 3 eV
of H$\alpha$\ and 0.5 to 1 eV of 63 $\mu$m emission per average atom
through the shock.  Once again assuming that the radiation by material
joining the dense shell from inside is similar in spectrum to the
radiative shock, we conclude that the luminosity in any of the above
lines should be about (3 to 10)$\times 10^{36}$\ erg s$^{-1}$.  The
[S{\small II}] was observed by Giacani \etal (1997), with a
distribution (and presumably intensity, though it was not discussed)
similar to that of H$\alpha$.  The 63 $\mu$m line was measured with
the ISO LWS by Reach and Rho (1996); from the regions they sampled
they estimate a total luminosity of about $10^3 L_{\rm \odot}$, in
excellent agreement with our estimate.

Interestingly, Reach and Rho considered a radiative shock model, but
rejected it as unable to provide the observed emission.  There appear
to be three contributing reasons for our difference: they compared the
theory of a single shock viewed face on with their maximum observed
surface brightness, $10^{-3}$\ ergs cm$^{-2}$ s$^{-1}$ sr$^{-1}$,
rather than an estimated average which for $10^3 L_{\rm \odot}$\ would
be a factor of 10 lower; in addition, the average surface brightness
of a spherical emitting surface is four times that of the local normal
value for one side; and, finally, we have an additional
factor of 2.5 to include the effects of current shell formation.  The
net effect is a factor of 100 which is just the difference between
their estimate of the rate per unit area that the shock would have to
be encountering ions and the $10^8$\ atoms cm$^{-2}$\ s$^{-1}$ in our
model.  This caused them to believe that an unacceptably high ram
pressure was somehow present where the [O{\small I}] emission
occurred, and that that was somehow connected to an encounter with
dense molecular material.  In our model this is unnecessary: the
correct total emission is found, and like the radio continuum and
optical emission, there are local enhancements over the average of
more than a factor of 10 due to rippling and edge brightening of the
thin shell.

\subsection{The Gamma Rays}
\label{section:gamma}

It is fairly straightforward to estimate the maximum gamma ray
intensity from cosmic ray ions interacting with the mass in the dense
shell.  Compression of the nonrelativistic component of ambient CR
ions, or acceleration of ions in the remnant can potentially lead to a
CR pressure in the shell comparable to the magnetic pressure, their
sum being the shell pressure, approximately $2.8 \times 10^{-9}$\ dyn
cm$^{-2}$.  This amounts to an enhancement of the ion pressure by a
factor of 1400 over ambient.  From the review by Bloemen (1989), and
models in Esposito \etal (1996), we infer that this enhancement would
lead to a gamma ray production rate above 100 MeV, from $\pi^o$\
decay, of about $2.2 \times 10^{-22}$\ photons s$^{-1}$\ per H atom in
the shell.  For a shell mass of about 450 \msun, as found in our 2d
modeling later in the paper, and a distance of 2.5 kpc, the maximum
flux would be $12 \times 10^{-8}$\ photons cm$^{-2}$\ s$^{-1}$, about
a fourth of the observed value, concentrated in the east or northeast.
(The total mass interior to W44, in our model, is about 1000 \msun,
but in our simplest picture we do not expect the more diffuse portions
in the interior to be subjected to this high cosmic ray pressure.
With some leakage and the associated additional compression, however,
more of the remnant's mass may be exposed.)  One may suppose that
cosmic rays are leaking out of the remnant altogether, and irradiating
the nearby dense molecular material, though if they do so at greatly
reduced pressure, vastly more mass will be required for interaction.
(Esposito \etal, 1996, found that irradiation of the entire Wooten
cloud would require an average cosmic ray enhancement over ambient of
a factor of about 80.)

If diffusion prevents the CR ion pressure from limiting the
compression to the low value implied above (about 40), the
bremsstrahlung component of the gamma rays can be enhanced by an
similarly large factor.  Generalizing on the methods of Section
\ref{section:vdL}, the ratio of Gamma ray bremsstrahlung to radio
synchrotron enhancement factors turns out to be $x^{-\alpha_{\rm
eff}}$, where $x$\ is the compression factor and $\alpha_{\rm eff}$\
is the effective spectral index, which we have taken as 0.5.
Comparison with our previous results for the van der Laan mechanism
yields an overall gain of
\begin{equation}
G_{\rm \gamma}(x) \approx 800  (x/100)^{1.5}. 
\label{equation:gainGamma} 
\end{equation}
Again following the review by Bloemen (1989), and models in Esposito
\etal (1996), we infer that this enhancement, for $x = 100$, would
lead to a gamma ray production rate above 100 MeV of about $6 \times
10^{-23}$\ photons s$^{- 1}$\ per H atom in the shell.  For the shell
mass and distance given above, the flux would be about $4 \times
10^{-8}$\ photons cm$^{-2}$\ s$^{-1}$, about 30\% of the upper limit
from $\pi^o$\ decay and 7\% of the observed gamma ray flux.  Recalling
from our discussion of the synchrotron radiation that the observed
flux required approximately 4 times more CR electrons than anticipated
from solar neighborhood estimates, and noting that electrons in the
same energy range would be responsible for both the synchrotron and
bremsstrahlung, we conclude that the gamma ray estimate must be raised
by a similar factor, to 16 $\times 10^{-8}$\ photons cm$^{-2}$\
s$^{-1}$, comparable to our upper limit from $\pi^{\rm o}$\ decay.
Both mechanisms therefore require more target mass than our model
provides, by about the same factor.

An alternative approach employed by deJager and Mastichiadis (1997),
solves for the CR electron populations needed to produce both the
synchrotron and bremsstrahlung.  Taking the ratio, they obtain a
relationship between the required magnetic field and density in the
interaction region if both are to be produced.  Interpreting the
interaction region as the shell in our model, their result for the CR
population required to produce the synchrotron is consistent with our
previous conclusions.  Their result for the required shell density for
generation of the gamma rays can be summarized as $n_{\rm dJM} \approx
25\ {\rm cm}^{-3} (B/10\mu{\rm G})^{1.33}$, while in our model we
anticipate $n_{\rm shell} \approx 20\ \mbox{cm}^{-3} (B/10\mu{\rm
G})$.  At the shell fields of 180 to 260 $\mu$G of our model, the
gamma ray production would require a shell density about 7 times that
available.  We repeated their analysis with a slightly steeper index
and found the disparity to be slightly less, but still a factor of 4,
consistent with our conclusion above that the electrons required in
our shell to produce the synchrotron fail by about a factor of 4 to
produce the observed gamma rays.

The contribution expected from inverse Compton radiation was also
calculated by deJager and Mastichiadis, assuming a radiation field
that is enhanced by the glowing dust in the neighboring molecular
cloud.  They concluded that the contribution was less than that from
bremsstrahlung when the mean density in the interaction region exceeds
about 10 cm$^{-3}$, a very low value compared to that found in our
shell.  Inverse Compton will also arise from CR electrons within the
remnant but not yet swept into the dense shell, but that too appears
negligible.

The net result of this investigation is that the dense shell in our model 
should produce gamma rays at roughly 25 to 50\% of the rate actually observed 
in the vicinity by EGRET.

\subsection{Evolution in a Density Gradient}
\label{section:Komp}

\noindent{\it{Approximation to Kompaneets}}

In planning the use of a 2d hydrocode to provide a model including the 
partial shell formation of W44, we assumed the ambient density to have the 
exponential distribution: 
\begin{equation}
\rho_{\rm o}(z)= \rho_{\rm *} \, \mbox{e}^{-z/h}\,,
\label{equation:strati}
\end{equation}
where $\rho_{\rm *}$ is at the explosion site, and $z$ is the 
distance from it, measured down the density gradient.  Thus, 
to specify a model, the 1D input parameters (the explosion energy $E_{\rm o}$ and 
the ambient density $\rho_{\rm o}$) have to be supplemented by the stratification 
scale-height $h$.  The parameter space for initialization of a run is 
therefore three-dimensional, with age a fourth parameter. As high-resolution 
simulations in two dimensions are expensive, an extensive numerical survey of 
that space is not possible. Therefore, we searched for appropriate input 
parameters with the help of analytical calculations based on the Kompaneets 
Approximation (Kompaneets, 1960).

The basic idea of the Kompaneets Approximation is to assume that the 
post-shock pressure $p_{\rm \scriptscriptstyle S}$ is uniform over the surface of 
the remnant, while the evolution occurs in an exponentially stratified 
medium. For a strong non-radiative shock in a $\gamma = \frac{5}{3}$ gas, 
$p_{\rm \scriptscriptstyle S} = \frac{3}{4} \rho_0 v_{\rm \scriptscriptstyle S}^2$, 
where $v_{\rm \scriptscriptstyle S}$ is the shock velocity and $\rho_0$ is the 
ambient gas density. With uniform post-shock pressure, the local shock speed 
is inversely proportional to the square root of the local ambient density. 
Thus the evolution of the shape of the remnant can be calculated without 
first knowing the time evolution. Once the shape is 
known, as a function of some size variable such as the semi-major axis, the 
volume can be calculated as well. Then, because the pressure is inversely 
proportional to the volume, the time can be found from an integration.

Maciejewski and Cox (1998) extend this approach in two ways, the first 
of which is to assume that the 
structure of the remnant is closely approximated by a prolate ellipsoid of 
semi major and minor axes $a$ and $b$.  They then readily find the semi-minor 
axis 
$b$, and the distance $s$ by which the center of the ellipsoid has displaced 
from the explosion site, as functions of semi-major axis $a$.  A surprising 
result of this exercise is that even remnants extending  over several scale 
heights remain nearly spherical;  the principal geometrical effect of the 
gradient is to shift the remnant center gradually away from the explosion 
site.  (See also Dohm-Palmer and Jones, 1996, for a thorough discussion of this
behavior based on 2d hydrodynamic modeling.)  

Assuming the postshock pressure, explosion energy, and volume have the same 
relationship as in the 1D Sedov solution determines the dependence of the 
shell expansion velocity on the SNR volume, and an expression for the time 
derivative of the semi-major axis.  Numerical integration then yields the 
time evolution $t(a)$, at which point, all of the quantities parameterized by 
the semi-major axis $a$ can be expressed as functions of time.

\noindent{\it{Cooling and shell formation times}}

Maciejewski and Cox (1998) also evaluate the progression of the cooling
time over a remnant in a density gradient, in almost exact parallel to the 1-d case.  
With the 
Kahn approximation, the incremental cooling time $\Delta t_{\rm cool}$\ of a 
recently shocked parcel depends on the post shock pressure and the local 
preshock density.  Adding that to the time at which a parcel is shocked 
determines its time to finish cooling.  Minimizing $t_{\rm finish}$\ over 
parcels (for a particular direction in the 2d case) determines the cooling 
time on that part of the remnant.  Finally, multiplying by 7/6ths supplies an 
estimate of the time for shell formation.  By performing this calculation for 
the remnant's dense end, tenuous end, and equator, the combination of Kahn 
and Kompaneets approximations provides a useful tool for the exploration of 
shell formation over the surface. 

Even a modest density gradient leads to a great spread in 
cooling and shell formation times for different parts of the remnant's 
surface, making it possible to select input parameters which will lead to any 
desired degree of partial shell formation.

We examined a wide range of input parameters.  In particular, we used our 
analytical method to generate plots on the $\rho_{\rm *}$-$h$ plane for assumed 
$E_{\rm o}$ and linear size of the remnant. The plotted quantities included the 
time required to reach the assumed size, the post-shock pressure, the cooling 
and shell formation time-scales for the dense, equatorial, and tenuous 
directions, and the shock velocities in those directions. (Although the 
Kompaneets model, like the Sedov model, cannot be used very reliably to 
estimate quantities after shell formation has occurred, we make the usual 
approximation that shortly after shell formation the shell velocity is about 
3/4ths of the shock velocity predicted for the non-radiative evolution.)  In 
the limit of large scale heights, the results are consistent with our 1D 
calculations.

\noindent{\it{W44's parameters}}

In this way we were able to locate regions of parameter space which promised 
to satisfy the observational constraints discussed for the 1-d model,
adding the further condition that shell formation must be well underway 
on the dense end, just commencing 
at the equator, and not yet present on the open more tenuous end.  This must 
be reconciled with a substantial radial velocity for the receding part of 
the 21 cm shell. 

These constraints turned out not to be sufficient to determine the scale 
height uniquely.  In the end we chose to examine the most asymmetric 
(smallest scale-height) case that appeared to be viable;  in it the cooling 
times on the high and low density ends differ considerably, but the expansion 
velocity into the densest medium is still large (models with still smaller 
scale-heights have too low an expansion velocity on the dense end).  The 
selected input parameters and estimated characteristics of the remnant at 
20,000 years are shown in Table~\ref{2Dparameters} while Figure 
\ref{schematic} summarizes the configuration and goal of the evolution. 

\begin{table}
\begin{center}
\caption{2D Model Parameters and Predicted Characteristics}
\begin{tabular}{|cccc|}
\multicolumn{4}{c}{}\\ \hline\hline
 \multicolumn{2}{|c}{$\rho_{\rm *}$ (g cm$^{-3}$)} &  
\multicolumn{2}{c|}{$1.317\times10^    {-23}$}\\
 \multicolumn{2}{|c}{$n_{\rm *}$    (  cm$^{-3}$)} &  \multicolumn{2}{c|}{ 
6.2}\\ 
 \multicolumn{2}{|c}{$h$      (  pc       )} &  \multicolumn{2}{c|}{ 22}\\
 \multicolumn{2}{|c}{$E_{\rm o}$    (  ergs     )} &  
\multicolumn{2}{c|}{$10^{51}$ 
}\\
\hline
    time  (yr)         &       dense end      &  equator      &      tenuous 
end\\
$t_{\rm cool}$      &         12,000       &  17,000       &      25,500 \\
$t_{\rm shell}$     &         14,000       &  20,000       &      30,000\\ 
\hline\hline
\multicolumn{4}{|c|}{At 20,000 years:}\\
\multicolumn{4}{|c|}{$a$ = 11.9 pc, $p_{\rm \scriptscriptstyle C}$ = 
$1.49\times10^{-9}$ dyn
cm$^{-2}$ }\\
\hline
  speed  (km s$^{-1}$) &           dense end      &  equator      &      
tenuous end\\
shock   &      178            &  233          &      305\\
shell   &      133            &  175          &      NA\\ 
\hline\hline
\end{tabular}
\label{2Dparameters}
\end{center}
\end{table}

\subsection{Sgro-Chevalier and Approximate Scalings}
\label{section:scaling}

As it is unlikely that the hydrodynamic models will produce a remnant of 
precisely the size desired, it is useful to be aware of the Sgro-Chevalier 
scaling law (Sgro, 1972, Chevalier, 1974),
that each evolution represents a family of evolutions with the same value
of the parameter En$_{\rm o}^2$. This scaling was based on the fact that both
ionization and cooling have temperature dependent two-body rates, 
and extends itself to include
models with differing electron and ion temperatures, thermal conduction, and
 dust sputtering.  In our case, for example, the hydrodynamic runs delivered
models with radii of 11 pc, which would need to be placed at a distance of 2200
pc to have the correct angular size. The Sgro-Chevalier family member 
with a radius of 12.5 pc and distance of 2500 pc 
has the same temperature and velocity 
structure and the same distribution of ionization,
 but has an initial density of 5.5 cm$^{-3}$, an energy of E$_{51}=1.3$, an
 age of 22,700 years,  x-ray surface brightness and count rate 
(at the same angular size) reduced by a 
factor of 0.88, and unchanged synchrotron flux density.  The shell mass is 
higher by the same factor as the energy, 1.3. 

A more general but more approximate scaling, searching for a remnant at the same evolutionary stage as a specific hydrodynamic model but with a different 
radius, can be based on Equations \ref{equations:shell}, 
\ref{equation:fluxForm}, and noticing that for 
collisional equilibrium plasmas with temperatures in the range
3 to 8 $\times 10^6$\ K, the ROSAT counts per emission measure function is
approximately proportional to $T^2$, particularly with significant absorption.
The latter implies that the x-ray band luminosity is approximately proportional
to $R^3 p^2$.  The inferred scaling is that at a given evolutionary state, the 
various model parameters scale with $n_{\rm o}$\ and $En_{\rm o}^2$\  as:
$R_{\rm shock} \propto n_{\rm o}^{-1} (En_{\rm o}^2)^{2/7}$, 
$t \propto n_{\rm o}^{-1} (En_{\rm o}^2)^{3/14}$,
$v \propto (En_{\rm o}^2)^{1/14}$,
$T \propto (En_{\rm o}^2)^{1/7}$,
$S_{\rm x-Ray} \propto n_{\rm o} (En_{\rm o}^2)^{4/7}$,
$F_{\rm x-Ray} \propto \theta^2 n_{\rm o} (En_{\rm o}^2)^{4/7}$, and
$F_{\rm synch} \propto \theta^2 (En_{\rm o}^2)^{3/7}$,
where $S$\ and $F$\ refer to surface brightness and flux, and $\theta$\
to angular size.
Thus, scaling a model from one with a radius of 11 pc to one with 
a radius of 12.5 pc and constant $E_{51}$\ requires
 an initial density of 4.6 cm$^{-3}$, and describes a remnant with an
 age of 23,700 years,  x-ray surface brightness and count rate
(at the same angular size) reduced by a
factor of 0.53, and a synchrotron flux density lower by
0.77.  The shell mass is
higher by 1.09, and unlike the Sgro-Chevalier scaling, the velocities
and temperatures are reduced by factors of 0.96 and 0.92 respectively.
With the slightly lower temperature the ionization and therefore the
details of the spectrum will be slightly different as well.

We will return to these results in discussing the hydrodynamic models. The
main feature, however, is that all changes are moderate, essentially within
uncertainties, except for the change in the X-ray surface brightness which
is lowered significantly in the constant energy scaling.
\subsection{Summary of the Analytical Section}
\label{section:AnalSum}

According to our analytic approximations, an explosion of $ 10^{51}$\ ergs 
into a medium with scale height of 22 pc and a density of  
approximately 6 cm$^{-3}$\ at the explosion site, when viewed at an age of 
20,000 years has the following successes as a model of W44:
\begin{enumerate}
\item  Its radius is consistent with the angular size and distance.
\item  It will have formed a neutral hydrogen shell on one side (left rear 
with our orientation), with approximately the observed column density.
\item  The expansion velocity of the  H{\small I} shell will be close to 
that observed.
\item  With thermal conduction at roughly its unquenched level, the central 
temperature and density are in agreement with those found from the x-ray 
emission.  The dependence of the values on the actual magnitude of the 
conductivity is weak.
\item  The temperature near the edge will be too low to contribute 
substantially to the x-ray emission, so that spectral softening with radius, 
particularly when
combined with the high observed column density of intervening absorbers, can 
provide the appearance of center filled x-ray emission.
\item  The  van der Laan mechanism---synchrotron enhancement by compression 
of ambient cosmic ray electrons and magnetic field---can provide the observed radio 
continuum emission, but only with about a factor of 4 higher 
ambient CR electron population than estimated
for the solar neighborhood.
Additional complicating factors cannot be ruled out, and we discussed the 
possibility 
of diffusive acceleration boosting the energies of the ambient cosmic ray 
electrons in the gas subshock or further acceleration in the converging flows 
bounding the dense shell.
\item  On-going shell formation and the radiative shock on the denser half 
lead to roughly 2000 L$_{\rm \odot}\,$\ of high contrast filamentary H$\alpha$\ 
from the dense fast moving shell, consistent with that measured after allowing for
extinction, plus about 800 L$_{\rm \odot}\,$\ of H$\alpha$\ 
from diffuse fluorescence in the surrounding 3 pc or so of stationary ambient 
gas.  The shell radiation is predicted to include also the full spectrum characteristic
of a fast radiative shock, including amounts of [S{\small II}] and 63 $\mu$m 
[O{\small I}] emission
completely consistent with those actually observed.
\item  The gamma ray emission expected from the dense shell is 25 to 50\% of
that seen in the vicinity with EGRET.
\end{enumerate}

\section{ Physics Assumptions, Justifications, and 
Implementations in the Hydrocodes}
\label{section:PhysAss}

This section surveys the assumptions and methods of the hydrocodes
employed in the next two sections
 to examine the spatial configuration of the remnant
model's characteristics, its appearances in 21 cm, radio synchrotron, and
x-ray emission, and its x-ray spectrum.  Particular attention is paid to
justification of the inclusion of thermal conduction in the hot interior of
the remnant.

\subsection{Pressure}
\label{section:codeP}

The thermal pressure is that of an ideal monoatomic gas with solar 
composition, assuming full ionization of hydrogen and helium, and equal 
temperatures for electrons and ions.  Nonthermal pressure is represented via 
an isotropic term proportional to $n^2$, whose principle effect is to limit 
the compression of the cold dense shell to a realistic value.  Although this 
is referred to as magnetic pressure and the density dependence
justified by flux freezing and compression of the tangential component, 
the codes do not perform 
magnetohydrodynamic calculations.  Cosmic ray acceleration by the shock is 
assumed to have a negligible effect on the shock or remnant structure.  

\subsection{Thermal Radiation}
\label{section:codeTR}

In the 1-d code,  nonequilibrium ionization and the corresponding  cooling 
rates and spectra are followed in each parcel with the  Raymond and Smith 
codes (Raymond and Smith 1977, Raymond 1995), accelerated with 
modifications using tables of rate coefficients  provided by Edgar (1994).  
The  abundance set adopted was that of Grevesse and Anders (1988).  Radiative 
transfer between parcels was not included.

Because the 2d code does not follow the ion states with time, it is 
necessary to estimate the radiation rate with a cooling function.  As 
proposed by Edgar (Edgar and Chevalier, 1986), a 
pseudo non-equilibrium radiative cooling table was adopted.  It contains the 
non-equilibrium radiative cooling coefficient  versus temperature for an 
optically thin plasma that was shock heated to $10^8$ K (though the choice of 
upper temperature has little effect) and then isobarically cooled.

The model x-ray spectra were also generated with the Raymond and Smith code, 
based on the temperature and non-equilibrium ionization state of each parcel 
previously calculated by the 1-d hydrocode.  The spectrum was attenuated 
using the absorption curves of Morrison \& McCammon (1983), and then  
convolved with the ROSAT PSPC and Einstein SSS response functions.  

\subsection{Dust}
\label{section:codeDust}

For the most part, dust is assumed to be non-existent within the hot portions 
of the remnant.  As shown by Smith \etal (1996), this  assumption has little 
effect on the inferred cooling timescale of the gas parcels.  The elemental 
abundances do, however, affect the spectrum.  As a result, we discuss 
the results of a 1-d run with dust destruction included 
explicitly, using the rate model of Smith \etal (1996).

\subsection{Thermal Conduction}

\label{section:CondImpl}

The thermal conduction treatment is similar to that  of Slavin and Cox 
(1992), a single temperature model  with an electron thermal conductivity which is 
classical in regions of weak temperature gradient but bounded by the 
saturation limit when the gradient is large.  As shown in Cui  and Cox 
(1992), by the times of interest here this provides a description
 which is close to that of 
models with different  electron and ion temperatures and their  corresponding 
electron and ion conductivities.  

In the present study, however, a single temperature model is not without 
dangers:
examination of the Cui and Cox thermal  structures  shows that there 
is disparity at  the center of the remnant.  In their case 4, in which the 
electron and ion temperatures are followed separately with only coulomb 
collisions to equilibrate them but each with its own conductivity, there is a 
persistent core of hotter ions and lower density (about a factor of 3 in the 
inner third in radius), compared to the single temperature case.  Normally, 
this disparity is  unimportant; we do not generally imagine these  hydrocode 
models to be useful in the central parts of the remnant where the  structure 
is surely affected by the details of the  explosion and the pre-supernova 
stellar evolution.  In Cui and Cox, for example, it is probably a relic of 
initializing with a Sedov structure.  But, when we are purporting to 
model the  central x-ray emission, we cannot help being concerned about this 
point.  

The next concern derives from the
 considerable uncertainty regarding how much conduction actually 
occurs in the hot interior of an SNR, relative to standard formulae, and how 
the answer depends on time.  Although we regard this as a sufficiently 
difficult problem that observations of objects like W44 will probably be 
called upon to solve it, there are some points worth making. 
 For example, it is important to realize that a very small amount of 
transport, compared to the classical heat flux, 
will make a very large difference to the central 
structure of a remnant (section \ref{section:TCtheory}),  even if that 
transport is active only when the remnant is rather young.

The most common arguments against the presence of significant heat transport 
in remnants have to do with the strong suppression of conductivity transverse 
to magnetic fields.  In 2d MHD models, a uniform ambient field is 
distorted outward in the bubble, in most places providing a significant 
tangential component, guiding what conduction there is to the magnetic caps.  
Arguing that there has been considerable turbulence in the center, or that 
the field was largely irregular in the first place, encourages the idea that 
all conduction paths along the field will be very long, suppressing 
conduction to a large degree.  

Of course, the self-same turbulence that was imagined to have tangled the 
field would also have intermixed the entropy components, accomplishing 
largely the same purpose as conduction.  At the very least, it would have 
smeared a tangled mixture of different temperature components across the 
interior structure.  In addition, field line wandering (Jokipii and Parker, 
1969) then allows transport among seemingly isolated regions.

In very young remnants, still interacting with their own wind material, there 
is a lot of radial mixing and the dominant field component is radial.  This 
would seem not to suppress conduction substantially.  On the other hand, for 
W44, entropy mixing among roughly the innermost 30-100 M$_{\rm \odot}$\ is 
needed.  

In point of fact, all conceptions of thermal  conduction in this environment 
are vastly too  simplistic: the collision mean free paths are very  long; the 
stellar ejecta are surely an important  complication; nonlaminar flow with 
mixing and turbulence has almost surely occurred; and the strict  
adiabaticity of the Sedov solution is surely the worst  possible approximation.  
Our use of classical  conductivity may seem naive, but, when  safeguarded for 
saturation, it provides a reasonable, physically based estimate for the rate 
of transport.  It  could, in fact be an underestimate if turbulent diffusion is 
important in remnant interiors.

 Two papers which have discussed magnetic suppression of conduction and limits
on that suppression include  Rosner and Tucker (1989),
and Tao (1995).

In summary, no existing model offers a better  idea.  This one is important 
to explore, but such  exploration must  be followed by  careful  
discussion of the discrepancies so we can have some  indication of what other 
processes might be  important.  

\noindent{\it{ Transport of ions}}

A related issue is that in remnant interiors the transport of energy is very 
likely accompanied by the  convection or diffusion of elemental abundances 
and of the ionization structure.  This too could alter the x-ray spectrum and 
surface brightness of the center, presumably moving toward greater 
uniformity.

\noindent{\it{Implementation of conduction}}

In the 2d hydrocode, expressions for the classical conductivity, the 
saturation flux, and the interpolation between them were taken from Slavin 
and Cox (1992).  The conduction equation is solved at the beginning of each 
base-grid time step on a uniform grid with the resolution corresponding to 
the resolution of the finest grid used in the simulation.  An implicit, 
fractional-steps method of Yanenko (1971) is applied.  The implementation is 
similar in 1d, except that the originally proposed formula for interpolation 
between classical and saturated fluxes ($F_{\rm total} = F_{\rm class}/[1 + 
|F_{\rm class}|/F_{\rm sat}]$) of Cowie and McKee (1977) was used.  Except
as noted, the classical conductivity assumed was the full
unquenched value as given in \S 
\ref{section:TCtheory}.

\subsection{Synchrotron Emissivity for Use With Hydrocode Results}
\label{section:synchImpl}

Estimation via the methods of Section \ref{section:vdL}
of the  radio synchrotron 
emission expected due to the van der 
Laan mechanism requires somewhat unusual accounting, evaluating for each
parcel in the model its initial volume, compression factor, and gain.  The
specifics are:  
the emissivity of the ambient medium at 100 MHz was taken as
$10^{-40} \ {\rm W/(m^2\ Hz \ sr)}$;
the compression factor was taken as $x = \rho / \rho_{\rm o}$\ (where
$ \rho_{\rm o}$ is the initial density of the parcel); 
the ambient emissivity was multiplied by the initial volume of the parcel, 
taken to be  M$ / \rho_{\rm o}$, dividing
by 10$^6$ to accommodate the mixed units;
that initial emission rate was multiplied by the gain at a given frequency, 
$G(x) \approx (\pi/4)  x^2$ for an effective spectral index of 0.5, and then
4$\pi$ to eliminate steradians.  The final 
result is the parcel's spectral emission (in Watts  Hz$^{-1}$).

To obtain the total spectral luminosity, the emission is summed 
over the parcels in the remnant.  To turn this into a contribution to the 
observed flux density, it is divided by $4 \pi d^2$ with the distance $d$\ in 
meters, to get Watts (m$^2$ Hz)$^{-1}$ , and divided by $10^{-26}$\ to convert 
to Janskys.

\section{Two-Dimensional Model}
\typeout{Two-Dimensional Model}
\label{section:2d}
\subsection{Background}
\label{section:2dBkgd}
\noindent{\rtx{Purpose}}

The 2d modeling provides a fascinating look at the structure of the remnant 
as well as quantitative representations of the H{\small I} distribution, across 
the remnant and in radial velocity, and of the radio continuum emission 
distribution anticipated from the van der Laan mechanism.  It also allows 
us to examine the x-ray emission from the distribution of density and 
temperature in the interior to see whether lack of sphericity might invalidate 
our more detailed 1d model of the x-ray surface brightness distribution and 
spectrum.

\noindent{\rtx{Code Description}}

The hydrodynamic and thermal conduction equations are solved 
with the help of the AMRA code (Plewa \& M\"uller 1998),
which combines the AMR 
(adaptive mesh refinement) algorithm of Berger and Colella (1989) 
and the PPM method of Colella \& Woodward (1984).  It has also 
been recently used in a numerical study of the evolution of ejecta 
fragments by Cid-Fernandes \etal (1996).  The specifics of our 
implementation of thermal conduction, nonthermal pressure, and 
thermal radiation are discussed in section \ref{section:PhysAss}.  
(The effective magnetic field strength assumed was quite low, 
corresponding to 2.1 $\mu$G at 20 cm$^{-3}$\ and proportional to density,
so the nonthermal pressure was felt only in the dense shell and not very
strongly even there.)

The present application employs spherical coordinates 
($r,\theta,\phi$) and rotational symmetry with respect to the polar 
($\theta=0$, $ 180^{\circ}$) axis. The initial grid extends 
over radii $0.0\le r\le r_{\rm{max}}$ and polar angles 
$0^{\circ}\le\theta\le180^{\circ}$, containing 128 and 15 uniformly 
distributed points in $r$ and $\theta$, respectively. The adaptive 
mesh refinement takes place in two stages, as needed, with total 
factors of up to 20 and 6 improvement in r and $\theta$\ resolution.  
Thus, the resolution of the finest grid is equivalent to that of a 
uniform grid of 2560$\times$90 points extending over the whole 
computational domain, while requiring about a factor of 15 less 
CPU time.  This resolution was chosen after 1d experiments showed 
that with 2560 zones in radius the fully formed shell will be resolved 
into not less than 10 cells. At lower resolution we would not have been 
be able to describe the velocity distribution of the cold material.  Even 
with the adaptive mesh, the CPU time used for the simulation presented 
below amounts to $\sim 10$ days of the CRAY~J90.

\noindent{\rtx{Initial conditions}}

The code is initialized at a remnant age of 300 years, with the 
explosion energy in expansion of the ejecta.  The mass of the ejecta 
is 5 \msun; the density and velocity profiles are those used by Cid-
Fernandes \etal (1996).  The outer expansion velocity is about 5000 
km s$^{-1}$\ and therefore the radius of the ejecta at initialization is 
about 1.5 pc. 

The values assumed for the explosion energy (10$^{51}$\ ergs), 
ambient density  at the explosion site (6.2 cm$^{-3}$), and scale 
height of the ambient density distribution (22 pc) are those found from 
our analytical study as recorded in Table~\ref{2Dparameters}.

\subsection{Results}
\label{section:2dResults}

Each cell, or parcel, consists of an annulus in $\phi$, with volume 
$2\pi r^2 \sin\theta\delta r\delta\theta$.  At a specific evolutionary time 
the output for each annulus consists of the average $r$ 
and $\theta$, along with the 
corresponding density, temperature, mass, and the two components of 
velocity.  From these, the thermal and magnetic pressures can be 
inferred, along with the x-ray emission (assuming collisional equilibrium) 
and radio continuum emissivities.

\subsubsection{Structure}
\label{section:2dStructure}

Figures~\ref{2Ddensity}, \ref{2Dtemperature}, and 
\ref{2Dpressure} show the density, temperature, and pressure 
structures of the simulated remnant at an age of 20,000 years.  In the 
transparent side view of Fig \ref{2Ddensity}, the densities in all 
cells are shown, versus their locations along the major axis.  The 
locus of values, between 3 and 10 cm$^{-3}$, making up the 
straight diagonal line extending beyond the remnant shows the 
exponential density distribution of the ambient medium. The diameter
(major axis) 
is seen to be just over 22 pc, essentially one scale height in the ambient 
medium. 

There are deviations from spherical symmetry in the low density 
interior, but they are quite modest.  The central density is about  
15\% of ambient at the equator, its minimum being about 1.0 cm$^{-3}$,
as predicted by the analytical modeling.  In addition the 
distribution is very flat, exceeding the minimum by less than a factor 
of two over a large region in the interior.  

The cells with very high density are those in the cold dense shell.  
As we discuss below, the thermal pressure was too large at high 
density in this run, and the magnetic pressure was too low.  With 
these conditions, the densest ones should have been compressed by 
about a factor of 120 over ambient, and are, as shown.  Complete 
shell formation, shown by the locus of highest density cells, has 
reached from the dense end to the equator, just as predicted, beyond 
which the maximum compression fades sharply.  

The many cells with compressions somewhat less than maximum 
show the well resolved transition of material which has begun 
cooling but has not yet reached its limiting density.  At the tenuous 
end, shock compression by only the adiabatic factor of 4 is 
apparent, while not far away where radiation is just beginning to 
affect the structure, that factor of 4 is followed by further 
compression.  The final distribution of grid resolution is also apparent 
in the figure, where the regions of highest density have the 
greatest concentration of cells in r and $\theta$.

The temperature structure in Figure \ref{2Dtemperature} is shown 
as a 3d surface plot with the densest end on the near left side.  It 
shows a central temperature of about $6 \times 10^{6}$\ K, 
consistent with our expectation from the conduction analysis
(Section \ref{section:TCtheory}), with a 
slightly lower shoulder just left of center where the density was 
higher.  The trough in temperature behind the shock on the tenuous 
end is a surprise, indicating that the shock there is already becoming 
radiative where our analysis had predicted a cooling time of 25,500 
years (Table~\ref{2Dparameters}).  This is consistent, however, 
with indications (below) of a relatively low pressure on the tenuous 
end.  As discussed in Maciejewski and Cox (1998), a pressure difference between 
the two ends is apparently also responsible for the reduced 
displacement observed for the remnant center.  Apart from the 
aforementioned shoulder and the radiative shock girding the tenuous 
end, the temperature structure is nearly symmetric and parabolic.  
Both the central value and the general distribution are quite similar to 
1d results presented in  Section \ref{section:1d} below.

The pressure in the interior of the remnant (Figure \ref{2Dpressure}) 
is very flat in the center with a 
value of $1.6 \times 10^{-9}$\ dyn cm$^{-2}$,  
but drops abruptly by 5\% about halfway to the 
edge.  On the tenuous end, this drop leads to a broad shallow valley 
at 85\% of the central pressure.  On the dense end the pressure 
continues to drop with increasing slope, to a deep trough.  

Very close to the edge on the dense end, it is apparent that the low 
pressure is due to rapid radiative cooling.  The structure further in, 
which can also be discerned in the density details of Figure 
\ref{2Ddensity}, resembles a substantial rarefaction moving inward 
or a compressional wave moving outward.  Such waves are fairly 
commonly found in hydrodynamic models although none is obvious 
in the particular 1d model presented in the next section.  
When present, they have very 
long reverberation periods and thus appear for a long time as nearly 
stable features.  We explored several possible sources for this structure
in our 2d run and found that the high flat central pressure, its weak but 
abrupt edge, and its asphericity are the residual of the reflected 
reverse shock.  
(An experimental 1d run with the same initialization as the 2d run 
produced a very similar structure.)  Because a 
reverse shock is required to raise the pressure of the center, and it will 
surely be reflected to some degree, there should be a substantial time during 
which this sort of structure is present within a remnant.  But detailed 
conclusions based on its presence in a particular remnant model should 
be regarded with some suspicion.  When it is found, where, and with what 
strength, depends on the initialization.  Perhaps a fruitful approach would 
be to search the x-ray emission pattern of a remnant for indications of such a 
structure and, if found, to use the structure to evaluate models of the 
remnant's early state.

Stating our pressure results more optimistically, the interior
pressures in our standard 1d and 2d runs differ by less than 30\% 
anywhere that the temperature exceeds $2 \times 10^6$\ K.  The 
differences will have negligible impact on the interpretation of the x-
ray data.

The pressure structure at the very edge of the remnant bears an imprint of both the 
cooling history and the way we treat cooling in 2d.   (Note that the pressure 
spikes shown are an artifact of the way in which the figure was prepared, 
not a characteristic of the model.  
They are due to incomplete sampling of the densest region in 
converting the model data from spherical to cartesian coordinates for the 
plot.  This makes the shell conveniently transparent, 
but inaccurately suggests that it was poorly resolved in the model.)  
At the tenuous end, the pressure maximum is in the immediate 
post shock gas, prior to the onset of cooling.  Further inward 
where cooling has begun, the pressure drops precipitously.  
The structure at this end is entirely resolved.
At the high density end, the radiative shock is unresolved and the 
high pressure region is in the already cold dense shell, itself trailed 
by a deep pressure trough of cold but not yet compressed material.  
Comparing the maximum pressures at the two 
ends, we found $3.4 \times 10^{-9}$\ and $2.4 \times 10^{-9}$\ 
dyn cm$^{-2}$, for the dense and tenuous ends, respectively.  Moving 
away from either end the edge pressure drops rather rapidly.  On the 
tenuous end this appears to be due to the onset of cooling, which 
coupled with the inverse temperature dependence of the cooling 
coefficient, temporarily stalls the shock, possibly leading to a period 
of instability and secondary shock formation (Falle, 1975, 1981).  
This is most exaggerated in 
an extended region of the remnant's surface over which the post 
shock pressure seems to be very low, just toward the tenuous end 
from the equator.  In this region the radiative shock appears not to 
be resolved (the initial compression exceeds 4), but cooling is so 
rapid that the post shock pressure is small and the shock velocity 
temporarily low.  On the dense half, the pressure in the cold shell 
drops a factor of 
two between the end and the equator.  On the whole, these 
pecularities in the pressure structure provide the biggest deviations 
from the expectations based on analytic and 1d models.   

In summary, the magnitudes of central density, temperature and 
pressure are all as expected.
Similarly, shell formation over half the surface has occurred as 
planned.  There were some new but qualitatively understandable 
features brought to light, particularly the strong 
variation of edge pressure over the surface at this epoch, and the 
seemingly premature onset of cooling on the tenuous end.  We are 
left not knowing whether the detailed interior pressure distribution 
of the 2d case is a better or worse approximation than the 1d 
structure presented in Section \ref{section:1dStructure}, but are relieved that the differences are modest.




\subsubsection{H{\small I} Shell}
\label{section:2dHI}

In the particular orientation of our 2d model, the 
symmetry axis is tilted 40 degrees from the line of sight with the 
dense end on the far side, and toward the northeast, giving us a 
glancing view into the open end of the half shell.  With this 
projection, most of its dense material (and associated strong 
magnetic field) is in the northeast, where the average radio 
continuum brightness is highest, and the dense gas is moving 
predominately away, so that the ranges of expansion velocities 
observable via 21 centimeter emission are occupied only on the 
receding side, as observed.  

In preparing the figures of this section, the mass of each $r$, 
$\theta$\ annulus of the calculation is represented by a proportional 
number of dots, randomly distributed within the annular volume.  
The spatial coordinates and velocity components of each dot are then 
tracked through the rotations to the final projection.  (In later 
sections on the radio continuum and x-ray appearances, the process 
is similar but the number of dots per annulus is chosen in proportion 
to emission rather than mass.)  The dot density in the shading thus 
represents the projected surface density (or surface brightness in the 
emission pictures).  Unless otherwise noted, the assumed distance
of W44 was that required to give an angular diameter of 35$^\prime$,
about 2200 pc, closer even than our revised distance estimate
as quoted in Table \ref{obstable}.  Methods for scaling the results
were presented in Section \ref{section:scaling} and  their application
will be discussed
further in the conclusions.

Figures~\ref{2Dhydrogen} and \ref{2Dvelocity} present our results 
for the distributions of mass and velocity in the dense shell, in 
formats for direct comparison with the observations of Koo and 
Heiles (1995).  They include only cold material more dense than 100 
cm$^{-3}$, assuring a sufficiently short ($\sim$\ 1000 year) 
recombination time to achieve neutrality (the ionization state not 
actually having been tracked in the model).

Figure~\ref{2Dhydrogen} consists of 4 panels, two of which are 
projected column density, the others being position velocity 
diagrams.  The upper right panel shows the  total column density 
map of the dense half shell.  As expected, it is concentrated toward 
the northeast, sharply bounded and edge brightened there, but fades 
out gradually to the southwest.  

This view, however, is not available to observers because of the 
large amount of low velocity hydrogen toward W44.  The 
corresponding view presented by Koo and Heiles (1995) is limited 
to the velocity range $v_{\rm LSR} = 135$\ to 170 km s$^{-1}$\ and 
displayed as their Figure 1a.  Their figure and the corresponding model column 
density map for that velocity range (including the systemic velocity 
of +43 km s$^{-1}$) are shown in the lower left (oblique view) panel of 
Figure~\ref{2Dhydrogen}.  The model view changes considerably when 
limited to this velocity range.  It is much more diffuse, lacking the sharp 
bright northeastern edge, smaller, more symmetric, and has a hole in 
the center from omission of highest velocities.  With these changes, 
it bears a very strong resemblance to the H{\small I} observations.  

The model's restricted velocity view has a total diameter of about 
25$^{\prime}$\ (a mass 
weighted diameter of about 18$^{\prime}$) versus the shell's total diameter of 
about 35$^{\prime}$.  For comparison, Koo and Heiles quote the mass 
weighted radius at $v_{\rm LSR} = 140$ km s$^{-1}$\ as 7.5$^{\prime}$.  The 
corresponding mass weighted 15$^{\prime}$\ diameter is smaller than our 
model.  As will become apparent below, the model is in good 
agreement with the data envelope in position velocity diagrams, but 
the H{\small I} data are more filled in velocity space, leading to a lower mass 
weighted size.

Taking model data from within the apertures outlined in the upper 
right panel of Figure~\ref{2Dhydrogen}, the upper left and lower 
right panels show the corresponding position velocity diagrams.  In 
contrast to the total surface density plots, the half shell of the model 
is immediately apparent.  The model remnant would be observable 
in 21 cm almost exclusively on the rear (receding) surface, as is 
W44.  (If our model were absolutely correct in its degree of shell 
formation and orientation, however, we would predict the 
observability of a small quantity of rapidly approaching H{\small I} in the 
northeast.)  

In Figure~\ref{2Dvelocity}, portions of the position velocity 
diagrams with W44's systemic velocity of +43 km/s added, are 
replotted with the same restricted velocity range as in Figs. 1c and 
1d of Koo \& Heiles (1995).  In addition, the H{\small I} data were kindly 
supplied to us by Koo, and are shown for comparison.  

The correspondence between the observations and the model is 
again very good: both show the highest velocity of the bulk of the 
material to be about 180 km/s, but with some moving as rapidly as 
200 km/s.  The highest velocities are found toward the center, with 
lower values near the edge and a more hollow center, indicative of 
an expanding shell.  The
radial scales correspond very well on the whole, with the model 
distribution outlining the data and having a diameter of about 25$^{\prime}$\ at 
140 km s$^{-1}$, about 2/3rds of the actual remnant diameter.

As noted above, the observations seem to have a wider range of 
velocity at a given radius than does the model, with a lumpier 
distribution of material.  
The model is highly 
idealized, neglecting small scale irregularities in the ambient medium 
and any turbulence that may be present.  As we discuss and 
demonstrate below, the radio continuum maps show strong 
filamentation that is associated with surface irregularities which, in 
addition to their effect on the surface brightness profile, will tend to 
broaden and disturb the velocity structure.  For example, a bright 
radio continuum filament indicates a tangential view of that portion 
of the remnant surface.  When such a filament is seen projected well 
away from the remnant edge, the line of sight velocity there will be 
very small, because the surface motion is turned crossways.  
Any remnant whose radio continuum map shows an intricately 
textured surface like that of W44 will have a much broader and more 
disturbed radial velocity distribution in 21 cm than any smooth shell 
model.

Considering that the model made no attempt to include any 
irregularities, it is somewhat surprising that its velocity width at the 
remnant center is as broad as it is.   This appears to be due to the 
ongoing rapid development 
of the shell.  There is material which has cooled and been 
compressed, but has not yet merged with the densest gas in velocity 
space.  This is indicative of the strong velocity convergence in the 
neighborhood of the densest regions, as discussed in 
Section \ref{section:RCDisc}
with regard to suppression of cosmic ray diffusion from the shell.  
There may also be a contribution from the instability of radiative 
shock waves at high velocity (Falle, 1975, 1981).



Koo and Heiles (1995) reported that the mass associated with the 
shell and having $v_{\rm LSR} \geq 130$ km s$^{-1}$\ is 73 $\pm$ 3 
M$_{\rm \odot}$.  The mass of the simulated shell having velocities 
within this range is 149 \msun at an age of 20,000 years.  The 
dependence of the model results on remnant age are shown in 
Table~\ref{2Dprogression}, both for the mass within that velocity 
range and for the total mass at density above 100 cm$^{-3}$.  (The 
results depend only slightly on the density criterion for inclusion in 
the neutral shell.)  Apparently the model shell's cold mass is about a factor 
of two larger than the observed H{\small I} mass, a conclusion that does not 
depend 
strongly on the age at which it is evaluated.

Several neglected factors are probably responsible for this differential.  
For example, the shell is partly ionized.  In Section 
\ref{section:Alphaetc} we found that (because of photoionization) 
each atom joining the shell had to recombine about 3 times.  We 
expect the ionized gas in the shell to be roughly equal to that 
collected during the past 3000 years (3 recombination times at 100 
cm$^{-3}$), lowering our estimate of the neutral mass with 
$v_{\rm LSR} \geq 130$ km s$^{-1}$ by up to 50 \msun.  Similarly,
the presence of the OH masers implies that the shell is at least partly
molecular.  If 200 of the 450 \msun in the shell were H$_2$, there 
would be no discrepancy.
Finally, the fact that the observed remnant has surface structure that 
broadens the velocity distribution in its position velocity diagrams 
makes it difficult to compare the amounts of mass in an observed 
velocity range.  The distortion could easily 
have skewed the velocity distribution downward near the cutoff 
velocity, reducing the measured shell mass.

\begin{table}
\begin{center}
\caption{Time Dependence of 2D Model Characteristics}
\begin{tabular}{|cccccc|}
\multicolumn{6}{c}{}\\ \hline\hline
Age (years)                     &   16,000  & 17,000      &   19,000     &
  20,000  &     21,000\\ \hline 
Cooled Mass  (\solm)   &  230     &     310      &      475     &      
565  &       655\\
Mass in Shell with        &        &      &       &      
     &      \\
n $\ge$ 100/cc (\solm)   &      135     &     210      &      370      &
450      &      535\\                                                     
Shell Mass with                 &    &    &     
&       &     \\
V(LSR) $\ge$ 130 km/s (\solm)   &      60     &     88      &      134
&      149      &      157\\\hline
Radio Continuum      &     104    &        179      &        346        
&     423       &    503\\ 
100 MHz (Janskys)    &   &   &   &   &   \\ \hline
ROSAT Hard Band (counts/s)    &   &   &   &   &   \\
NH = 1.0E22              &      8.5     &     7.3      &      5.5        &      
4.8      &      4.2\\
NH = 1.5E22              &      4.8     &     4.1      &      3.1        &      
2.7      &      2.4\\
NH = 2.0E22              &      2.9     &     2.5      &      1.9        &      
1.6      &      1.4\\ \hline
ROSAT Soft Band (counts/s)    &   &   &   &   &   \\
NH = 1.0E22              &      29.0    &    24.9    &      18.9      &      
16.6    &      14.7\\
NH = 1.5E22              &      7.5     &     6.5      &      4.9        &      
4.3      &      3.8\\
NH = 2.0E22              &      2.5     &     2.1      &      1.6        &      
1.4      &      1.2\\ \hline\hline
\end{tabular}
\label{2Dprogression}
\end{center}
\end{table}

\subsubsection{Radio Continuum}
\label{section:2dsynch}

This section discusses the radio continuum emission expected due to 
the van der Laan mechanism, following the prescription in Section 
\ref{section:synchImpl}.  Before doing so, however, we had to 
overcome one difficulty.
Our 2d model assumed an 
unrealistically low value for the magnetic pressure.  In addition, cooling 
was suppressed below $10^4$\ K, yielding an excessive thermal pressure.  
Together these led to results for density in the shell that were unreliable 
(though the two effects roughly cancelled and the density structure turned 
out to be reasonable. 
For this section, we corrected the 
compression factor in each annulus of the calculation, under the 
assumptions that: 
 the hydrocode provides a correct value for the annulus's total 
pressure, the initial density is the 
ambient density just outside the shell at the same z (that is, that 
material in the dense shell has not moved far in z), the ambient 
tangential field outside the remnant is $3 \mu$G independent of external 
density, and that thermal and cosmic ray pressures are negligible
within the shell 
(yielding a reasonable but optimistic assessment).  Setting the 
magnetic pressure equal to the total computed pressure in the 
annulus determines the corrected compression factor for calculating the gain.  
Following the rest of the prescription provided the flux densities shown in Table
\ref{2Dprogression}.
As found in the analytic section, this slightly optimistic estimate of the 
van der Laan flux density is consistent with the observation of 500 
Janskys at 100 MHz.  But once again there is the hidden assumption that
the background can be produced by CR electrons in the mean ambient field,
an assumption that fails by approximately a factor of 4 when applied
to estimates in the solar neighborhood.

The 2d model also makes possible a projected image of the synchrotron
 emission,
prepared in much the same way as 
was done for the H{\small I} structure.  Under the assumption that the true 
radio continuum emissivity is proportional to our van der Laan 
modeling, and neglecting the directionality and polarization of the 
emission relative to the magnetic field, our map for the radio 
continuum is shown for our standard projection in the upper right 
panel of Figure~\ref{2Dsynch}.  A similar image in the lower left 
panel shows the view that would be had from a direction 
perpendicular to the major 
axis; the half shell of the model is obvious. The figure 
also shows three horizontal and three vertical slices through the 
emission.  The central slices quantify the strong degree of edge 
brightening apparent on the northeast.

The observed radio continuum distribution, illustrated beautifully in 
Smith \etal (1993) and in Giacani \etal (1997), is qualitatively similar but 
differs in two significant characteristics.  The observations in fact 
were the stimulus for our projected half shell model, giving the 
distinct impression of a view into an open hemisphere, 
brighter in the northeast than southwest because of the double 
contribution from the two sides in projection.  Our model, however, 
has a blurred edge extending further around the shell, and thus 
projects into a more rounded configuration with less of a step in 
overall brightness.  The second major 
difference is that the observations show quite pronounced 
filamentation, some of which appears to be associated with the sharpness
of the equatorial edge.  Hester (1987) has shown in the case of the optical 
emission of radiative shocks that a high degree of filamentation can 
derive from spatial irregularities in very thin emitting structure.  In 
order to test this as a possible source for the filamentation of the 
radio continuum in W44, we prepared an image of the 2d model in 
which the individual emission elements were radially displaced in a 
sinusoidal pattern resembling the corrugated surface of an acorn 
squash.  The results are shown in Figure~\ref{2Dcrinkledsynch} 
and very clearly show that a modest degree of surface crinkling can 
lead to striking filamentation.  The horizontal and vertical slices 
show the filamentation contrast.

In modeling the filamentation in this way, the areas of especially 
bright radio continuum are assumed to derive from tangencies along 
the surface from regions which are extremely thin.  As such, they 
should also be filaments in H{\small I}, optical, and all other emissions 
associated with the dense shell and radiative shock, as is observed.  

Because the ambient cosmic ray electrons have a flatter energy distribution 
at lower energies, the radio continuum 
spectrum of W44's filaments should be flatter than that of the 
remnant's more diffuse component.  This comes about 
because, qualitatively, there are two layers of radio emission, a very 
thin bright one from fully compressed material in the dense shell, 
and a somewhat thicker layer from cooled material in the process of 
compression.  The thinner layer dominates the filaments because it is 
most edge brightened by tangency, while the two layers contribute more 
equally to the diffuse component.  On average, then, the filaments have 
suffered greater compression and their emission derives from electrons 
which were initially at lower energy, where the spectrum is flatter.
Potential applications of this 
phenomena include the possibility of inferring the very low energy 
interstellar cosmic ray electron energy distribution from filament spectra 
of very high pressure (small, high density) supernova remnants.



%

\subsubsection{X-Rays}
\label{section:2dX}

The x-ray surface brightness distribution and count rate for the 2d 
model were calculated from equilibrium Raymond and Smith 
spectra, folded through ROSAT PSPC response functions for two bands 
(soft R4+R5+R6, 0.4 - 1.56 keV, and hard R7, 1.05-
2.04 keV), 
and attenuated by photoelectric 
absorption by the foreground material. The total count rates for both 
bands, at three absorption column densities, for five ages are shown 
in Table \ref{2Dprogression}.  The observed band ratio of about 1 for 
these two bands (Rho \etal, 1994) points to an 
absorption column  between 1.8 and 1.9 $ \times 10^{22}$ cm$^{-2}$.  At an 
age of 20,000 years, the corresponding total predicted ROSAT PSPC count rate 
is then also close to the observed 4.2 s$^{-1}$\ found by Harrus \etal (1997).  

The distribution of X-ray surface brightness for this case is shown 
in Figure~\ref{2DXray}.  It is brightest in the center, approximately 
gaussian in appearance, 
and highly symmetric about the center.  The full width at half 
intensity is about 19$^{\prime}$, just over half the full  diameter.  There is 
nothing about the structure that appears to be particular to the two 
dimensionality of the calculation or lead one to suspect there was a 
significant density gradient in the ambient medium.  The center of 
the emission is shifted slightly away from the explosions site, but is 
well centered within the remnant.  Of course, when comparing with 
the observations, one has to bear in mind the strong sensitivity of 
the count rates to the absorption column density.  It is quite possible 
that variability in the absorption screen leads to artificial structure in 
the observed x-ray surface brightness.  Scattering of the X-rays by
intervening dust may also be a factor.

The appearance the model remnant would have without absorption  is 
shown in Figure~\ref{2DXrayBare}.  From 
this result, it is clear that at this particular age, 
the outer parts are everywhere too cool to contribute to the emission 
and the absence of the outer shell emission is not primarily due to 
absorption.  The fact that the pressure drops somewhat away from 
the remnant center also contributes to this behavior. 

>From these results we are confident that 1D modeling of the 
nonequilibrium ionization structure and x-ray spectrum presented
in the next section reasonably corresponds to a hypothetical 
2d model which followed the ionization. 

%

\section{One-Dimensional Model}
\typeout{One-Dimensional Model}
\label{section:1d}
\subsection{Background}
\label{section:1dBkgd}
\noindent{\rtx{Purpose}}

A variety of 1d hydrocode models were made during the course of this 
investigation.  They were used to verify the approximate cooling time 
formula, to explore the lag time between cooling and shell formation 
described in Section \ref{section:ShellForm}, to explore the time development 
of shell mass and radio continuum emission, to measure the resolution
needed in the 2d model, to examine the dependence of 
the central conditions (density and temperature) on the assumed degree of 
suppression of the conductivity, to explore the degree of dust destruction 
expected within the hot interior of the remnant, to simulate inclusion of
additional heavy element abundances in the ejecta, to explore the 
dependence of the interior pressure structure on the SNR initialization 
characteristics as per Section \ref{section:2dStructure}, and to calculate 
non-equilibrium x-ray spectra for comparison with the surface brightness 
distribution and with the available spectral information. 

\noindent{\rtx{Code Description}}

Our code architecture employs a Lagrangian mesh, based on the scheme
discussed in Richtmyer and Morton (1967) and described in detail
in Smith and Cox (1998).  The method is similar to that 
of Cui and Cox (1992) and Slavin and Cox (1992).  The ionization 
and emission characteristics are discussed in Section 
\ref{section:PhysAss}.

\noindent{\rtx{Initial conditions}}

A simulated supernova explosion is initialized as a 2 pc radius sphere 
containing 2 solar masses of ejecta with a velocity linearly proportional to 
the distance from the center. 

The values assumed for the explosion energy (10$^{51}$\ ergs), and
ambient density at the explosion site (6.2 cm$^{-3}$) are those found
in our analytical survey and recorded in Table~\ref{2Dparameters}.
All results reported are for an age of 20,000 years.  With these
parameters and age, the model SNR is found in an adolescent stage of
shell formation.  It approximates edge conditions only near the
equator of the 2d model, and is therefore not directly useful in
studying the radio continuum or shell development.  We have, however,
examined these characteristics at later times and in models that were
higher in density to verify behaviors of the 2d results.

Results of six models appear below, differing in assumptions about
the level of thermal conductivity (full, reduced, zero), dust
destruction (utter, thermal spallation only) and abundances in the
ejecta (solar, 5 solar).  Their different characteristics are
summarized in Table \ref{table:1dRuns}.

\begin{table}
\begin{center}
\caption{Parameters of 1d Runs}
\label{table:1dRuns}
\begin{tabular}{|cccc|}
\multicolumn{4}{c}{}\\ \hline\hline
\multicolumn{4}{|c|}{$n_{\rm o}$ = 6.2  atoms cm$^{-3}$, $E_{\rm o}$ = 
1.0 $\times 10^{51}$ ergs, t = 20,000 years}\\ \hline\hline
Model  & Conductivity  & Dust Destruction &  Abundances  \\  \hline 
   A      &       full &          utter   &    solar     \\
   B      &       zero &          utter   &    solar     \\
   C      &       full &          utter   &   enhanced   \\
   D      &       zero &          utter   &   enhanced   \\
   E      &     1/10th &          utter   &    solar     \\
   F      &       full &       sputtering &    solar     \\  
\hline\hline
\end{tabular}
\label{1dRunTable}
\end{center}
\end{table}

\subsection{Results}
\label{section:1dResults}

At a specific evolutionary time the output for each parcel consists of the 
inner and outer radii and velocities of each parcel, plus averages of density, 
temperature, thermal and nonthermal pressures, and nonequilibrium 
ionization.  In models following gradual dust destruction, or enrichment in 
ejecta, the gas phase elemental abundances are also recorded for 
each parcel. From these, the x-ray spectral emissivity of the hot interior, 
shell mass and its velocity distribution, and radio continuum emission can 
be calculated as described in Sections \ref{section:codeTR} and
\ref{section:synchImpl}.

\subsubsection{Structure}
\label{section:1dStructure}

Figures~\ref{1Ddensity}, \ref{1Dtemperature}, and
\ref{1DThermalpressure} show the density, temperature, and thermal
pressure structures of Models A and B at an age of 20,000 years.  The
differences, due entirely to thermal conduction, are immediately
apparent.\footnote{ {\atx{The glitches at a radius of 6 pc in the
Model B results are at the contact discontinuity for that run and
arise from the discontinuity in velocity at initialization.  For Model
A, the 2 \msun of ejecta is all interior to 2.5 pc and the
discontinuity has been erased by thermal conduction.}}}  In agreement
with the 2d run, the radius is just over 11 pc and the interior
electron density in Model A is about 0.92 cm$^{-3}$\ (nuclear density
about 0.84 cm$^{-3}$), and very flat.  This central density is 13\% of
the ambient density, just as predicted in Section
\ref{section:TCtheory}.  For Model E (not shown), with its thermal
conductivity reduced by a factor of 10, the central electron density
is still fairly flat but has a value of only 0.51 cm$^{-3}$.  The
central density reduction by a factor of 1.8 with a conductivity
reduction of 10 is also exactly that expected from the analytical work
in Section \ref{section:TCtheory}.

The temperature structure in the interior of Model A also reconfirms
the 2d model, dropping only gradually with radius from a central value
of 6$\times 10^6 $\ K, in the range consistent with x-ray
observations.  In the outer parts, thermal conduction has delayed the
progress of shell formation somewhat and provided a slightly larger
remnant.  Models A and B both have similar temperature and density
ranges in the outer parts, so that, as in Harrus \etal (1997), it is
possible to find an x-ray model with a similar total luminosity, but
it will be flatter or even shell like in surface brightness, and may
exhibit a stronger radial temperature profile.

The pressure in the remnant interior is very flat in either case, as
expected, with a value of about 1.5$\times 10^{-9}$\ dyn cm$^{-2}$
for Model A.  As previously mentioned in Section
\ref{section:2dStructure} the pressure distribution of Model A is
smoother than that of the 2d model and the difference was traced to
the different initializations.




\subsubsection{X-ray Surface Brightness Distribution}
\label{section:1dXSurf}

Figures~\ref{fig:image}a and b compare the predicted radial flux
distributions from Models A and B (after absorption) with observations
in the ROSAT PSPC soft (R4+R5+R6, 0.4 - 1.56 keV) and hard (R7, 1.05-
2.04 keV) bands.  The model spectra were attenuated by a constant
N$_{\rm{H}} = 1.89\times 10^{22}$ cm$^{-2}$ (found by fitting the
Model A spectra to observations, below), although Rho \etal (1994)
found best fit values for collisional equilibrium isothermal models
varying between 1.57 and 2.10 $\times 10^{22}$ cm$^{-2}$ across the
remnant.  The model remnant was taken to be at a distance of 2.2 kpc
for these plots so that the angular diameter of the shock would be
$35^{\prime}$.  This has no effect on the magnitude of the surface
brightness but determines the angular scale and PSPC count rate.

Obtaining a meaningful radial distribution from the asymmetric ROSAT
image involved several steps.  The data were cleaned and calibrated
using the method described in Snowden \etal (1994), and following the
examples given in Snowden (1994).  Emission from a bright G star on
the left side of image was then removed and, using Jones \etal's
(1993) choice for the nominal center ($\alpha,\delta = 18^h 53^m 30^s,
1^\circ 15^{\rm{'}} (1950)$) the data were averaged over a $90^\circ$\
angle in each of the four cardinal directions (north, south, east,
west).  The persistence in Figures~\ref{fig:image}a and b of the
surface brightness to greater radii in the north is real: the X-ray
distribution is longer north-south than east-west and the point chosen
for the nominal center is south of its centroid.


Overall, the count rate and surface brightness distributions of Model
A and our 2d model are in excellent agreement.  Compared to
observations, they account for several key features: the surface
brightness is highest at the center, the limb is not brightened, and
the total count rate of Model A is just 12\% higher than observed (see
Table~\ref{table:emission}).  Model B, however, without conduction,
has a factor of 2 too little emission overall and is edge brightened.
(A model similar to B, by Harrus \etal (1997) was somewhat more
successful.)  The only apparent failings of the conductive models are
that their emission distribution is dome-like while the observations
resemble a peaked volcanic cone,\footnote{ {\atx{Note that sharpness
of the central brightness peak varies among this class of supernova
remnants, with W44 and W28 having sharper peaks than 3C391 or W63
(Rho, 1997).}}}  and that the 2d model is too symmetric, not providing
the north---south elongation apparent in the image.  The next few
paragraphs record various experiments which explored model
complications that might sharpen the central peak of the distribution.

\noindent{\it{Emission from Thermalized Ejecta?}}

At 1 cm$^{-3}$, a radius of 4 pc contains 8 \msun.  Thus, the mass in
the central few parsecs of W44 should clearly be heavily contaminated
by the stellar ejecta and enriched in metallicity.  Instabilities and
turbulent mixing potentially offer the possibility of diluted
enrichment through much of the roughly 100 \msun in the hot interior.
With an enrichment that decreased with radius, the observed radial
profile could be matched.  What is surprising, however, is how little
enrichment could be tolerated in conductive models before the
prediction exceeds the observed brightness.  For example, the factor
of only 5 enhancement over solar abundances in the 2 \msun of ejecta
in Model C increases the line emissivity by a factor of 5 in the inner
2.5 pc $\sim 4'$, and the central surface brightness by about a factor
of 2.  More than that would be too much unless compensated by modeling
a less dense remnant at greater distance or by a reduction of thermal
conductivity.  Interestingly, the nonconductive models do not suffer
from this limitation; they are so hot in the interior that their
(already weak) central emission is mostly due to bremsstrahlung and
does not increase substantially.  Our tentative conclusion is that the
elemental abundances of the ejecta must be accommodated much more
realistically before the very centermost region can be addressed.  A
major part of getting that right will be to manage to hide most of the
metallicity, in clumps that were never fully heated, in overdense
regions which have already managed to cool, or in dust grains.

\noindent{\it{Cloudlet Evaporation}}

Setting aside concerns over the ejecta, we imagined a hybrid model,
including thermal conduction in the interior 
(represented by our hydrocode results) and cloudlet
evaporation as parameterized by White and Long (1991).
In particular, we looked for possibilities in which evaporation
contributes to the brightness only in the central few parsecs.
We expected that 
somewhat lower mass loss from evaporating clouds would be
required than estimated by Rho \etal (1994) or Jones \etal
(1993). This possibility is not clearly successful, however.  Figure 4
of White and Long (1991) shows that the effect\footnote{ {\atx{The
environment for their work was a model which included thermal
conduction only in the the evaporation of clouds, while our picture
assumes thermal conduction everywhere.  Perhaps the two views do not 
mix as easily as we have assumed in this discussion.}}}  of an
evaporating population of clouds is to increase the column emission
measure for projected radii less than about 0.65 to 0.85 of the shock
radius (depending upon the chosen evaporation parameters).  Only in
the extreme case of an infinite evaporation time scale and very large
ratio of preshock cloud mass to preshock intercloud mass is the
contribution in the center significantly greater than that at medium
impact parameters, and even this extreme case is not significantly
more centrally peaked than our models, owing to their flat central
distributions of density and pressure.  Their model as it stands does
not seem to contribute usefully to peaking of the distribution.

\noindent{\it{Restriction of Dust Destruction to Sputtering}}

Restricting dust destruction to thermal sputtering in the hot gas was
considered in Model F, using the method described in Smith \etal
(1996) to calculate the rate.  We found that dust destruction was
greater in the center of the remnant, providing a relative enhancement
of the abundances of the hot gas, but that the effect was not strong
enough to match the peaking of emission in the center of the remnant.
Additionally, since the plasma with depleted abundances is less
efficient at X-ray emission, the total flux was reduced to about
$50\%$ of that of Model A with the same absorption column density.  A
more careful study including dust destruction processes in the shock
and revision of absorption based on the new model spectrum might be
more successful overall.  In any case it is somewhat dangerous to
assume that the elemental abundances everywhere in the hot interior
are undepleted.  Thermal sputtering alone is insufficient to destroy
the dust completely.

\noindent{\it{How Can a Remnant be Shell-Like in Radio and Centrally 
Bright in X-rays?}}

Like W44, our Model A does not appear limb brightened.  And, as with
our 2d model, this absence of limb brightening is not dependent on
preferential absorption of softer x-rays emitted by the cooler material
further out, although that effect certainly shapes the profile
somewhat.  The density rises toward the outside, and there is
considerable emission from material there, but not at x-ray
wavelengths.  The reason that our model of W44 is centrally brightened
in x-ray emission is that the explosion occurred in a sufficiently
high density that the X-rays emitted by the interior are still bright
when the outer part of the remnant becomes too cool to detect.  Using
our rule of thumb that, at the shell formation time, the central
density in the presence of thermal conduction should be about 0.13 of
the ambient density, explosion site densities in excess of a few
cm$^{-3}$\ should lead to radiative remnants (shell like in radio
continuum) with observable x-ray emission from the interior.  Without
thermal conduction, 
x-ray emission is confined to an interior thick
shell-like region, as in our Model B, with temperature and density
similar to Model A. Harrus \etal (1997) showed that such a model
would not necessarily show its shell like
character in projection, although our Model B does.  Although Model A
works better, it still fails to address the central peak without
further complications.

\subsubsection{X-ray Spectrum}
\label{section:1dXSpect}

\noindent{\it{X-ray Spectrum}}

The observed spectra from the ROSAT PSPC and {\it Einstein}\ SSS are
compared with Model A in Figure~\ref{fig:spectra}a and b.  Because the
SSS had a $6^{\prime}$\ diameter field of view, the model SSS spectrum
was calculated using a cylindrical slice through the center of the
remnant while the PSPC spectrum used the entire remnant.  Because our
models are somewhat underluminous in the center and the Einstein SSS
and ROSAT PSPC have different fields of view, the normalization for
each instrument fit was allowed to vary independently.  The total fit
to both datasets has a reduced $\chi^2$\ of 2.0.  Most of the
$\chi^2$\ comes from the Einstein SSS data because the model has too
little emission in the Si {\sc xiii}\ complex around 1.85 keV.  The
best fit normalizations turned out very well. The model was
underluminous compared to the Einstein SSS observations, by 36\%.  For
the PSPC, however, Model A predicts $25\%$ more emission than is seen.
Since the SSS had a smaller field of view than the PSPC, this result
suggests the central region is brighter than this model predicts.  The
best fit N$_{\rm H}$\ was $1.89 \times 10^{22}$\ cm$^{-2}$, a value
we've chosen to use for all other comparisons, and which concurs with
the 2d model conclusions.


The spectrum for Model C with its enhanced abundance ejecta is shown
in Figure~\ref{fig:metalspectrum}.  The column density fit to the PSPC
and SSS data is barely changed (N$_{\rm H} = 1.88\times10^{22}$\
cm$^{-2}$), but the PSPC response to the model now implies the model
is $35\%$\ too bright.  The model also overestimates the Einstein SSS
emission, predicting $23\%$\ more emission than observed.  The overall
spectral fit to the data is slightly worse; the reduced $\chi^2 =
2.5$, and again the worst part is the fit to the Einstein SSS at 1.85
keV.


\noindent{\it{Ionization Equilibration}}

Rather than a thorough exploration of collisional ionization
equilibration (CIE) of the diffuse interior in the presence of
substantial conduction, we content ourselves with a comparison of the
x-ray spectra of Model A, with and without the assumption of CIE, to
learn what instrument characteristics are required to distinguish
between them.  Although we have the complete nonequilibrium ionization
(NEI) balance saved for each parcel, we can calculate the
corresponding CIE model as well.  We did two different model fits.  In
the first case, we allowed both the overall normalization and the
absorbing column density N$_{\rm H}$\ to vary.  In this case, the best
fit value (with reduced $\chi^2 = 1.89$) had N$_{\rm H} =
2.03\times10^{22}$\ cm$^{-2}$.  The PSPC model was 10\% underluminous,
while the Einstein SSS model was underluminous by a factor of 2.
Figure \ref{fig:equilAndNon} compares this equilibrium model with the
nonequilibrium spectra as seen by the ROSAT PSPC and the Einstein SSS.
For the second case we held the column density fixed at N$_{\rm H} =
1.89\times10^{22}$\ cm$^{-2}$\ and fit the data.  In this case the fit
was extremely poor (reduced $\chi^2 = 4.7$).  The PSPC model,
unsurprisingly, is now overluminous by 14\% while the Einstein SSS
model emission is still a factor of 1.82 underluminous.

The Einstein SSS spectrum shows three strong peaks, which were
identified by Jones \etal (1993) to be from Mg {\sc xi}\ (1.34 keV),
Si {\sc xiii}\ (1.85 keV), and S {\sc xv}\ (2.45 keV).  These ions are
all part of the helium isosequence, and each corresponds to the strong
helium triplet of forbidden, intercombination, and resonance lines.
(See Gabriel \etal (1991) for a fuller discussion of these lines in
the context of O {\sc vii}.)  The largest difference between the CIE
and NEI models is that in equilibrium, the dominant stage is
helium-like.  In the NEI model, the gas has been over-ionized by the
shock, and slowly recombines from the hydrogenic ion to the
helium-like one.  In this case the forbidden line of the triplet will
be pumped by the recombination. A resolving power $E/\Delta E > 60$\
would be needed to separate these lines.  The hydrogenic ion, however,
also has emission lines--specifically, the Ly$\alpha$\ lines, which
are at 1.47 keV, 2.01 keV, and 2.62 keV respectively for Mg {\sc xii},
Si {\sc xiv}, and S {\sc xvi}.  In the recombining plasma, these lines
may be an order of magnitude stronger than at equilibrium.  Energy
resolution of only $E/\Delta E > 11$\ is required to distinguish them.
The Einstein SSS has a FWHM of $\sim 0.14$\,keV (Giaconi \etal 1979)
which would allow at best a marginal limit be placed on these lines.


\noindent{\it{Total Count Rate}}

Table~\ref{table:emission} shows the total count rates in the two
ROSAT PSPC bands, soft (R46) and hard (R7), for the observations and
the three different models. The observational values are extracted via
the method of Snowden \etal (1996), while the model values are
calculated using a response matrix generated through FTOOLS.  The
assumed N$_{\rm H}$\ was $1.89 \times 10^{22}$\ cm$^{-2}$\ for models
A, B, and C.  Using Xselect (Turner, 1996) and XSPEC (Arnaud, 1996) to
reduce the data gives an $\sim 10\%$ higher count rate for these bands
than the Snowden \etal (1996) method.  The reason for this discrepancy
is not completely understood, but is probably due to the improved
background subtraction done in the Snowden \etal (1996) method.

\begin{table}
\begin{center}
\caption{The observed and calculated emission in the soft and hard ROSAT 
bands.}
\label{table:emission}
\begin{tabular}{|ccccc|}
\multicolumn{5}{c}{}\\ \hline \hline 
     Band      & Observations      & Model A & Model B & Model C \\ \hline 
soft (R4+R5+R6)& $2.026 \pm 0.069$ & $2.22$  & $0.923$ & $2.37$  \\
hard (R7)      & $1.838 \pm 0.067$ & $2.12$  & $0.915$ & $2.28$  \\ \hline
\end{tabular} 
\end{center}
\end{table}

\noindent{\it{Reduction of Thermal Conductivity}}

As discussed in Section \ref{section:CondImpl}, there is some
controversy regarding the efficacy of thermal conduction in the
presence of a magnetic field.  In order to explore the effect of
lowering the thermal conductivity, we calculated Model E, identical to
Model A except with the thermal conduction constant $\beta$ reduced to
10$\%$ of the standard value of $6\times10^{-7}$\,cgs.  This model has
a higher central temperature ($T_c = 9.8\times 10^6$\,K), lower
central electron density ($n_c = 0.51$\,cm$^{-3}$), and steeper
gradients than does case A.  As a result, the emission spectrum
(Figure~\ref{fig:reducedTC}) is harder.  The final fit is poor, with a
reduced $\chi^2 = 3.4$.  The PSPC model is overluminous by 5\% while
the SSS model is underluminous by a factor of 2.13.  The best-fit
value has $N_H = 1.78\times10^{22}$\,cm$^{- 2}$, slightly less than
for the best fit to Model A.


\section{Discussion}
\label{section:Discussion}

The methodology of this investigation has been the application
of analytic theory as far as possible, followed up with
numerical hydrodynamic modeling to verify the details.
The main successes are summarized in the 
abstract, while the overview of the paper's organization and 
an index to the various results are provided in the 
introduction.  A number of analytical tools, some in review 
and some unique and new, were presented in
\S \ref{section:AnalMod}.  The synchrotron flux available from 
the van der Laan mechanism is discussed in \S
\ref{section:vdL}, with the main result in Equation 
\ref{equation:fluxForm}.  Proof that thermal conduction should 
lead to a fixed ratio of central to ambient density at the 
cooling epoch, and the dependence of that ratio on 
conductivity appear in
\S \ref{section:TCtheory}. The 
theoretical section is summarized in
\S \ref{section:AnalSum}, 
and a schematic of the basic model appears in 
Figure~\ref{schematic}.  The remainder of this section 
provides a brief summary as a backdrop for a fuller 
confrontation with a few key points. 
Each subsection presents successes first, followed by
caveats, and ideas for observations.

\subsection{Energy, Environment, Age}
\label{section:EEA}

The model we have presented for W44 is the remnant of a 
$10^{51}$\ erg supernova.  It occurred in a region with an 
explosion site density of about 6 cm$^{-3}$\ and a density 
gradient whose scale height equals one remnant diameter at 
the present time.  Our hydrodynamic studies assumed a 
smooth density distribution, but there must be irregularities to 
achieve the shell corrugation implied by the filamentary radio 
continuum emission pattern.  The age of the remnant is that of 
the associated pulsar, about 20,000 years.

None of observations we interpreted required the remnant to 
be interacting with dense molecular material in the vicinity. 
The model's dense end has a preshock density of only 10 cm$^{-3}$.
The remnant itself shows no significant morphological relationship 
between its radio continuum and the oft mentioned Wooten cloud.
It seems to be 
in a region with surrounding molecular clumps, but is not 
definitely encountering any.  That said, we stress two caveats.  
First, the shell corrugation is a consequence of a lack of 
smoothness of the ambient medium on 1 to 10 pc scales and 
we have not explored the degree of density contrast required 
to generate that corrugation.  Second, we found that the 
gamma ray flux could be raised to the observed level from the 
vicinity if the cosmic rays in the dense shell were interacting 
with 2 to 4 times as much mass as we predict.  It is not known 
that the gamma rays do derive from the remnant, but, if it can 
be shown that the eastern shell is the source and has the 
presently quoted flux, it may be necessary to revisit the 
possibility of dense inclusions or high contrast irregularities.

\subsection{The Hot Interior}
\label{section:HotInterior}

We assumed that thermal conduction in the hot interior 
occurs at its unquenched rate, limited by saturation 
(\S \ref{section:CondImpl}), with the result that the minimum 
interior density drops only to about 13\% of ambient and the 
central temperature is limited to about $6 \times 10^6$\ K.
These conditions produce 
the observed central x-ray 
emission
 (\S \ref{section:TCtheory}, \ref{section:2dStructure}, 
\ref{section:1dStructure}).    The remnant's
temperature is low in the outer parts, in part because it
has become radiative.  As a result, its center 
filled x-ray morphology is not accompanied by
a bright shell, even prior to 
absorption 
(\S\ref{section:2dX}, \ref{section:1dXSurf}).  After 
absorption it yields approximately the observed intensity 
and spectrum.

At normal abundances, the x-ray emission from the remnant 
interior is calculated to arise from approximately 30 \msun.  
One expects that the enriched stellar ejecta should 
have distorted the spectrum, the surface brightness 
distribution, and the overall intensity.  And yet, our basic 
model reproduces most of the spectral detail, missing only 
noticeably in the strength of the Si XIII complex around 1.85 
keV, and provides essentially the correct total count rate
(\S \ref{section:1dXSpect}).  

The only hints that the ejecta play a 
role are in the surface brightness distribution
(\S \ref{section:1dXSurf}).  That distribution has three 
characteristics of interest.  Unlike that provided by our basic 
model, the radial distribution has a relatively sharp 
central peak and is concave upward elsewhere.  In addition, 
the image is somewhat lumpy, 
a characteristic which may have contributed to the idea that 
there are discrete evaporating clouds associated with the 
emission.  Some of this distribution may arise from variability 
in the absorption screen, but we are inclined for reasons 
given in the text to believe that at least the central peaking is 
real (\S \ref{section:XObs}).  Our ad hoc experiment with 
moderately enhanced abundances in the stellar ejecta
(\S \ref{section:1dXSurf}) easily provided sufficient additional 
central emission, but far more can be done to clarify the 
situation.  The observed distribution and spectrum depend at 
least on the abundance distribution in the stellar envelope, 
on the wind activity of the star in shaping its environment prior 
to the explosion, on the degree to which dust formation 
extracts elements from the gaseous phase, on the instabilities 
in the wind bubble and explosion activity tending naturally to 
form lumps, on the potential for some of those lumps to cool 
and return to high density, on the radial mixing profile of the 
interpenetrating fluids of ejecta, wind and ISM, on the thermal 
conductivity within the stellar wind bubble and in the hot 
interior of the remnant, on the electron-ion thermal equilibration
rate, on the dust destruction rates, and on 
the stellar density profile and the irregularities in the medium 
which will influence the focussing and location of the reverse 
or reflected reverse shock at the present time
(\S \ref{section:2dStructure}).

Improved imaging and spectra of W44 X-rays can very much 
clarify the constraining information, particularly on the 
ionization history and current temperature of the gas
(\S \ref{section:1dXSpect}), but it would be premature to suppose 
that existing modeling efforts have been sufficient for 
meaningful comparison with data taken from the central third 
of the remnant.  A great deal more careful work lies 
ahead, with major difficulties and dangers in the 
sophistication needed in approximating the interrelated 
phenomena.  The good news is that many models of the 
details will be totally unacceptable, producing far more X-rays 
with very much the wrong spectra.  The payoff will be 
multiparameter constraints on the possible characteristics of 
the star, it's interaction with the surroundings, and on the 
degree of thermal conduction allowed or required.  As for the 
concavity of the angle averaged surface brightness 
distribution, modelers must take care to be sure that 
scattering of the X-rays by intervening gas is included in their 
final comparisons. 

\subsection{The Shock and Dense Shell}
\label{section:ShellShock}

The model remnant's shock shows cooling effects 
everywhere, but a dense shell has formed only on the higher 
density end (Figure~\ref{schematic} and
\S \ref{section:OverviewBasic}, \ref{section:Komp}, 
\ref{section:2dStructure}).  With that end tilted somewhat 
away from the line of sight, the half shell can be seen as a 
receding distribution of H{\small I} whose velocity structure 
agrees with observations
(\S \ref{section:2dHI}).  The radio 
continuum arises from the dense shell, via synchrotron 
emission in an approximately 200 $\mu$G field
(\S \ref{section:vdL}, \ref{section:2dsynch}), with its filamentary 
structure due to local tangencies of the corrugated shell with 
the line of sight
(\S \ref{section:2dsynch}).  The shell should 
also be a source of gamma rays, from both bremsstrahlung 
by the CR electrons and $\pi^o$\ production by the protons, at 
about 1/4th to 1/2 the rate measured by EGRET as arising 
from the vicinity
(\S \ref{section:gamma}). 

The radiative shock and ongoing shell formation generate the 
usual radiative remnant's spectrum, in UV, optical, and IR, 
with rates that correspond to recent observations of 
H$\alpha$, [S{\small II}], and 63 $\mu$m [O{\small I}]
(\S \ref{section:Alphaetc}).  Much of this emission is associated 
with the edge of the dense shell and should correlate well 
with the filamentary structure, though some could be more 
diffusely arranged in the remnant's equatorial belt where 
shell formation is currently strongest.  Both aspects are 
consistent with existing observations, though in the optical 
the observations are made through a variable screen of 
several magnitudes of foreground extinction (measured by 
the absorption column of the x-ray spectrum).  

Because the filaments arise geometrically from tangencies to 
the shell, any correlated line emission should appear at the 
local rest velocity (expansion is perpendicular to the line of 
sight), a characteristic that distinguishes filaments due to 
tangency from those due to actual filamentary (rope-like) 
structures.  This velocity null for the filaments has been 
confirmed in the associated OH maser emission (Frail, 1998, 
private communication), but should be checked in  
H$\alpha$\ because of possible disagreements over the 
origin of the maser emission.

Unfortunately, the large and variable optical depth to W44 
degrades the information available from optical intensity 
studies.  But a velocity resolved study of the H$\alpha$, or of 
other spectral components potentially associated with the 
shell and or evaporating clouds, could be useful in several 
ways.  We expect emission from the existing shell, from the 
immediate surroundings, and from the equatorial belt where 
shell formation is in progress.  The shell component should 
strongly resemble the HI distribution in velocity and space, 
but can likely be detected from the approaching (shell 
forming) side as well as from the receding side, and traced to 
the shell edge at low velocity, the latter not being possible in 
21 cm due to confusion with other material along the line of 
sight.  In addition to the important verification of the the zero 
radial velocity of the filamentary emission, a search is needed 
for a possible emission component associated with the x-ray 
bright spots, as would be expected from evaporating (or 
cooling and condensing) lumps.  Perhaps most interesting is 
the possibility of seeing emission from  gas which is just 
joining the inner surface of the shell.  If the cooling is 
sensitive and therefore somewhat erratic, the condensation 
flow could be irregular and have some components  
expanding at a noticeably higher speed than the shell they 
are overtaking.  

We found that the van der Laan mechanism, sweeping up of 
ambient cosmic rays and magnetic field and compression of 
both into the dense shell, can plausibly provide both the 
observed spectrum and intensity of the radio synchrotron 
emission
(\S \ref{section:vdL}), but there was a hidden 
difficulty with the formulation that amounted to assuming that 
the ambient CR electron density there is about 4 times 
greater than that estimated for the solar vicinity.  This is not 
necessarily a problem.  Attempts to measure the synchrotron 
emissivity of the general ISM have been difficult and seem to 
have both significant uncertainty and genuine variability.  In 
addition, W44 is in an active region at the base of the Aquila 
supershell (Maciejewski \etal, 1996) and could very well be 
encountering an enhanced CR population.  Other important 
processes potentially contributing to the cosmic ray electron 
population are discussed in
\S \ref{section:RCDisc}.  A caveat 
in our analysis is that our use of the viewing-angle averaged 
synchrotron radiation formula may have been misleading 
in the presence of the orderly tangential magnetic field seen 
in W44.

The extremely high angular resolution of the OH maser 
locations and their relationship to radio continuum ridges and 
bright spots (Claussen \etal 1997) appears to us to require 
the masers to arise from the leading edge of the dense shell, 
i.e. at the trailing edge of the cooling and recombination zone 
of the radiative shock, with masering occuring tangential to 
the shell's outer surface so that it is observed only along the 
filaments.  Consistently, the magnetic field measured via the 
maser emission is just that found for the shell in our model
(\S \ref{section:RCDisc}), and as noted above, their radial 
velocities are at the local rest velocity.  We are not aware of 
models for shocks in such low density gas to test for the 
presence of sufficient OH, or of maser pumping models for 
this nonequilibrium environment, but we note that the models 
by Hollenbach and McKee (1989) for shocks in denser gas 
do have a significant abundance of OH on leaving the 
photoionization/recombination plateau, and show a local 
peak in the OH emission at a column density behind the 
shock of about 10$^{20}$\ cm$^{-2}$, where the temperature 
is dropping rapidly through 1000 K.  This result may not be 
sensitive to the initial density (McKee, private communication 
1998); and the required column density is achieved in our 
model.  The anticipated neutral density in this region is a few 
hundred cm$^{-3}$, limited by the magnetic pressure.  The 
observations cry for studies of 100-150 km s$^{-1}$ shocks 
into densities of 10-100 cm$^{-3}$\ and their potential for creating 
molecular gas and generating OH maser activity tangential to 
the shock surface, possibly with the pumping arising in 
transience of the material (e.g. formation of the OH in excited 
states).  The solution could be elusive, however, based on 
the observation that there are many small maser spots but far 
more filamentary regions without masering, at least not 
beaming towards us.

Our predicted shell mass would also agree better with the HI 
data if much of the shell were molecular
(\S \ref{section:2dHI}).  
If there is a strong molecular line which is insensitive to 
reddening and associated with radiative shocks, perhaps 
from the fluorescence region at the outer edge of the dense 
shell, a line not present in the ambient medium or foreground 
molecular gas, it could provide an excellent probe of the 
shell's actual extent and velocity structure.  The model's 
overall shell mass is about 450 \msun of which perhaps half 
is molecular.  If it can be shown that the molecular mass in 
the shell is substantially larger than this, perhaps in structures 
related to the shell corrugation, it could also remove the 
apparent discrepancy between our model's gamma ray 
production and the EGRET measurements, while requiring 
only some adjustment of the ambient density distribution in 
the next generation of models. 

\subsection{Distance and Fine Tuning}
\label{section:Tuning}

The oft-quoted distance for W44 is 3 kpc, but we reviewed 
this assessment in
\S \ref{section:Dist}, concluding that the 
appropriate modern number is 2.5 to 2.6 kpc.  Our theoretical 
attempts to choose parameters that would produce the 
correct angular size at this distance failed slightly, forcing us 
to describe the model characteristics for a distance of 2.2 kpc.  
It is not impossible that the remnant is this close, and within 
the uncertainties of such modeling, the differences are 
inconsequential.  Nevertheless, it may be worth comment.  
Section \ref{section:scaling} introduced scaling laws for 
adjusting our results to remnants located at different 
distances.  Two scalings are presented, exact Sgro-Chevalier 
scaling, and approximate scaling for constant explosion 
energy.  For Sgro-Chevalier scaling to 2.5 kpc, we found that 
the only appreciable changes were a 12\% reduction in the 
x-ray surface brightness and luminosity, and a factor of 1.3 
increase in explosion energy and shell mass.  The increase 
in explosion energy is the only uncomfortable part, leading us 
to favor the more approximate scaling.  In that case, the only 
appreciable change is the reduction of the x-ray surface 
brightness and luminosity by a factor 0.53.  Recalling that 
Model A was about 25\% overluminous leads to  no worse 
overall agreement.  However, considering that a significant 
part of the total x-ray emission may be associated with the 
elementally enriched center as discussed above, this 
increase in distance is an improvement, allowing a greater 
degree of enrichment of the interior without having to resort to 
a reduced thermal conductivity for compensation.  With this 
model, the explosion site density would be only 4.6 cm$^{-
3}$\ and the age 23,700 years.  Both are intermediate 
between Model A and the parameters found by Harrus, \etal 
(1997).

\subsection{Relationship of the Model to Other Remnants}
\label{section:Others}

Finally we would like to remark on the possibility of 
generalizing this model to other remnants.  To have both a 
strong radio shell component and an interior which is 
luminous in thermal X-rays requires both a substantial 
compression of the dense shell, and a substantial interior 
density and pressure.  In
\S \ref{section:vdL} we found that the 
radio synchrotron surface brightness shortly after shell 
formation should be roughly proportional to Rp$_{shell}$, 
while in
\S \ref{section:scaling} we found the x-ray surface 
brightness of the interior proportional to Rp$^{2}_{interior}$.  
We have not explored our hydrodynamic models to learn how 
rapidly these surface brightnesses fade as the remnant 
continues to expand, but we expect a very considerable drop 
in the x-ray emission by the time it has grown very far.  The 
radio continuum is less sensitive in the first place, and will 
drop more gradually because it rises with cooled mass and 
because remnants in inhomogeneous media have shell 
formation occuring over part of their surface for a protracted 
period of time.  We naturally expect then that remnants with 
radio shells will be more common than those with x-ray 
luminous interiors, and can concentrate on the conditions 
needed for the latter.  At shell formation, when the outer edge 
is too cool to radiate in x-rays, the surface brightness of the 
interior should be roughly proportional to R$_{cool}^{-5}$.  In 
that case it is clear that the class of remnants to which our 
model applies should have a very steep dropoff of surface 
brightness with size at shell formation, as well as with size 
after shell formation.  Using
\S \ref{section:ShellForm} to relate 
shell formation radius to ambient density, we infer that the 
x-ray surface brightness of the interior at shell formation is 
approximately proportional to $n_o^2$.  As we have not 
explored the effects of any presupernova wind on the later 
internal structure of the remnant, this result may be 
somewhat naive, but our tentative expectation is that 
an x-ray bright interior for a supernova remnant of typical 
energy will occur only when the explosion occurs in an 
extended region of higher than average interstellar density.

For contrast, consider the Cygnus Loop.  Its radio continuum 
derives largely from compressed material behind roughly 100 
km s$^{-1}$\ radiative shocks moving into recently 
encountered large scale clouds of density roughly 8 cm$^{-
3}$, not so different from W44.  Its flux is similar to W44's, its 
distance about a third as great, so its radio continuum 
luminosity is lower by about a factor of 10.  The remnant 
volume is about 4 times that of W44, so the mean internal 
pressure is about 4 times lower.  The brightest x-rays are 
associated with the interaction with the large dense clouds, 
and are found just inside the radiative shocks.  The 
distribution of the more 
diffuse x-rays, however,  resembles that expected from
a Sedov structure with 
ambient density 0.16 cm$^{-3}$.  We therefore estimate the 
interior density to be about 1/30th that of W44 and the interior 
temperature about 7.5 times greater, 4 to 5 $\times 10^7$\ K.  
Continuing to assume that the abundances are unaffected by 
the ejecta, which provides the correct count rate for W44, and 
consulting our ROSAT PSPC counts per emission measure 
tables with negligible absorption for the Cygnus Loop and 
the amount found above for W44, we conclude that the 
observable (after absorption) surface brightness of the Loop 
interior should be lower by roughly a factor of 10.  That 
reduction appears modest until compared with the radiation 
from the 30 times denser edge.   At a 
temperature of 2.4 $\times 10^6$\ K, the ``shell" emission is 
about 2 orders of magnitude brighter in projection across the 
center of the remnant and much brighter yet as seen near the 
edge.  Although there are radiative shocks on part of the
surface of the Cygnus Loop, much of the evolution is in low 
density and is very far from having the low edge temperatures 
associated with incipient cooling and shell formation.
Only an instrument with sensitivity to very hot gas 
could discern the high temperature interior from 
the softer ``shell" emission.

There are two ways of creating much larger remnants with x-ray 
emitting interiors.  One, explored by Shelton (1998) in the context
of halo and interarm supernovae, simply involves exploding in a 
very low density (0.01 cm$^{-3}$) and waiting for the edge temperature
to drop too low to show up as an x-ray shell.  The other
was explored in Smith and Cox (1998) in 
connection with modeling the Local Bubble.  A second 
supernova within the cavity of an earlier remnant can reheat 
the interior and bring it back into x-ray emission.  And that 
emission can last a substantial length of time, possibly 
making unlikely events yield not uncommon objects in the 
sky.  The two approaches are not entirely dissimilar in that
the Local Bubble model starts with a higher ambient density
and uses the first explosion to create the low density into which
the second explosion initially evolves.  One difference
is that the multiple explosion model has denser surroundings
which slow the shock and lead more quickly to a low temperature
edge.

\subsection{Closing Remarks}
\label{section:Closing}

This study was undertaken to show that the characteristics of 
a heavily observed and very exciting supernova remnant, 
W44, could be understood in the context of a very simple 
model, a model which did not invoke evaporating clouds in 
the center, or blast wave accelerated clouds to make the 
observed H{\small I}, or interaction with a dense molecular cloud 
environment, or diffusive shock injection and acceleration of 
cosmic rays.  The required features were a relatively dense 
ambient environment, a significant density gradient, shell 
corrugation due to moderate irregularities in the ambient 
medium, betatron acceleration of the ambient cosmic ray 
population, and a significant amount of thermal conduction in 
the hot interior.  It was hoped that the observed intensity of 
x-rays from the hot interior would then finally give us a relatively 
conclusive measure of the magnitude of the thermal 
conductivity.

The model was wildly successful, but not without its 
disappointments.  We were able to show that compression of 
the ambient cosmic rays could lead to approximately the 
correct radio continuum flux and spectrum, and to 
a significant contribution to 
the gamma ray intensity, but could not show that there is no 
room left for diffusive or other forms of acceleration.  We 
found that the intensity and spectrum of the interior x-ray 
emission were consistent with unquenched thermal 
conduction, but could not convince ourselves that the 
metallicity contamination by the ejecta and preexplosion 
evolution could be neglected. As a result we are less
confident in our determination of the magnitude of the
thermal conductivity in the hot interior of the remnant. 
On the positive side, the 
need for clouds and clumps in the remnant interior vanished, 
as did any hint of interaction with dense molecular clouds.  
The shell emission characteristics were in beautiful 
agreement with the measurements, and comparison with the 
OH maser observations led to the ideas that the shell could 
be largely molecular and that the trailing edge of the cooling 
region is the site of the masering activity.

\section{Acknowledgements}

We would like to acknowledge useful advice, conversations, and
assistance from, with, and of Namir Kassim, Bob Benjamin,
Rob Petre, Jeonghee Rho,
Dale Frail, Linda Sparke, Pat Slane, Blair Savage,
and John Raymond during various times of the
long course of this project.  We are also grateful to Bon-Chul Koo for
providing the H{\small I} data for comparison with our model.

This work was supported in part by NASA Grant NAG5-3155 to the
University of Wisconsin-Madison.  RLS and RKS were also supported by grants 
from the National Research Council while at the NASA/Goddard Space 
Flight Center, in the Laboratory for High Energy Astrophysics, and the
Laboratory for Astronomy and Solar Physics, respectively.

TP and MR benefited from the Polish Committee for Scientific Research
grant 2.P03D.004.13. The 2-D simulations reported in this paper were
conducted at MPG Computer Center in Garching and Interdisciplinary
Center for Computational Modelling (ICM) in Warsaw.

This research has made use of data obtained through the High Energy
Astrophysics Science Archive Research Center Online Service, provided
by the NASA-Goddard Space Flight Center.

\section{References}

\vspace{1cm}
\setlength{\parindent}{0em}

{\scriptsize

Arnaud, K.A. 1996, Astronomical Data Analysis Software and 
    Systems V, eds. Jacoby G. \& Barnes J., p17, ASP Conf. 
    Series volume 101\\ 

Berger M.J., Colella P. 1989, J. Comput. Phys, 82, 64\\

Bloemen, H. 1989, ARAA, 27, 469\\

Caswell, J.L., Murray, J.D., Roger, R.S.,
      Cole, D.J., \& Cooke, D.J. 1975, A\&A, 45, 239\\

Chevalier, R.A. 1973, PhD thesis, Princeton University\\


Cid-Fernandes, R., Plewa, T., R\'o\.zyczka, M., Franco, J., 
     Terlevich, R., Tenorio-Tagle, G. \& Miller, W. 1996, MNRAS, 
    283, 419\\

Clark, D.H., \& Caswell, J.L. 1976, MNRAS, 174, 267\\

Clark, D.H., Green, A.J., Caswell, J.L. 1975,
      Australian Journal of Physics Astrophysical Supplement,
      37, 75\\

Claussen, J.M., Frail, D.A., Goss, W.M., \& 
      Gaume, R.A. 1997, ApJ, 489, 143\\

Clemens, D.P. 1985, ApJ, 295, 422\\

Colella P., Woodward P.R. 1984, J. Comput. Phys., 59, 264\\

Cowie, L.L. \& McKee, C.F. 1977, ApJ, 211, 135\\

Cox, D.P. 1970, PhD thesis, UC San Diego\\

Cox, D.P. 1972a, ApJ, 178, 143\\

Cox, D.P. 1972b, ApJ, 178, 159\\

Cox, D.P. 1986, ApJ, 304, 771\\

Cox, D.P. 1996, ``Guide to Modeling the Interstellar Medium," in 
      Astrophysics in the Extreme Ultraviolet, eds S. Bowyer \& R. 
      F. Malina, Kluwer, p. 247\\

Cox, D.P. \& Anderson, P.R. 1982, ApJ, 253, 268\\

Cui, W. \& Cox, D.P. 1992, ApJ, 401, 206\\

Dame, T.M. 1983, PhD thesis, Columbia University\\

Dame, T.M., Elmegreen, B.G., Cohen, R.S., \& Thaddeus, P. 1986,
ApJ, 305, 892\\

deJager, O.C. \& Mastichiadis, A. 1997, ApJ, 482, 874\\

DeNoyer, L.K. 1983, ApJ, 264, 141\\

Dickel, J.R., \& DeNoyer, L.K. 1975,
      AJ, 80, 437\\

Dickel, J.R., Dickel, H.R., \& Crutcher, R.M. 1976, PASP, 
88, 840\\

Dohm-Palmer, R.C. \& Jones, T.W. 1996, ApJ, 471, 279\\

Edgar, R.J. 1994, personal communication\\

Edgar, R.J. \& Chevalier, R.A. 1986, ApJ, 310, L27\\


Ellison, D.C. \etal 1994, PASP, 106, 780\\

Esposito, J.A., Hunter, S.D., Kanbach, G., \& Sreekumar, P. 1996, ApJ, 461, 
820\\

Falle, S.A.E.G. 1975, MNRAS, 172, 55\\

Falle, S.A.E.G. 1981, MNRAS, 195, 1011\\

Frail, D.A., Giacani, E.B., Goss, W.M., \& Dubner, G. 1996, 
     ApJL, 464, L165\\

Frail, D.A., Goss, W.M., Reynoso, E.M., Giacani, E.B., 
      Green, A.J., Otrupcek, R. 1996, AJ, 111, 1651\\

Gabriel, A.H., Bely-Dubau, F., Faucher, P., \& Acton, L.W 1991, ApJ, 378, 438\\

Giacani, E.B., Dubner, G.M., Kassim, N.E.,
      Frail, D.A., Goss, W.M., Winkler, P.F., \& Williams, B.F.
      1997, ApJ, 113, 1379\\

Giaconi,~R. \etal  1979, ApJ, 230, 540 \\


Goss, W.M., Caswell, J.L., \& Robinson, B.J.
       1971, A \& A, 14, 481\\


Grevesse, N., \& Anders, E. 1988, AIP Conference
      Proceedings No. 183, ``Cosmic Abundance of Matter'', ed. C.J. Waddington,
1\\


Kovalenko, A.V., Pynzar', A.V., \& Udal'tsov, V.A.
      1994, Astr. Rep., 38, 78\\

Harrus, I.M., Hughes, J.P., \& Helfand, D.J.
     1996, ApJ, 464, L161\\

Harrus, I.M., Hughes, J.P., Singh, K.P., Koyama, K.,
      \& Asaoka, I. 1997, ApJ, 488, 781\\

Hester, J.J. 1987, ApJ, 314, 187\\


Hollenbach, D. \& McKee, C.F. 1989, ApJ, 342, 306\\

Jokipii, J.R. \& Parker, E.N. 1969, ApJ, 155, 777\\

Jones, L.R., Smith, A., \& Angellini, L. 1993, MNRAS, 265, 
631\\

Jones, T.W., Rudnick, L., Jun, B.-I., Borkowski, K.J.,
       Dubner, G., Frail, D.A., Kang, H., Kassim, N.E., \& McCray, R. 1998, 
PASP, 
       in press\\

Kahn, F.D. 1975, In Proc. 14th Int. Cosmic Ray Conf. (Munich) 
      11, 3566\\

Kahn, F.D. 1976, A\&A, 50, 145\\

Kassim, N.E. 1992, AJ, 103, 943\\

Kompaneets, A.S. 1960, Soviet Phys. Dokl., 5, 46\\

Koo, B.-C., \& Heiles, C. 1995, ApJ, 442, 679\\

Kovalenko, A.V., Pyznar, A.V., \& Udal'tsov, V.A. 1994, 
     Astro. Rep., 38, 78\\

Kundu, M.R., \& Velusamy, T. 1972, A\&A, 20, 237\\

Long, K.S., Blair, W.P., White, R.L., \& Matsui, Y. 1991, ApJ, 
      373, 567 \\


Maciejewski, W., Murphy, E.M., Lockman, F.J., \& Savage, B.D. 1996, ApJ, 
469, 
      238\\

Maciejewski W. \& Cox, D.P. 1998, ApJ, submitted\\



McKee, C.F. 1988, personal communication\\ 



Morrison, R., \& McCammon, D. 1983, ApJ, 270, 119\\


Plewa, T. \& M\"uller, E. 1998, in preparation\\

Raymond, J.C. 1976, PhD thesis, University of Wisconsin-Madison\\

Raymond, J.C. 1995, personal communication\\

Raymond, J.C. \& Smith, B.W.  1977, ApJ S, 35, 419.\\


Reach, W.T., \& Rho, J. 1996, A \& A, 315, L277\\

Reich, W., F\"{u}rst, E., Reich, P., \& Reif, K.
       1990, Astron. \& Astrophys. Suppl. Ser., 85, 633\\


Rho, J.-H. 1995, PhD thesis, University of Maryland\\

Rho, J.-H. 1997, oral presentation at the Minnesota Meeting on 
Supernova Remnants (summarized in Jones \etal 1998)\\


 
Rho, J., Petre, R., Schlegel, E.M., \&  Hester, J.J. 1994,
ApJ, 430, 757\\


Richtmyer, R.D., \& Morton, K.W. 1967, ``Difference Methods for 
    Initial Value Problems, Second Edition'' (New York: 
    Interscience Publishers)\\

Rockstroh, J.M., \& Webber, W.R. 1978, ApJ, 224, 677\\

Rosner, R., \& Tucker, W.H. 1989, ApJ, 338, 761\\

Sato, F. 1986, AJ, 91, 378\\\

Schmidt, M.  1965, In {\it{Galactic Structure}}, eds. A. Blaauw 
       \& M. Schmidt, University of Chicago, Press, p. 513\\


Sgro, A. 1972, PhD thesis, Columbia University\\

Shelton, R.L. 1998, in preparation\\

Shelton, R.L., Smith, R.K., \& Cox, D.P.
        1995, BAAS, 27, no. 3 (186.5904)\\


Slavin J.D. \& Cox, D.P. 1992, ApJ, 392, 131\\


Smith A., Jones, L.R., Watson, M G., Willingale R., Wood N., 
    \& Seward, F.D. 1985, MNRAS, 217, 99\\

Smith, R.K., \& Cox, D.P. 1998, in preparation\\

Smith, R.K., Krzewina, L.G., Cox, D.P., Edgar, R.J., \& Miller, 
    W.W. 1996, ApJ, 473, 864\\

Snowden, S.L., McCammon, D., Burrows, D.N., \& Mendenhall, 
    J.A. 1994, ApJ, 424, 714\\

Snowden, S.L. 1994, Cookbook for Analysis Procedures for 
    ROSAT XRT/PSPC Observations of Extended Objects and the 
    Diffuse Background, U.S. ROSAT Science Data Center, 
    Goddard Space Flight Center\\



Tao, L. 1995, MNRAS, 275, 965\\


Turner, T.J. 1994,``ROSAT data analysis using xselect and 
    ftools'', OGIP Memo OGIP/94-010, U.S ROSAT Science Data 
    Center, NASA Goddard Space Flight Center\\

van der Laan, H. 1962a, MNRAS, 124, 125\\

van der Laan, H. 1962b, MNRAS, 124, 179\\

Webber, W.R., Simpson, G.A., \& Cane, H.V. 1980, ApJ, 236, 448\\

White, R.L. \& Long, K.S. 1991, ApJ, 373, 543\\

Wolszczan, A., Cordes, J.M., \& Dewey, R.J. 1991, ApJ, 372, 
    L99\\


Wootten, H.A. 1977, ApJ, 216, 440\\

Yanenko, N.N. 1971, The method of fractional steps: The solution 
    of Problems of      Mathematical Physics in Several Variables. 
    M. Holt (ed.), Springer-Verlag, New York\\

}

\setlength{\parindent}{4em}

\newpage

\section{Figure Captions}

\suppressfloats

\begin{figure}
\caption{
Schematic of the Model for W44.  Shows
 annotated section through remnant along plane
containing the look direction.  Inset shows projection of half shell
on the sky.
\hspace{11cm}}
\label{schematic}
\end{figure}

\begin{figure}
\caption{
Number density of each cell in the 2d model, projected on the
symmetry axis.  The ambient density distribution, partial shell
formation, factor of 4 compression at the tenuous end's shock, and
the moderately uniform and non-negligible central density are all
apparent, as are the regions of finer gridding in the adaptive mesh.
The remnant age is 20,000 years.
}
\label{2Ddensity}
\end{figure}
\begin{figure}
\caption{
Surface plot of the 2d model's temperature distribution at 20,000
years.  The temperature structure is approximately parabolic, fairly
uniform in the central half (in radius) with a steady drop in the outer
half to near zero at the edge.  On the tenuous end, the post shock
temperature of the shock can be seen, along with the trough behind
it showing its transition to radiative.
}   
\label{2Dtemperature}
\end{figure}
\begin{figure}
\caption{
Surface plot of the 2d model's pressure distribution at 20,000 years.
The constant central pressure, sloping shoulders in the outer half,
strong variations in edge pressure over the surface and the deep
trough of cooled but so far uncompressed material can be seen.
(The spikes are an artifact of the figure preparation, not a characteristic
of the model.)
}
\label{2Dpressure}
\end{figure}       
\begin{figure}
\caption{
({\it{top-right}}) H{\small I} mass from the simulated SNR viewed
at 40$^{\rm{o}}$ from the symmetry axis.  The axis is also tilted
toward the southwest.
({\it{bottom-left}}) The shaded contours represent the H{\small I} brightness
observed
for W44 between $v_{\rm LSR} = 135$ and 170 km s$^{-1}$ (Koo
and Heiles
1995).  The dots represent H{\small I} mass from the model
for the same velocity interval, assuming a 43 km s$^{-1}$
systemic velocity.
({\it{top-left}}) Position velocity diagram for the vertical aperture
indicated in the top right panel.
({\it{bottom-right}}) Position velocity diagram for the horizontal
aperture indicated in the top right panel.
}
\label{2Dhydrogen}
\end{figure}
\begin{figure}
\caption{
({\it{top}}) Position-velocity diagram of the fast, receding H{\small I}
gas.  The shaded contours represent H{\small I} brightness 
for a cut along
$\alpha_{\rm{1950}} = 18^{\rm{h}}53^{\rm{m}}30^{\rm{s}}$,
from Koo \& 
Heiles (1995).  The dots represent H{\small I} mass for the vertical
cut indicated in the previous figure.  A systemic velocity of
43 km s$^{-1}$ is assumed.
({\it{bottom}}) Same as {\it{top}}, for a horizontal cut through
$\delta_{\rm{1950}}
= +01^{\rm{o}}18'$.
}
\label{2Dvelocity}
\end{figure}
\begin{figure}
\caption{Map of the synchrotron emission from the simulated SNR.
In the upper right panel, the remnant is viewed at the same
orientation as in Figure~\ref{2Dhydrogen}, whereas the lower left
panel shows a view perpendicular to the symmetry axis.  The upper left
and lower right panels show the integrated intensity profiles for the
horizontal and vertical slices shown. }
\label{2Dsynch}
\end{figure}
\begin{figure}
\caption{Map of the synchrotron emission from the simulated SNR,
in which a pattern of radial crinkling has been superposed to
illustrate the formation of apparent filaments.  The remnant is
viewed at the same orientations as in Figure~\ref{2Dsynch}.}
\label{2Dcrinkledsynch}
\end{figure}
\begin{figure}
\caption{The 2d model x-ray surface brightness distribution.  Includes
both hard and soft ROSAT PSPC
bands, at a column density of 2$\times 10^{22}$\
cm$^{-2}$.}
\label{2DXray}
\end{figure}
\begin{figure}
\caption{The 2d model x-ray surface brightness distribution with no
absorption.  The 2d model's thermal X-rays are not edge
brightened (shell like) even without absorption.}
\label{2DXrayBare}
\end{figure}
\begin{figure}
\caption{  Density distributions of 1d models at 20,000 years.
With thermal conduction (Model A, solid line), the
electron density is almost 1 cm$^{-3}$\ in the center and changes slowly
with radius except near the edge.  Without thermal conduction
(Model B, dashed line) the density is extremely small at the center
and increases rapidly with radius.  The structures near 11 pc
show the early stages of shell formation.}
\label{1Ddensity}
\end{figure}
\begin{figure}
\caption{  Temperature distributions of 1d models at 20,000 years.
With thermal conduction (Model A, solid line), the temperature
is $6 \times 10^6$ K in the center and the changes only slowly
with radius except near the edge.  Without thermal conduction (Model B,
dashed line) the temperature is very high in the center
and changes rapidly with radius.  The bump at $\sim$ 6 pc is an artifact of
the code associated with the contact discontinuity.  The structures near 11
pc  show the early stages of shell formation.}
\label{1Dtemperature}
\end{figure}
\begin{figure}
\caption{Thermal pressure distribution of 1d models at 20,000 years, with
(Model A, solid line) and without
(Model B, dashed line) thermal conduction.}
\label{1DThermalpressure}
\end{figure}
\begin{figure}
\caption{a:  The ROSAT PSPC Soft Band (R4+R5+R6, 0.4 - 1.6 keV)
count rate
for Model A (with thermal conduction, solid line), Model B
(without thermal conduction, dashed line), and the observed rates
toward the north (cross), east (star), south (diamond), and west
(triangle).  Models A and B were attenuated with an absoption
column density of
N$_{\rm{H}} = 1.89 \times 10^{22}$cm$^{-2}$.
b: Same as a, but for the ROSAT PSPC Hard Band (R7, 1.0 - 2.0 keV).}
\label{fig:image}
\end{figure}
\begin{figure}
\caption{ROSAT PSPC (a) and Einstein SSS (b) spectra fit with Model A. 
\hspace{4cm}}
\label{fig:spectra}
\end{figure}
\begin{figure}
\caption{ROSAT PSPC (a) and Einstein SSS (b) spectra fit with Model C.
\hspace{4cm}} 
\label{fig:metalspectrum}
\end{figure}
\begin{figure}
\caption{ROSAT PSPC (a) and Einstein SSS (b) views of Model A
comparing collisional ionization equilibrium (dashed) and full
non-equilibrium ionization (solid).  The non-equilibrium model is the
same as shown in Figure~\protect{\ref{fig:spectra}}.}
\label{fig:equilAndNon}
\end{figure}
\begin{figure}
\caption{ROSAT PSPC (a) and Einstein SSS (b) spectra fit with Model
E, identical to Model A but with the thermal conductivity reduced by
10.}
\label{fig:reducedTC}
\end{figure}

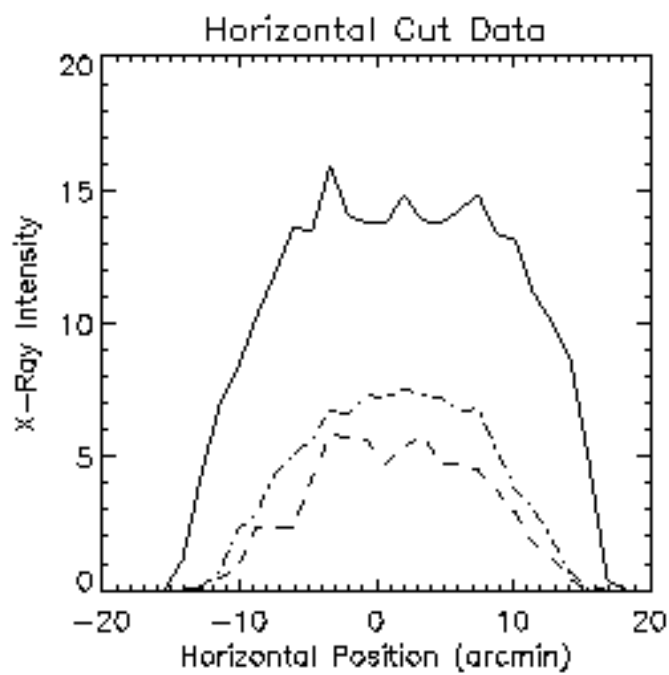
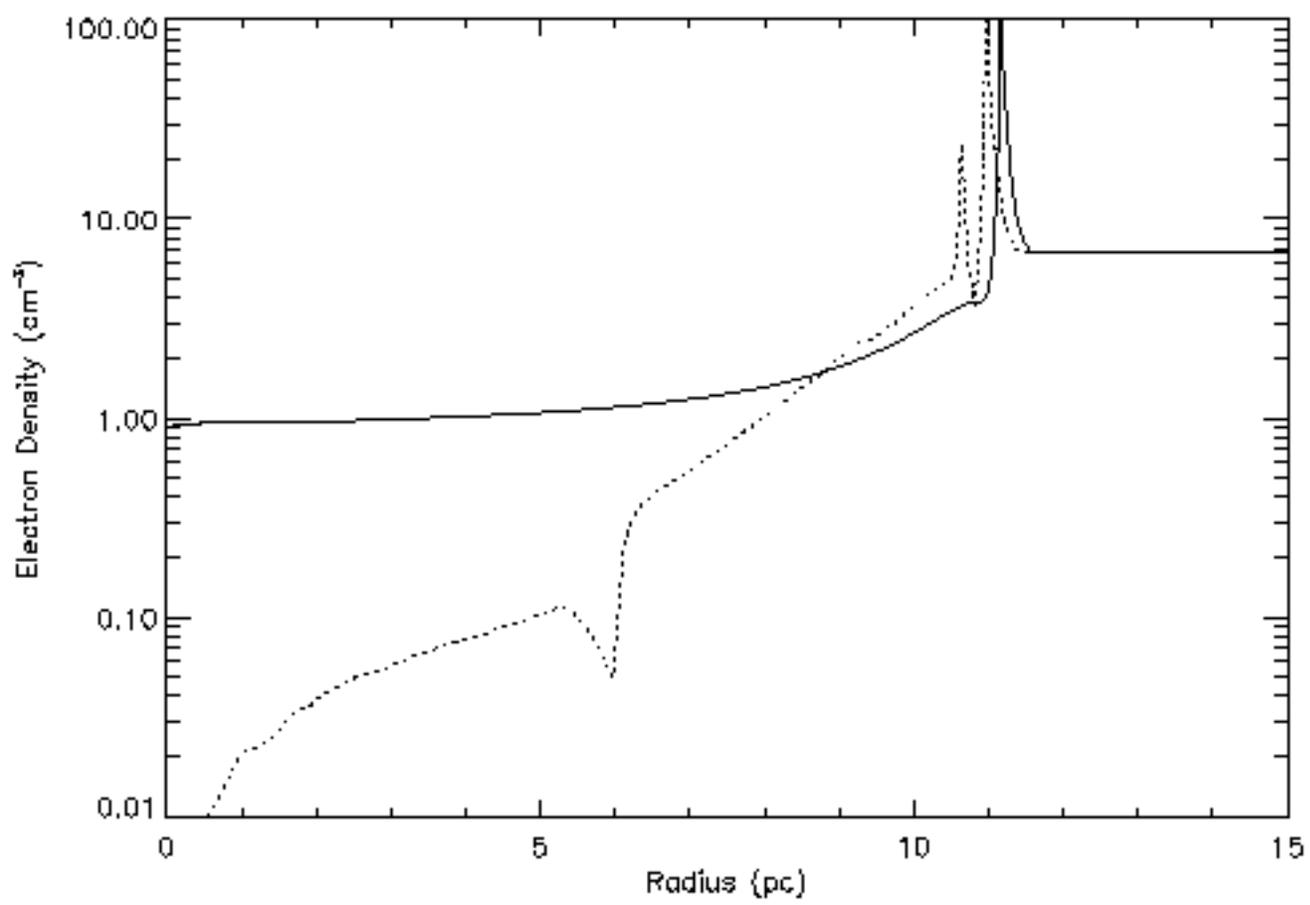
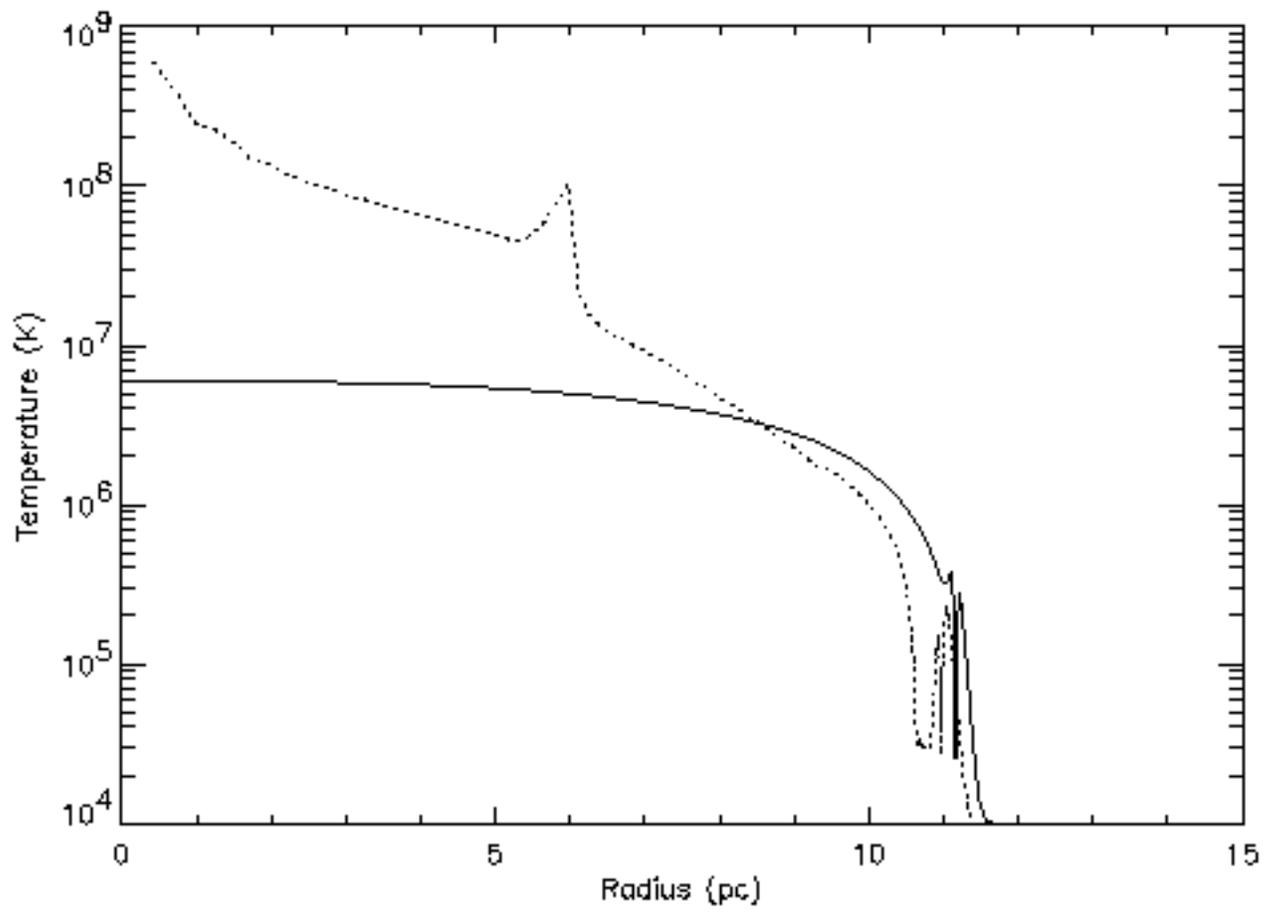
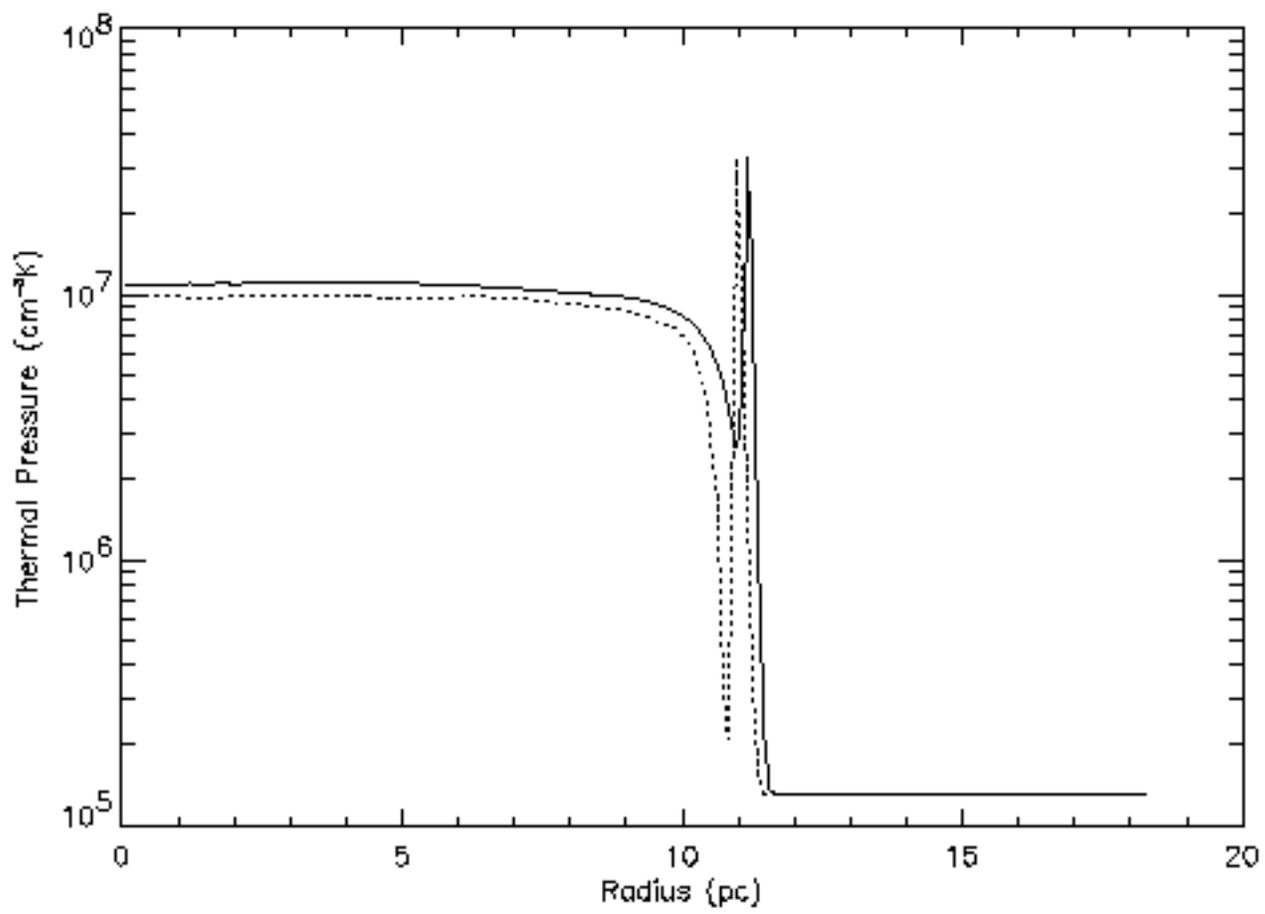
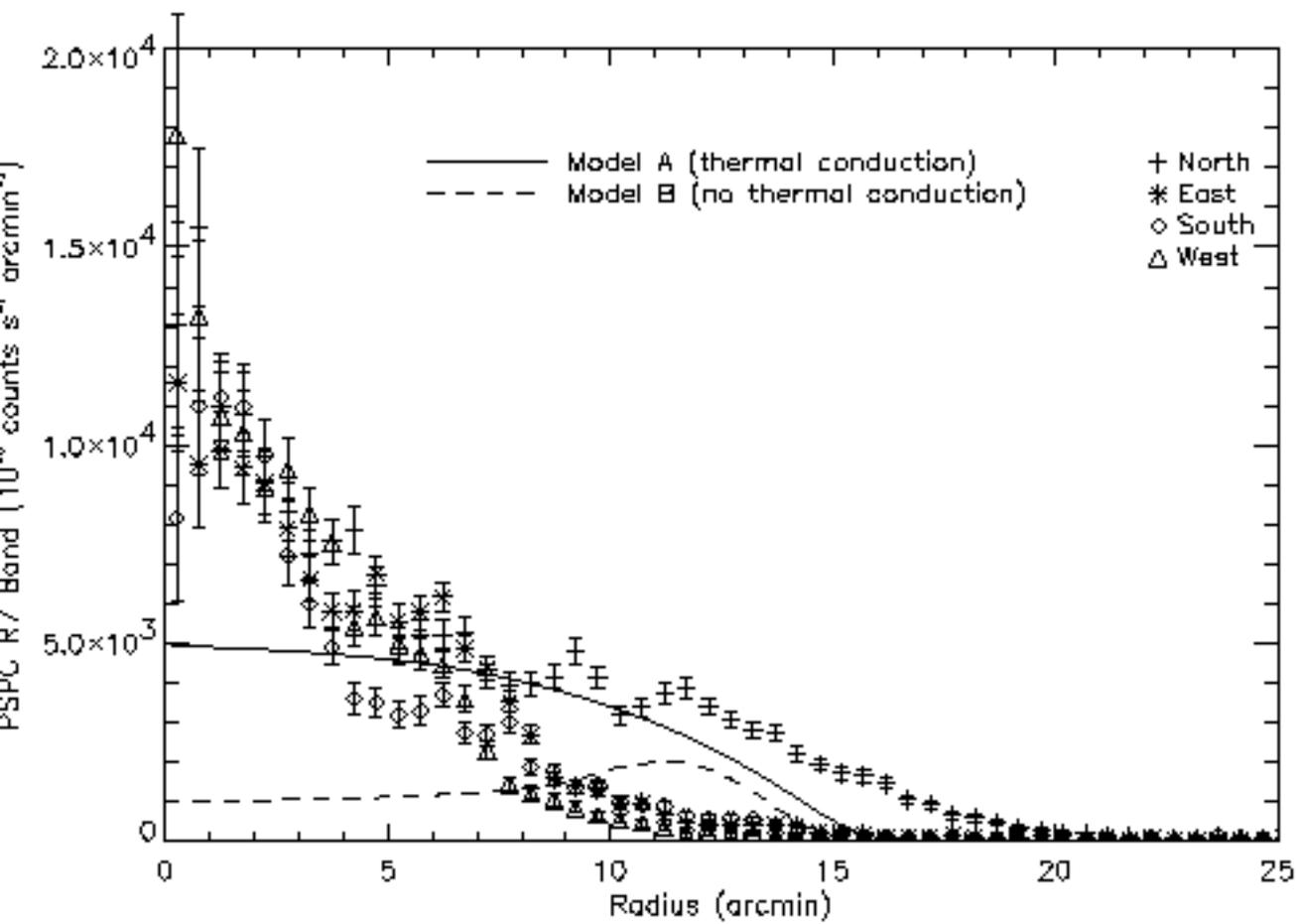
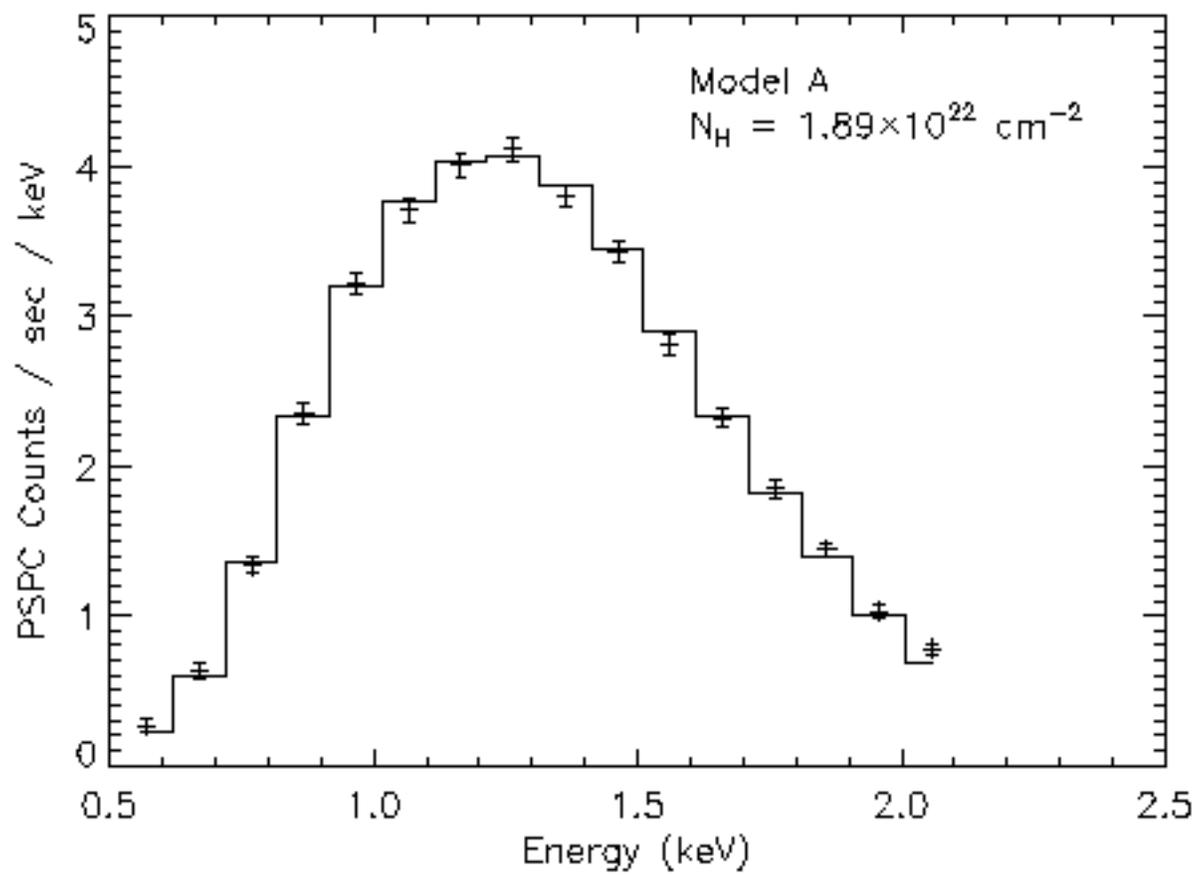
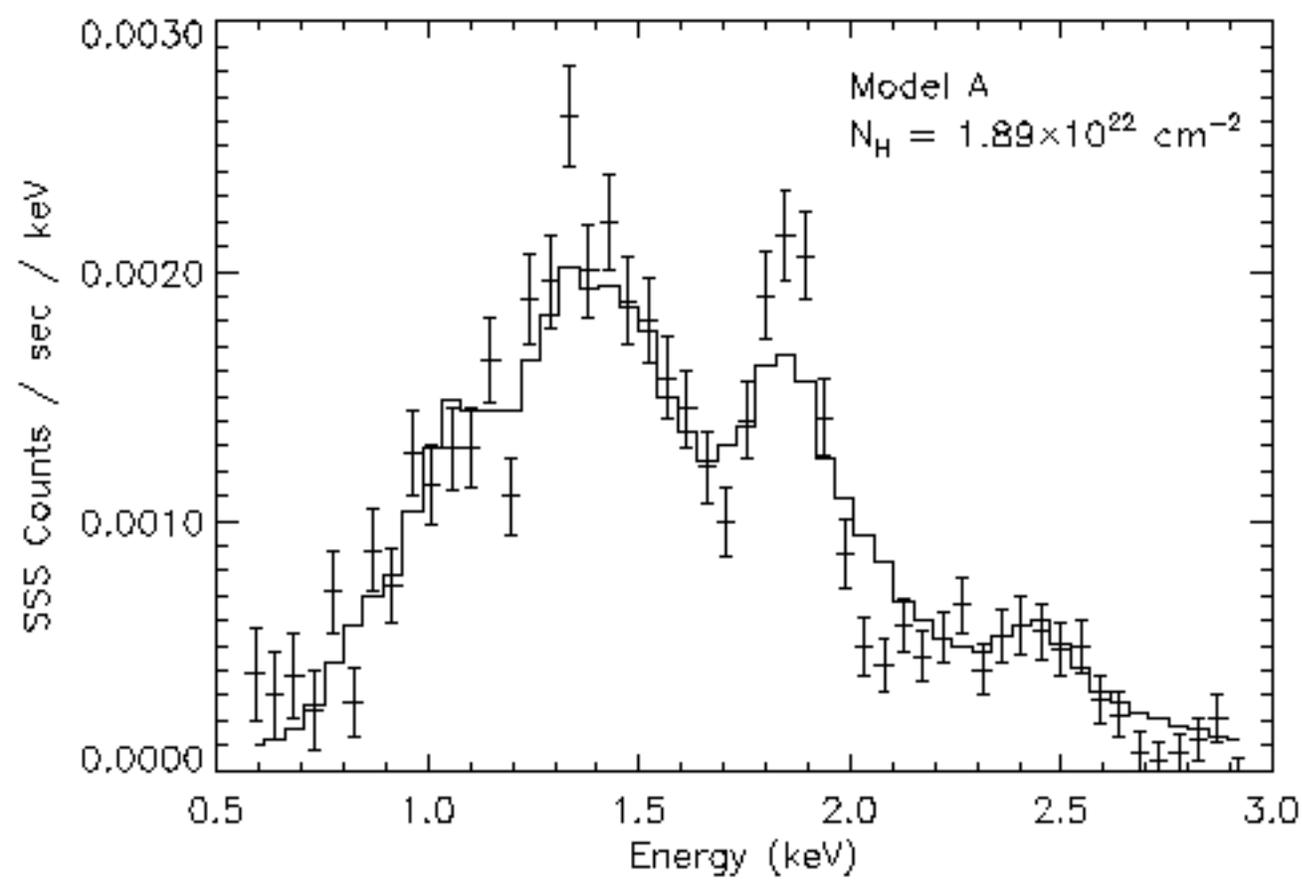
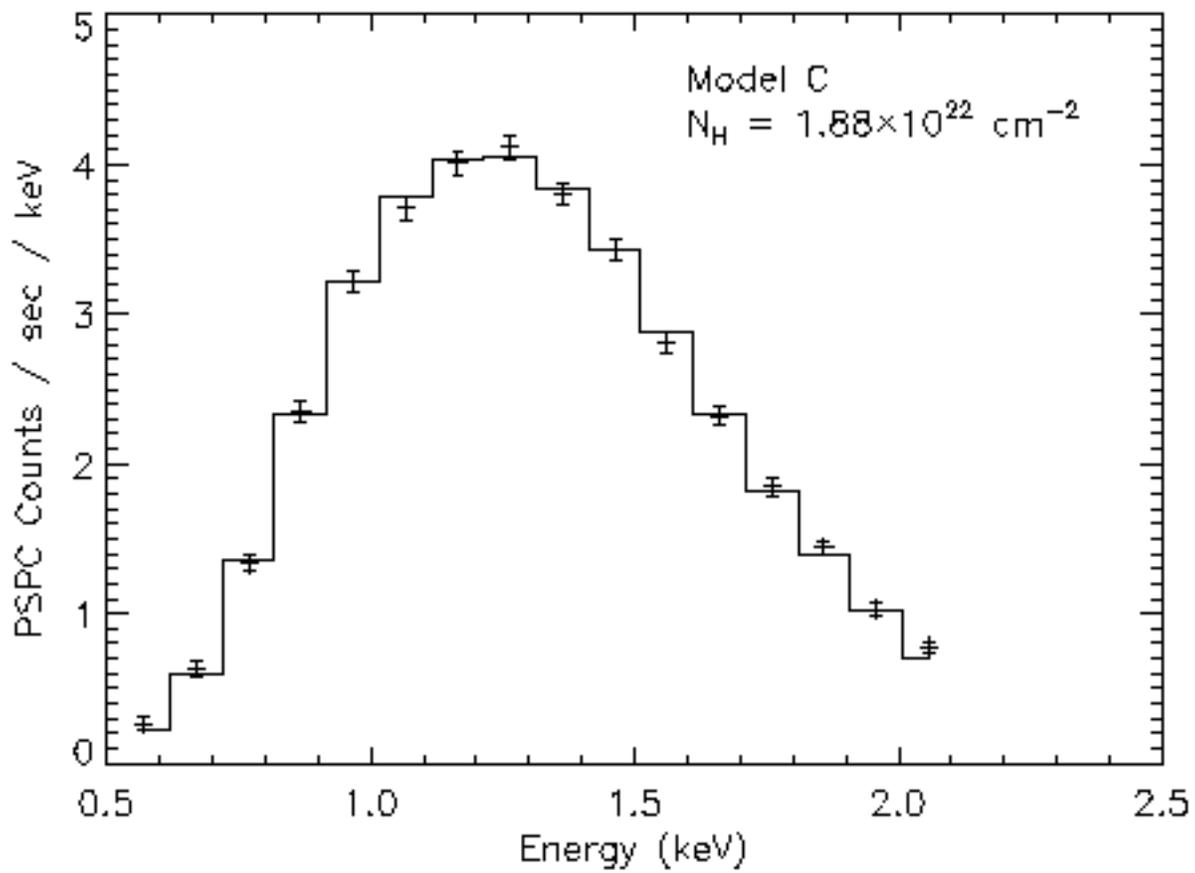
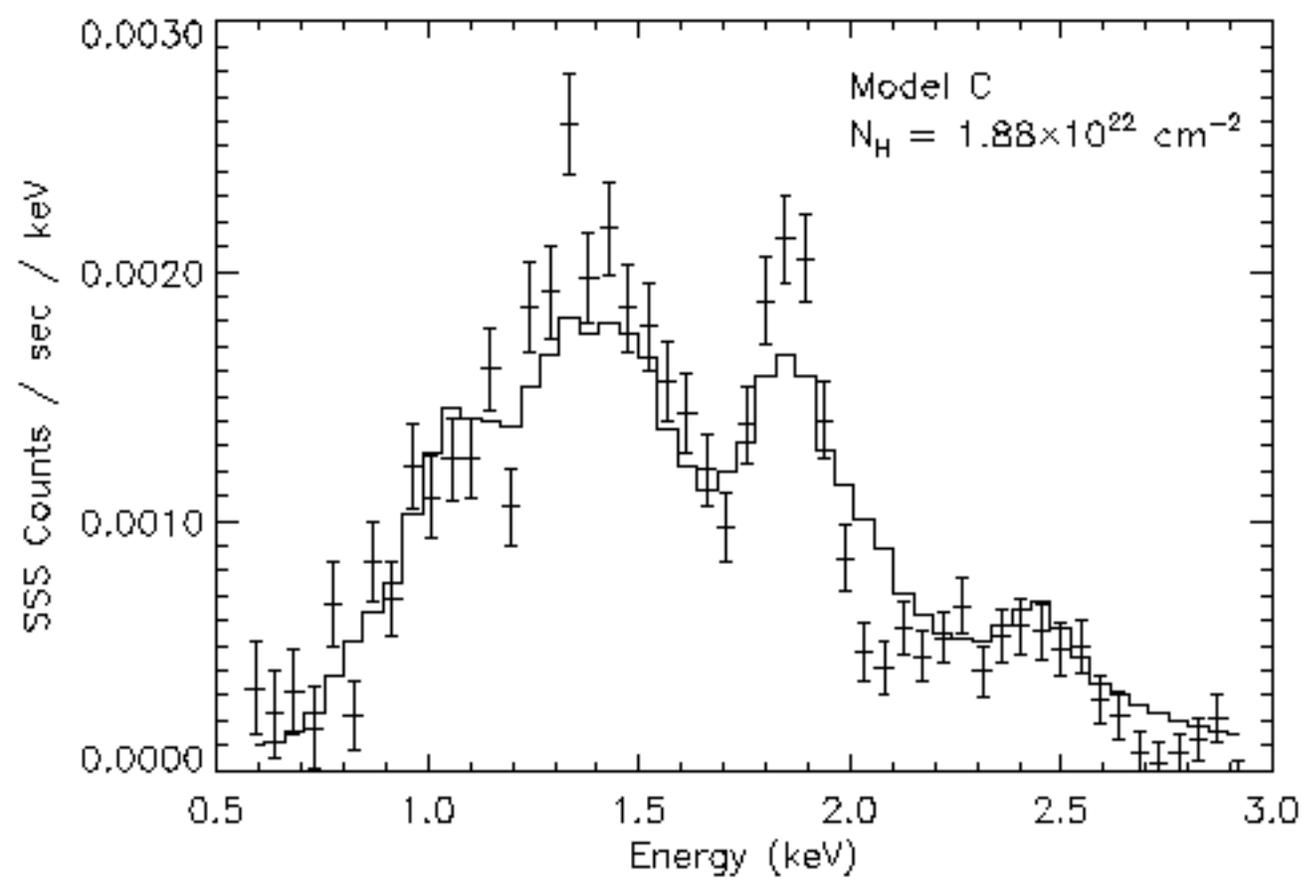
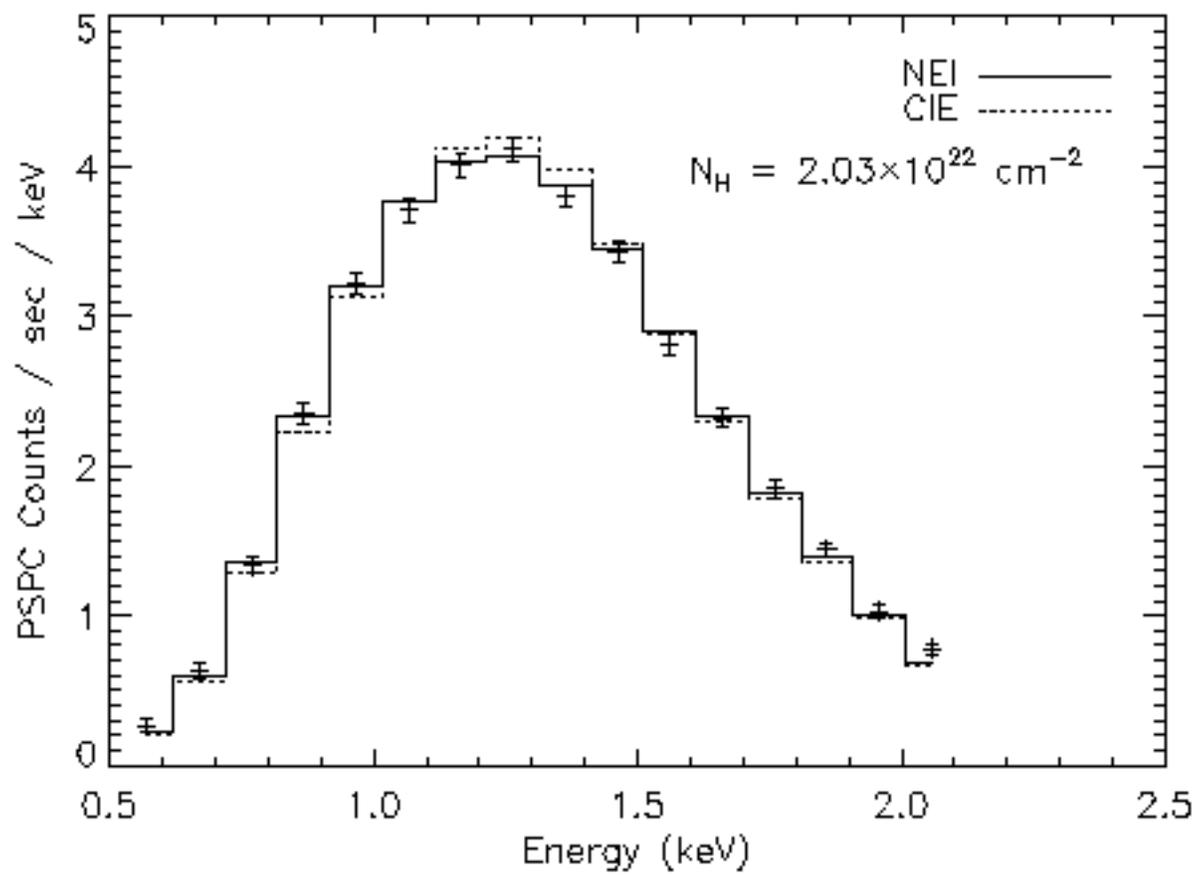
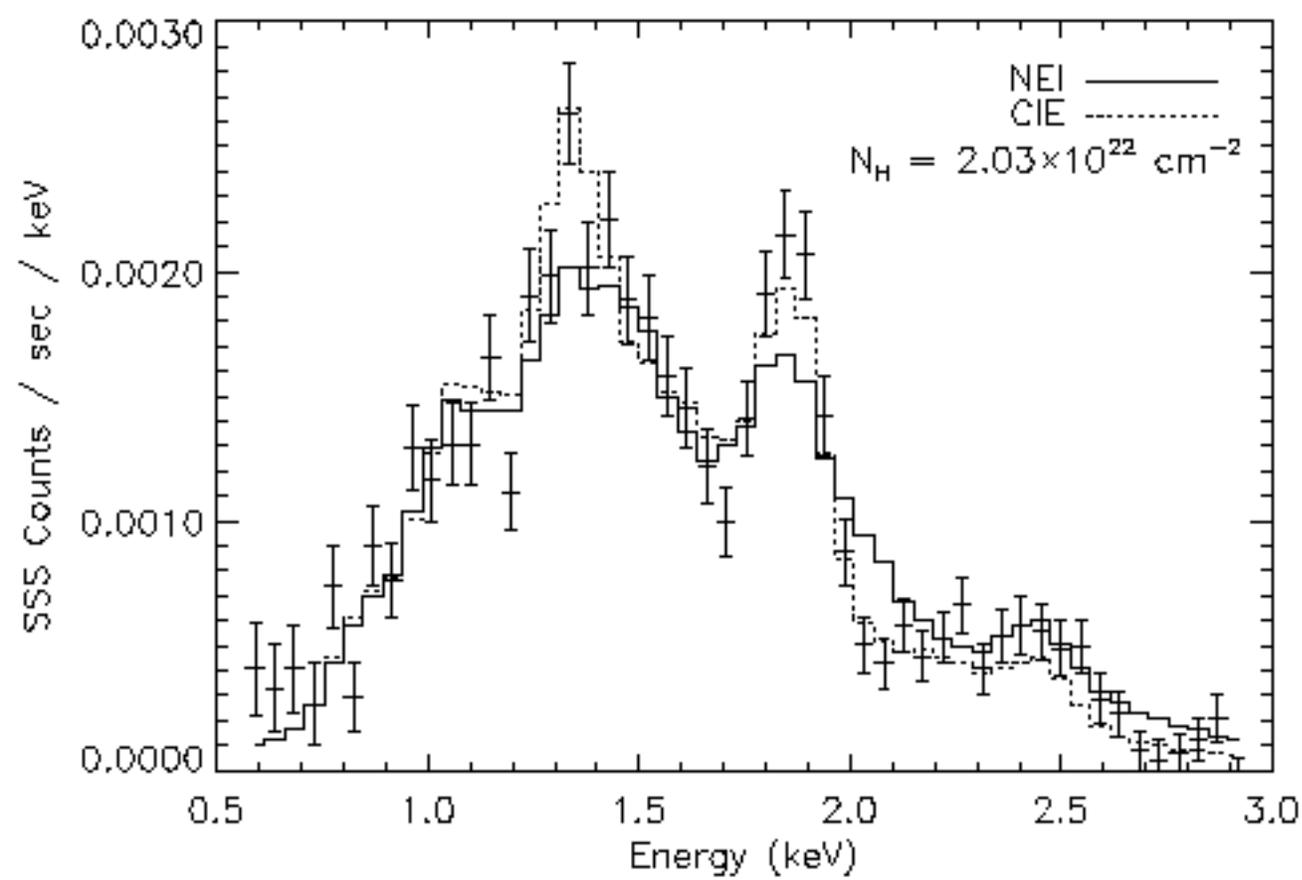
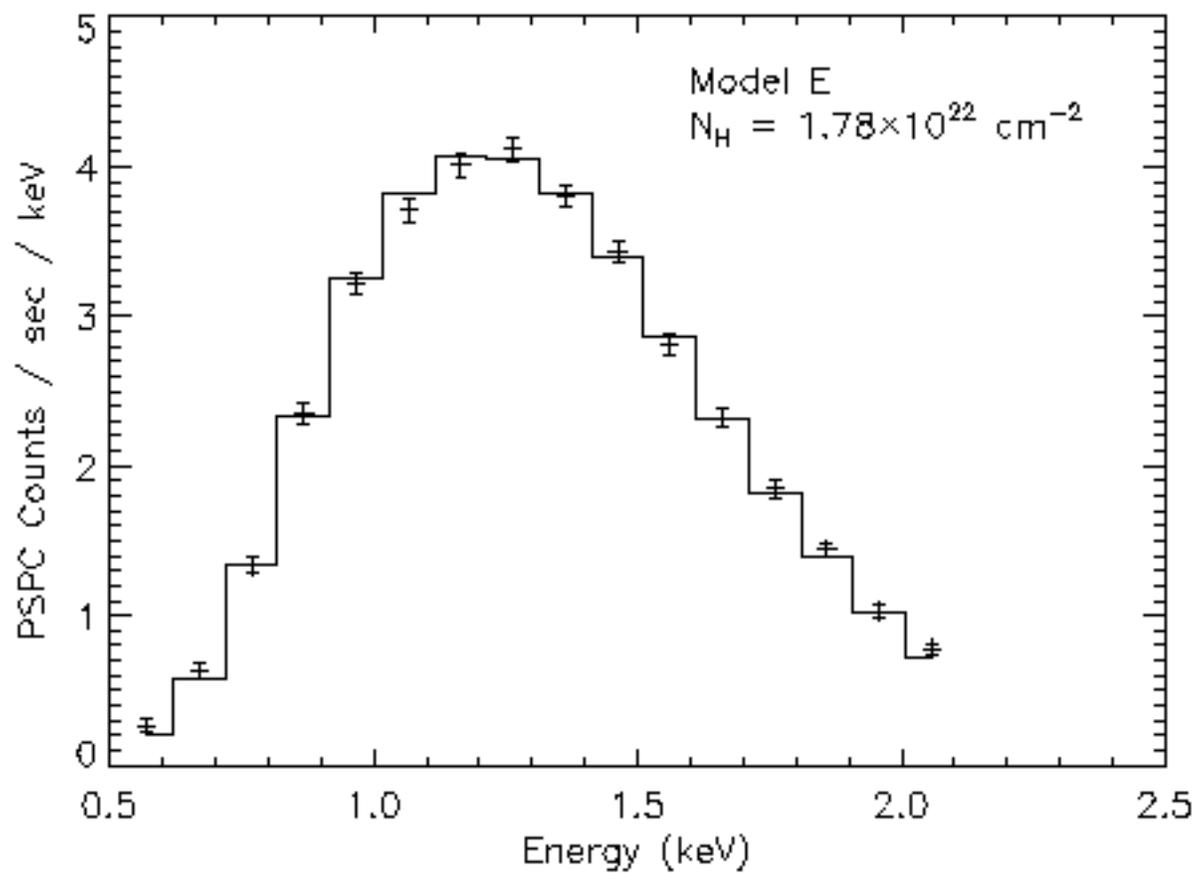
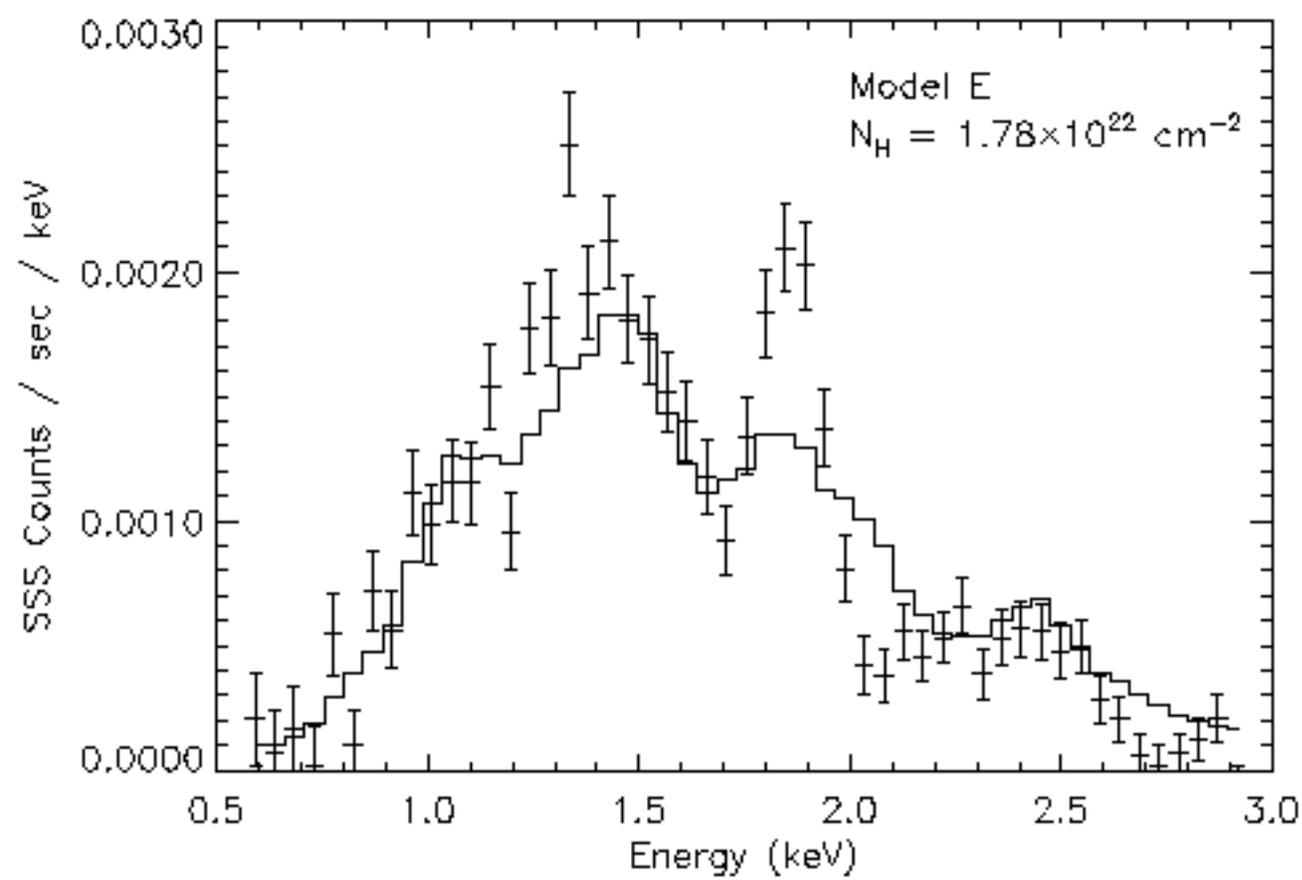
\end{document}